\documentclass[a4paper,11pt]{article}
\pdfoutput=1 

\usepackage{color}
\usepackage[dvipsnames]{xcolor}
\usepackage{jheppub} 

\usepackage{here}

\preprint{RUP-26-31}

\title{Quasinormal modes, complex null geodesics and singularity resolution in regular AdS black holes}

\author[a]{Heng-Yu Chen,}
\author[b]{Yasuaki Hikida}
\author[c]{and Yasutaka Koga}

\affiliation[a]{Department of Physics, National Taiwan University, Taipei 10617, Taiwan }
\affiliation[b]{Department of Information and Computer Science, Osaka Institute of Technology, Kitayama, Hirakata, Osaka 573-0196, Japan}
\affiliation[c]{Department of Physics, Rikkyo University, Toshima, Tokyo 171-8501, Japan}

\emailAdd{heng.yu.chen@phys.ntu.edu.tw}
\emailAdd{yasuaki.hikida@oit.ac.jp}
\emailAdd{yasutaka.koga@rikkyo.ac.jp}
 
\abstract{The black hole singularity is hidden behind its horizon, and it is difficult to determine whether it is resolved by quantum gravity or stringy effects. For an AdS-Schwarzschild black hole, it is known that the two-point function of the dual conformal field theory has a singularity associated with a null geodesic that bounces off the black hole singularity. In this paper, we investigate how the resolution of the black hole singularity is encoded in the dual conformal field theory by considering regular AdS black holes as regularized geometries. 
Using WKB analysis, we compute the quasinormal modes of regular AdS black holes and read off the dual two-point functions from the modes. 
We find that the singularity corresponding to the bouncing null geodesic for singular black holes is replaced by a smooth bump associated with a complex, rather than real, null geodesic.
The location of the complex geodesic singularity is supported by analyzing the structure of the operator product expansion for the two-point function. Our results provide a holographic signature of black hole singularity resolution and clarify how information about the regular interior is encoded in boundary correlation functions.}

\begin{document}
\maketitle
\flushbottom

\section{Introduction}

General relativity predicts the existence of singularities inside black hole horizons, and their existence may indicate an inconsistency in the gravitational theory.
It is thus widely believed that such a singularity is resolved in a consistent theory of quantum gravity, such as superstring theory. However, the mechanism responsible for singularity resolution remains unclear.
In this paper, we investigate issues related to the resolution of black hole singularities in the context of the AdS/CFT correspondence. It is known that the structure of an asymptotically anti-de Sitter (AdS) geometry can be examined from the dual conformal field theory (CFT). In particular, the two-point function of a scalar operator in the dual CFT has singularities associated with bulk null geodesics in the geometry, and we can extract information about the geometry from the singularity structure of the dual CFT correlator, see, e.g., \cite{Hubeny:2006yu,Maldacena:2015iua,Dodelson:2020lal,Dodelson:2023nnr}.
In our previous works \cite{Chen:2025cee,Chen:2025ywj,Chen:2025jbf} (see also \cite{Koga:2026tqx}), we examined exotic compact objects that resemble black holes outside the photon sphere but in which the region including the horizon is replaced by a horizonless geometry. In this paper, we instead consider a geometry with a horizon, where the black hole singularity is resolved by modifying the geometry inside the horizon.

\begin{figure}
 \centering
 \includegraphics[height=5cm]{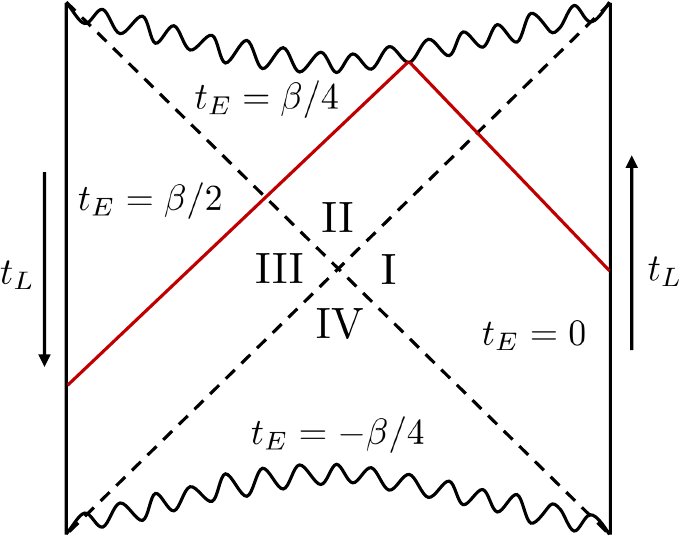}
 \caption{ The Penrose diagram for an AdS-Schwarzschild black hole. We consider the maximally extended geometry and use a complex time coordinate $t = t_L + i t_E$. In each region, $t_E$ takes a constant value, as indicated in the figure. The red line represents a null geodesic that bounces off the black hole singularity.
 }
 \label{fig:Penrose}
\end{figure}

Let us first consider the AdS-Schwarzschild spacetime with black hole singularity.
It is known that the singularity can be probed from the dual CFT even though it is hidden behind the horizon \cite{Kraus:2002iv,Fidkowski:2003nf,Festuccia:2005pi,Amado:2008hw}. 
There are two boundaries in the geometry, and we consider the two-point function of scalar operators $\mathcal{O}$ inserted on the opposite boundaries. For a scalar operator with a large conformal dimension $\Delta$, the two-point function can be evaluated in terms of the bulk spacelike geodesic connecting the two insertion points. When the bulk geodesic becomes null, the corresponding two-point function develops a singularity. The null geodesic bounces off the black hole singularity, and in this way one can infer the existence of the singularity from the CFT correlator, see figure~\ref{fig:Penrose}.
For finite $\Delta$, the two-point function can be computed from a bulk scalar field satisfying the Klein-Gordon equation on the background geometry and the singularities related to the null bouncing geodesics have been observed.
See, e.g., 
\cite{Ceplak:2024bja,Buric:2025fye,Afkhami-Jeddi:2025wra,Ceplak:2025dds,Dodelson:2025jff,Chakravarty:2025ncy,Jia:2025jbi,AliAhmad:2026wem,Araya:2026shz,Giombi:2026kdz,Grozdanov:2026cut,Jia:2026ryl,Hartnoll:2026vhu,Xiao:2026pir,Coviello:2026tol,Ceplak:2026ard,Coviello:2026hdc,Komargodski:2026oob,Batra:2026gbq,Buric:2026sit,Chakravarty:2026fdm}
for recent developments.

The singularity of the AdS-Schwarzschild black hole may be resolved by quantum gravity or stringy effects. We use a regular AdS black hole to represent a regularized geometry.
In the case of asymptotically flat backgrounds, there have been many works on regular black hole geometries, such as those by Bardeen \cite{Bardeen}, Hayward \cite{Hayward:2005gi}, Dymnikova \cite{Dymnikova:1992ux}, and others. The geometry is described by a metric which is very close to that of the singular black hole geometry outside the horizon but is regular at its core. The metric is not a solution of pure Einstein gravity. 
However, it has recently been shown that such a geometry can arise as a solution to a quasi-topological gravity theory incorporating higher-order curvature corrections \cite{Bueno:2024dgm}. This analysis has been extended to the case with a negative cosmological constant in \cite{Hennigar:2025yqm,Aguayo:2025xfi}. Due to the simplicity of its metric, we focus on the Hayward-type regular AdS black holes throughout this paper.
We first analyze a regular AdS black brane solution and then extend the analysis to the case of a regular AdS black hole with a spherical horizon.

We holographically compute the thermal two-point functions of scalar operators $\mathcal{O}$ using the gravitational description of regular AdS black holes.
We evaluate the two-point function by applying the thermal product formula of \cite{Dodelson:2023vrw,Dodelson:2025jff}, in which the two-point function is reproduced from the quasinormal modes (QNMs) of the dual black hole geometry. Thus, the task reduces to evaluating the QNMs of regular AdS black holes. We are interested in high-frequency modes, which provide leading contributions to the singularity of CFT two-point function. The asymptotic spectra of QNMs for asymptotically flat black holes have been computed from the properties of WKB wave functions for the Klein-Gordon equation \cite{Motl:2003cd,Andersson:2003fh}. The method has been generalized to the case of asymptotically AdS black holes in \cite{Cardoso:2004up,Natario:2004jd}. 
We extend these techniques to make them applicable to regular AdS black holes and compute the asymptotic spectra of QNMs. From the spectra, we obtain the CFT two-point functions dual to the regular AdS black holes.

A main conclusion of this work is that the two-point function has singularities associated with complex null geodesics turning at complex points. We claim that this is an indication of black hole singularity resolution.
To clarify its meaning, let us consider the thermal two-point function of scalar operator $\mathcal{O}$, where the two operators are inserted at different times but the same spatial point.
We first examine the case of the AdS black brane solution. We consider the maximally extended spacetime as in figure~\ref{fig:Penrose}. The time coordinate is now set to be complex as 
$
t = t_L + i t_E 
$
with $t_L , t_E \in \mathbb{R}$.
When crossing a horizon, we perform an analytic continuation that shifts the time coordinate by $\pm i\beta/4$, where $\beta$ is the inverse temperature.
For the operators inserted at the opposite boundaries, we set $t=t_L+i\beta/2$.
We thus consider the two-point function of the form
\begin{align}\label{eq:Ct0}
C(t)= \left \langle \mathcal{O} \left(t - i \frac{\beta }{2} ,0 \right) \mathcal{O}(0 ,0) \right \rangle_\beta .
\end{align}
The subscript $\beta$ indicates that we are considering the thermal two-point function with inverse temperature $\beta$. We then perform an analytic continuation to extend the region into $\mathrm{Im} \,t>\beta/2$. 
Crossing the horizons induces a shift in time $t$ by $in\beta/2$, where $2n$ is the signed net number of horizon crossings, with the sign of each crossing determined by its direction.
Thus the boundary two-point functions are realized when $t=t_L+in\beta/2$ with $n\in\mathbb{Z}$. On the other hand, it was found
in \cite{Dodelson:2025jff} that $C(t)$ has singularities at $t=t_{mk}$ $(m , k \in \mathbb{Z})$, satisfying 
\begin{align} \label{eq:realslice}
\mathrm{Im} \,t_{mk}= (1 +n )\frac{ \beta}{2} 
\end{align}
with $ n \in\mathbb{Z}$. Thus, the singularities lie on the real sections, $\mathrm{Im}\,t=(1 +n)\beta/2$, and hence can be observed in the CFT two-point function.

We then move to the case of the regular AdS black brane.
As mentioned above, we find that the CFT two-point function has singularities associated with bulk complex null geodesics. Concretely, the CFT two-point function is shown to have singularities at $t=t_{mk\ell}$ with
$m,k, \ell \in \mathbb{Z}$ for small enough $m,k$. 
Because the geodesic is complex, the singularities are generically located away from the real sections as
\begin{align}
    \mathrm{Im}\,t_{mk\ell} \neq (1 + n)\frac{\beta_+}{2}
\end{align}
with $n\in\mathbb{Z}$, except when $m+k$ is even and $\ell=(m+k)/2$.
Here, $\beta_+$ is the inverse temperature determined by the outer horizon of the regular black brane. 
In particular, the bouncing null geodesic in the singular black brane corresponds to the case with $(m,k) = (1,0)$ and the singularities at $t=t_{10\ell}$ do not lie on the real sections, $\mathrm{Im}\,t=(1 +n)\beta_+/2$.
As argued in \cite{Dolan:2011fh}, singularities of this type are replaced by bumps in the CFT two-point function.
We show that a similar argument also holds for a regular AdS black hole with a spherical horizon.

We confirm this result by examining the operator product expansion (OPE) structure of the CFT two-point function. The operator product of two scalar operators can be expanded in terms of multi-stress-tensor operators and double-trace operators. This leads to the decomposition $G=G_\text{T}+G_{[\mathcal{OO}]}$, where the contributions from the multi-stress-tensors and the double-trace operators are denoted by $G_\text{T}$ and $G_{[\mathcal{OO}]}$, respectively.
For the AdS black brane solution,
$G_\text{T}$ has been computed in \cite{Ceplak:2024bja} by applying the method of \cite{Fitzpatrick:2019zqz}, see, e.g.,  \cite{Ceplak:2024bja,Buric:2025fye,Ceplak:2025dds,Araya:2026shz,Giombi:2026kdz}  for related works.
We apply a similar method to the regular AdS black brane and confirm the result obtained from the thermal product formula with the asymptotic spectrum of QNMs.

The organization of this paper is as follows.
In the next section, we explain how to compute the CFT two-point function  holographically and apply it to the case of AdS black brane.
We argue that a singularity in the CFT correlator is related to the null geodesic bouncing at the black hole singularity.
In section~\ref{sec:RBB}, we introduce the regular black brane solution and examine its properties. We then explain the method to compute the asymptotic spectrum of QNMs by the WKB analysis and apply the method to the case of the regular black brane. In section~\ref{sec:singularity}, we show that a singularity in the two-point function is related to a complex null geodesic.
In section~\ref{sec:RBH}, we extend the analysis to the case of a regular black hole with a spherical horizon.
In section~\ref{sec:OPE}, we confirm our results by examining the thermal two-point function  based on its OPE structure. Section~\ref{sec:conclusion} is devoted to conclusions and open problems. 
In appendix~\ref{app:Stokes}, we explain how to compute the Stokes and anti-Stokes lines and draw them for the examples of regular black holes.
In appendix~\ref{sec:Jansen}, we numerically compute QNMs for regular black holes by following the analysis in \cite{Jansen:2017oag}.
In appendix~\ref{app:ChargedBH}, we compute the asymptotic spectrum of purely imaginary QNMs for AdS-Reissner-Nordstr\"{o}m black hole by applying the WKB analysis developed in the main text.
In appendix~\ref{app:C}, we provide a list of nonzero components of $C^{i,j,k}_{b,c,d}$ appearing in section~\ref{sec:OPE}.

\section{Black hole singularity from CFT correlator}
\label{sec:2pt}

In this section, we explain how the thermal two-point function of CFT scalar operators is obtained from the information of QNMs of dual black hole geometry. We then apply the method to AdS black brane solution and observe a singularity in the CFT correlator corresponding to the null geodesic bouncing at the black hole singularity.

\subsection{Thermal two-point function and QNMs}

Except in section~\ref{sec:RBH}, we consider a $(d+1)$-dimensional geometry described by a metric of the form
\begin{align} \label{eq:metric}
ds^2 = - f(r) dt^2 + f(r)^{-1} dr^2 + r^2 d \vec x^2 ,
\end{align}
where $d \vec x^2$ denotes the flat metric on $\mathbb{R}^{d-1}$.
We begin with a black brane solution, where the metric function $f(r)$ is given by
\begin{align} \label{eq:fr}
    f(r) = \frac{r^2}{L^2} - \frac{m}{r^{d-2}} .
\end{align}
Here the AdS radius is denoted by $L$.
The metric approaches AdS spacetime in the Poincar\'{e} coordinates for $r \to \infty$.
The AdS boundary is located at $r\to \infty$ and given by $\mathbb{R}^{1,d-1}$.
The geometry has a horizon at 
\begin{align} \label{eq:horizon}
r = r_h =  (L^2m)^{1/d}, 
\end{align}
which satisfies $f(r_h) = 0$. The inverse temperature is 
\begin{align} \label{eq:invtemp}
\beta = \frac{1}{T} =  \frac{4 \pi L^2}{d r_h} =  \frac{4 \pi L^2}{d (L^2m)^{1/d}}
\end{align}
and the curvature singularity is located at $r = 0$.

We consider the maximally extended geometry, whose Penrose diagram is shown in figure~\ref{fig:Penrose}.
We can describe the entire geometry using a complex time coordinate
$
    t = t_L + i t_E
$
with $t_L , t_E \in \mathbb{R}$. 
We start from region I with $t_E = 0$, which lies outside the horizon at $r = r_h$. We analytically continue to regions II and IV by crossing the horizon, where we set $t_E = \beta/4$ in region II and $t_E = - \beta/4$ in region IV. We can move from region II to region III by crossing the horizon once more, where we set $t_E = \beta/2$.

We would like to see the signatures of the black hole singularity from dual CFT.
For this purpose, we evaluate the thermal two-point function of the dual CFT (see \eqref{eq:Ct0})
 \begin{align} \label{eq:Ct0p}
     C (t, \vec x) = \left \langle \mathcal{O} \left(t - i \frac{\beta}{2} , \vec x \right) \mathcal{O} (0, 0) \right \rangle_\beta .
 \end{align}
We consider a scalar operator $\mathcal{O}(t, \vec{x})$ dual to the bulk scalar field. The conformal dimension $\Delta$ is related to the mass of scalar field $M$ by
\begin{align}
    \Delta = \frac{d}{2}+ \nu , \quad \nu^2 = \frac{d^2}{4} + M^2 L^2 .
\end{align}
We insert the two operators at the origin $\vec x = 0$ in the spatial directions $\mathbb{R}^{d-1}$ as in \eqref{eq:Ct0}.

We consider the Fourier transform of the two-point function $C(t)$ as
\begin{align} \label{eq:Comega}
    C (\omega) = 2 \int_0^\infty dt \cos (\omega t) C(t) 
\end{align}
with $\omega \in \mathbb{C}$.
We compute $C(\omega)$ 
by solving the Klein-Gordon equation for the bulk scalar field $\phi$.
Decomposing the scalar field as
\begin{align}
    \phi = e^{- i \omega t + i \vec k \cdot \vec x}  r^{\frac{1-d}{2}} 
 \psi_{\omega  k} (r)  ,
\end{align}
the equation for the radial part reduces to
\begin{align} \label{eq:radial}
(- \partial_z^2 + V (z) - \omega^2 ) \psi_{\omega  k} = 0 .
\end{align}
Here the tortoise coordinate is introduced as
\begin{align} \label{eq:tortoise}
    z = \int_r^\infty \frac{dr'}{f(r')}
\end{align}
and the potential is given by
\begin{align} \label{eq:potential2}
    V(z)=f(r)\left[\frac{k^2}{r^2}+ \frac{\nu^2 - \frac{d^2}{4}}{L^2} +\frac{(d-3)(d-1)}{4}\frac{f(r)}{r^2}+ \frac{d-1}{2 r} \frac{d}{dr}f(r)\right] .
\end{align}
For the case of black brane solution with \eqref{eq:fr},
the potential becomes
\begin{align} 
    V(z)=f(r)\left[\frac{k^2}{r^2}+ \frac{\nu^2 - \frac14}{L^2} + \frac{(d-1)^2 m}{4 r^d}\right] .
\end{align}
We set $k=0$ throughout this paper as we are interested in the correlator of the form \eqref{eq:Ct0}.%
\footnote{Precisely speaking, the two-point function with $k=0$ is not the same as \eqref{eq:Ct0}, 
where the latter is given by an integral $C(t,\vec x) = \int dk e^{i \vec k \cdot \vec x} C(t,\vec k)$ with $\vec x = 0$. The dominant contribution comes from $k=0$, but sub-leading contributions would arise, say, in the power of singularities. Since we are interested only in the positions of singularities, we do not distinguish between them in this paper.}

In terms of the tortoise coordinate \eqref{eq:tortoise}, the horizon is located at $z \to \infty$, where the ingoing boundary condition is imposed. 
Near the AdS boundary at $z = 0$, the scalar field behaves as 
\begin{align} \label{eq:asymptotic}
    \psi_{\omega0} (z) \sim \mathcal{A} (\omega) z^{\frac12 - \nu} + \mathcal{B} (\omega) z^{\frac12 + \nu} .
\end{align}
Note that the condition $\mathcal{A}(\omega) = 0$ is the same as the one for QNMs in the AdS black brane geometry.
The thermal two-point function \eqref{eq:Comega} is then obtained as
\begin{align}
    C(\omega) = - \frac{ \text{Im} \, G_R (\omega)}{\sinh (\beta \omega/2)} , \quad 
    G_R(\omega) = \frac{\mathcal{B}(\omega)}{\mathcal{A}(\omega)} ,
\end{align}
where $G_R (\omega)$ denotes the retarded Green function of CFT scalar operator.
As argued above, the poles of the thermal two-point function in the complex $\omega$-plane should be given by $\omega = \omega_n$, where $\omega_n$ are QNMs. If $\omega_n$ are not pure imaginary, then there are other poles at $\omega = - \omega_n , \omega_n^* , - \omega_n^*$. In fact, it was shown in \cite{Dodelson:2023vrw} that the form of thermal two-point function can be fixed by the QNMs as
\begin{align} \label{eq:thermalproduct}
    C (\omega) = \frac{C(0)}{\prod_{n=1}^\infty (1 - \frac{\omega^2}{\omega_n^2})(1 - \frac{\omega^2}{(\omega_n^*)^2})} .
\end{align}
If there is a series of purely imaginary QNMs with $\omega_p$, then there are additional poles at $\omega=-\omega_p$, and the extra factor
\begin{align} \label{eq:thermalproductpm}
\frac{1}{\prod_{p=1}^\infty (1-\frac{\omega^2}{\omega_{p}^2})}
\end{align}
should be included in the two-point function \eqref{eq:thermalproduct}.
In this way, the computation of the thermal two-point function can be reduced to finding QNMs for dual black hole geometry.

\subsection{Singular AdS black brane}
\label{sec:singular}

We are interested in the singularity structure of the thermal two-point function. In order to see it, we come back to the original expression by
\begin{align} \label{eq:Fourier}
C(t)=\int_0^\infty\frac{d\omega}{\pi}\cos(\omega t)C(\omega) 
\end{align}
and use the asymptotic form of $C(\omega)$ for large $\omega$.
Let us assume that the asymptotic spectrum of QNMs takes the form
\begin{align} \label{eq:qnmform}
    \omega_n = r e^{-i \theta} n + s e^{-i \phi} 
\end{align}
for large $n$. Here we set $0 \leq r, s $ and $0 < \theta < \pi/2 , - \pi/2 < \phi < \pi/2$.
As argued in \cite{Dodelson:2025jff}, 
the two-point function can be then approximated as
\begin{align} \label{eq:thermalproduct2}
    C (\omega) \approx \frac{\omega^{2 \Delta - d}}
        {\sin \left( \frac{\pi e^{ i \theta}}{r} (\omega - se^{-i \phi }) \right) \sin \left( \frac{\pi e^{-i \theta}}{r} (\omega -  se^{i \phi}) \right)} .
\end{align}
The Fourier transform leads to
\begin{align} \label{eq:Ctsingular}
C (t) \simeq \sum_{m,k =0}^\infty  e^{\frac{2 \pi i s}{r} (k e^{i (\phi - \theta ) } - m e^{i (\theta - \phi)})} \left[ \frac{1}{(i (t - t_{mk}))^{2 \Delta - d + 1}} + \frac{1}{(- i (t + t_{mk}))^{2 \Delta - d + 1}}  \right] .
\end{align}
The singularities are given by 
\begin{align}\label{eq:singularpt}
t_{mk} = \frac{i \beta}{2} + m v_+ + kv_- , \quad
v_\pm = \pm \frac{2 \pi e^{\pm i \theta} }{r}
 \end{align}
with $m,k \in {\mathbb Z}_{\geq 0}$.

In the case of AdS black brane solution, it is known that the asymptotic spectrum of QNMs is given by \cite{Cardoso:2004up,Natario:2004jd}
\begin{align} \label{eq:qnmsads}
z_* \omega_n \simeq
\pi n + \left(\frac{\Delta}{2}-\frac{d+2}{4} \right) \pi
-\frac{i}{2}\log2
\end{align}
for large $n$. 
Here $z_*$ is defined by
\begin{align} \label{eq:z0}
z_* = \int_0^\infty \frac{dr}{f(r)} = \frac{\beta }{4}\frac{e^{\frac{i \pi }{d}}}{ \sin \left( \frac{\pi}{d}\right)} .
\end{align}
The asymptotic spectrum of QNMs can be computed by applying WKB analysis and the integral $z_*$ appears as a WKB period. 
The arrival time of null geodesic bouncing at the singularity is also computed by the (twice of) the integral \eqref{eq:z0}, i.e., $2 z_*$.
We thus have
\begin{align} \label{eq:realsection}
\text{Im} \, t_{mk} = \frac{\beta}{2} (1 + m + k) ,
\end{align}
which indicates that the singularities lie on the real sections, 
$\text{Im} \, t = \frac{\beta}{2} (1 + n) $.
The singularity located at%
\begin{align}
t - \frac{i \beta}{2} = t_{10} - \frac{i \beta}{2} =  2 z_*
\end{align}
has been identified with the one associated with a null geodesic bouncing at the singularity at $r = 0$ \cite{Dodelson:2025jff}. 
This result is consistent with the one obtained from the analysis with large conformal dimension $\Delta$ of CFT operator, where the geodesic approximation can be applied \cite{Kraus:2002iv,Fidkowski:2003nf,Festuccia:2005pi}.
From the current analysis, this identification comes from the relation between the WKB period appearing in the computation of QNMs and the arrival time of the bouncing null geodesic.

\section{Regular AdS black brane}
\label{sec:RBB}

As seen above, the null geodesic bouncing at the black hole singularity can be observed from dual CFT two-point function as a singularity.
It is natural to expect that such a singularity in the CFT correlation function disappears in the case of a regular AdS black hole without any singularity.
In this section we show this by applying the method developed in the previous section.
In the next subsection, we introduce a regular AdS black brane solution and examine its properties. In the rest of this section, we evaluate the QNMs of the regular black brane by applying WKB analysis. The results are confirmed by comparing the numerical computations explained in appendix~\ref{sec:Jansen}.

\subsection{Geometry}

In this paper, we assume that the quantum gravity and/or string effects are very small and only the region near the singularity is modified.
We thus consider a regular black hole which has horizons but not curvature singularities. 
As seen above, the singularity structure of thermal two-point function can be read off from the asymptotic spectrum of QNMs.
In order to obtain a solution to the radial equation \eqref{eq:radial} satisfying the QNM boundary conditions, we extend the radial coordinate $r$ to the complex plane, which allows us to examine the region inside the horizons.
For the analytic continuation, it is useful to use an analytic function for the metric function $f(r)$.
Regular AdS black holes were examined in 
\cite{Hennigar:2025yqm,Aguayo:2025xfi},
where several types of regular black hole geometries satisfying asymptotic AdS conditions were addressed.
Among them, we utilize the Hayward type with the metric function given by
\begin{align} \label{eq:Hayward}
f(r) =  - \frac{r^{d+2} - L^2 m r^2}{a r^d - L^2 (r^{d} + a m)} .
\end{align}
The metric function is meromorphic in $r$, which allows us to examine the geometry in detail.

Before analyzing QNMs, we examine the properties of the regular black brane solution described by the metric function \eqref{eq:Hayward}.
Setting $m =0$, the geometry becomes an AdS spacetime  with
\begin{align}
    f(r) =  \frac{r^2}{L^2 - a} \equiv \frac{r^2}{L_\text{eff}^2} .
\end{align}
Namely, the AdS radius is replaced by $L_\text{eff} = \sqrt{L^2 - a}$, which may be regarded as an effective AdS radius.
We can realize the geometry as a solution of quasi-topological gravity with higher derivative corrections to the Einstein gravity as in \cite{Bueno:2024dgm}. In the case, $a$ is regarded as a parameter organizing the higher derivative corrections.
We set $0 < a < L^2$ such that the geometry is an asymptotically AdS spacetime.

We then set $m \neq 0$. For  large $r$, the metric function can be expanded as
\begin{align}
    f(r) = \frac{r^2}{L_\text{eff}^2} - \frac{L^4 m}{L_\text{eff}^4 r^{d-2}} + \mathcal{O} (r^{2 -2d}) . 
\end{align}
Thus the geometry approaches the AdS black brane solution with \eqref{eq:fr}. 
The metric function has a zero at
\begin{align}
\label{eq:horizonRBB}
r = r_h =  (L^2m)^{1/d} = ((L_\text{eff}^2 + a) m)^{1/d} ,
\end{align}
which means that the geometry has a horizon at $r = r_h$.
The expression is the same as the one for the singular AdS black brane solution in \eqref{eq:horizon} in terms of $L$.
However, for the application of AdS/CFT correspondence, we need to change the scale, such as, $L_\text{eff} =1$, which effectively shifts the position of horizon.
The inverse temperature is computed as
\begin{align}
\label{eq:invtemprbb}
\beta = \frac{1}{T} =  \frac{4 \pi L^2}{d (L^2m)^{1/d}}=\frac{4 \pi (L_\text{eff}^2 + a)}{d ((L_\text{eff}^2 + a)m)^{1/d}} . 
\end{align}
Again the expression is the same as the one for the singular black brane solution in \eqref{eq:invtemp} in terms of $L$.

Compared with the metric function for the singular case in \eqref{eq:fr}, the most significant difference appears near $r=0$.
The metric function can be expanded near $r=0$ as
\begin{align}
    f(r) = - \frac{r^2}{a}  + \mathcal{O} (r^{d+2}) .
\end{align}
Changing the coordinates as $(t, r) \to \sqrt{a}(z , \tau)$, the metric near $\tau = 0$ becomes
\begin{align}
    ds^2 = a \left[ - \frac{d \tau ^2}{\tau^2} + \tau^2 (dz^2 + d \vec x^2 )\right] .
\end{align}
This is the metric of de Sitter (dS) spacetime in the Poincar\'{e} patch.
The curvature singularity located at $r=0$ is now replaced by the dS horizon at $\tau = 0$.

\subsection{Complex WKB analysis}

\label{sec:WKB}

We are interested in the thermal two-point function of CFT dual to a gravity theory on the regular AdS black brane.
As argued in section~\ref{sec:2pt}, the CFT two-point function can be read off once the asymptotic spectrum of QNMs for regular black hole is obtained. 
We solve the wave equation \eqref{eq:radial} subject to the ingoing boundary condition 
at the horizon $r=r_h$ and Dirichlet boundary condition $\mathcal{A} (\omega) =0$ for the expansion \eqref{eq:asymptotic} near $r \to \infty$.
As we need only the asymptotic spectrum of QNMs, we assume large negative $\text{Im} \, \omega $.
We rewrite the wave equation \eqref{eq:potential2} as
\begin{align} \label{eq:we}
    [\partial_z^2 + Q^2 (z)] \psi = 0 , \quad Q^2 (z) = \omega^2 - V(z) .
\end{align}
We extend the domain of $r$ (or $z$) to the complex plane.
For real $r$, there is a zero of the metric function $f(r)$ at $r = r_h$ in \eqref{eq:horizonRBB}.
For complex $r$, there are other zeros at
\begin{align}
\label{eq:horizonRBBcomplex}
r = r^{n}_h =  e^{2 \pi i n   /d}(L^2m)^{1/d}
\end{align}
with $n=1,2,\ldots , d-1$. The real horizon corresponds to $r_h \equiv r^0_h$ in this notation.
In the case of a singular AdS black hole, a singularity is located at $r=0$, where the metric function diverges or its inverse, $1/f(r)$, vanishes.
In the case of regular AdS black brane, there is no zero of $1/f(r)$ for real $r$ as the singularity is removed by modifying the geometry.
On the other hand, for complex $r$, the function $1/f(r)$ has zeros at
\begin{align} \label{eq:zeroRBB}
r = t_{k} = e^{(2 k + 1)\pi i/d} \left ( \frac{ a L^2 m }{ L^2 -a } \right ) ^{1/d},
\end{align}
where we choose the domain $k= -1,0,1,\ldots , d-2$.

We are considering the differential equation \eqref{eq:we} and we are going to solve it by applying the WKB approximation.
Two WKB solutions are given by
\begin{align} \label{eq:wkbsol0}
f_{1}^{(t)} = \frac{1}{\sqrt{Q(r)}} \exp \left[  i \int_t^r \frac{Q(r')}{f(r')} dr' \right] , \quad
f_{2}^{(t)} = \frac{1}{\sqrt{Q(r)}} \exp \left[ - i \int_t^r \frac{Q(r')}{f(r')} dr' \right] .
\end{align}
For large $|\omega|$, $\omega^2 $ dominates the function $Q(z)$, thus we can use the approximation $Q(z) \sim \omega$ as
\begin{align}
\label{eq:wkbsol}
f_{1}^{(t)} \sim \frac{1}{\sqrt{\omega}} \exp \left[  i \omega \int_t^r \frac{1}{f(r')} dr' \right] , \quad
f_{2}^{(t)} \sim \frac{1}{\sqrt{\omega}} \exp \left[ - i \omega \int_t^r \frac{1}{f(r')} dr' \right] .
\end{align}
We impose boundary conditions on the wave function at the AdS boundary, $r \to \infty$, and the black hole horizon, $r = r_h$. 
In order to impose the two boundary conditions, we move the wave function between the two points. 
We follow an anti-Stokes line, along which the two functions $f_1^{(t)}$ and $f_2^{(t)}$ are of the same order of magnitude, so that the coefficients of these functions can be reliably determined.

\begin{figure}[htbp]
 \centering
 \includegraphics[height=6cm]{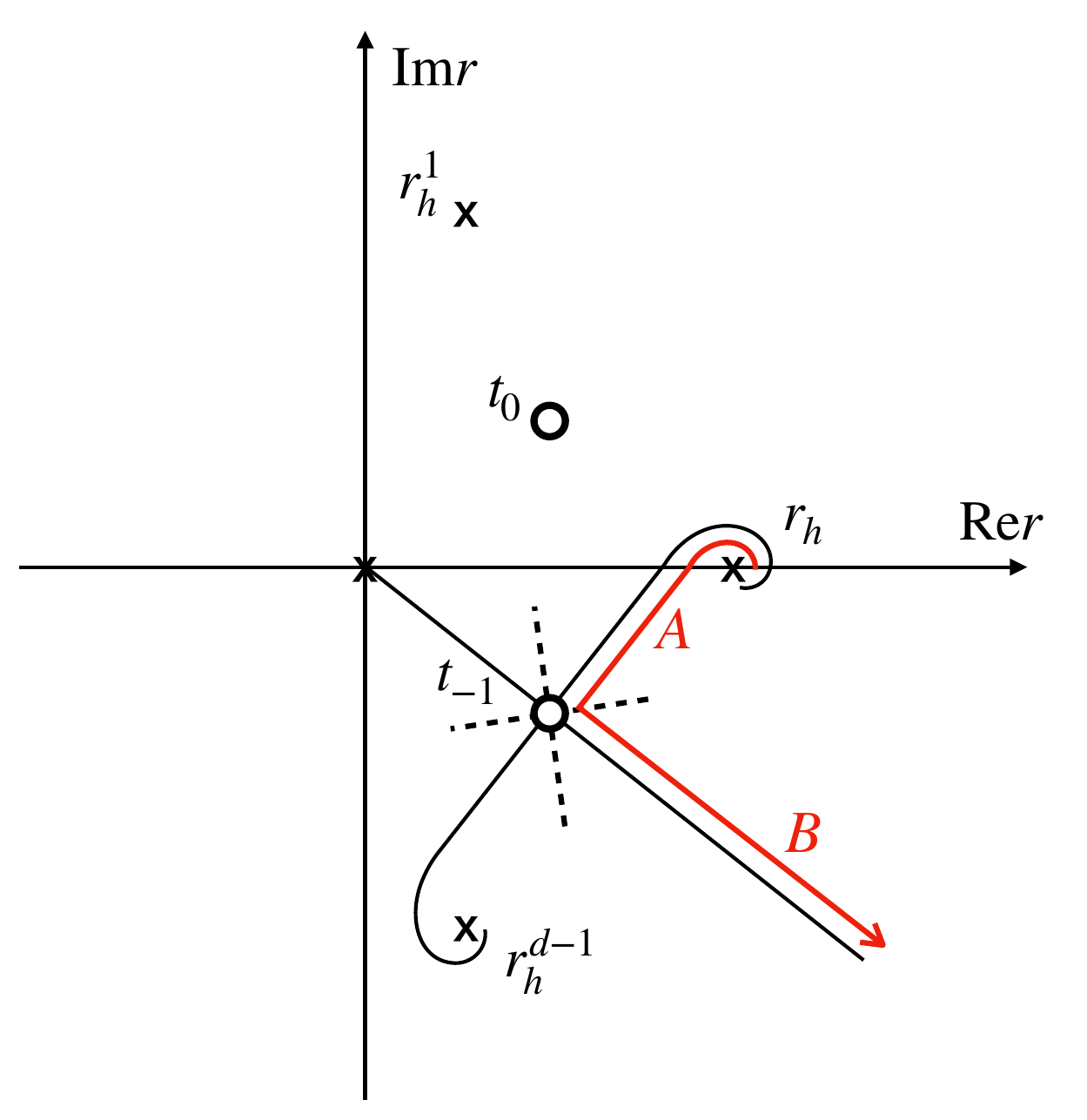}
 \caption{Stokes and anti-Stokes lines for regular AdS black brane. Here we assume that $\omega \propto \exp(- \pi i /d )$. }
 \label{fig:RBBg}
\end{figure}

We will consider two cases with $\omega \propto \exp(- i \pi  /d )$ and $\omega \in i \mathbb{R}$. For a while, we assume that  $\omega \propto \exp(- i \pi /d )$. In this case, 
 anti-Stokes lines can be drawn for the approximated expression with $Q(z) \sim \omega$ as in figure~\ref{fig:RBBg}. 
We will start from the black hole horizon at $r = r_h$ and then move along anti-Stokes line $A$ to a zero of $1/f(r)$ at $r = t_{-1}$. We then move along another anti-Stokes line $B$ to the AdS boundary $r \to \infty$. The approximation with $Q(z) \sim \omega$ is valid for a large part of $r$-complex plane.
However, there are two regions where the approximation cannot be used as $|V(z)| \gtrsim|\omega|^2 $.

The first region is near the AdS boundary with $r \gtrsim |\omega|$, where the potential and the metric function behave as
\begin{align}
    V (z) \sim \frac{f(r)}{L_\text{eff}^{2}} \left(\nu^2 - \frac{1}{4} \right) , \quad 
    f(r) \sim \frac{r^2}{L_\text{eff}^2} .
\end{align}
The radial equation for large $r$ reduces to
\begin{align} \label{eq:deads}
    \left[\partial_z^2 + \omega^2 -  \frac{1}{z^2} \left(\nu ^2 - \frac14 \right)  \right]\psi = 0 ,
\end{align}
where $z = L_\text{eff}^2/r$. The equation is the same as the one with pure AdS spacetime and analytic solution is available.

The second region is with $r \sim t_k$, where the metric function diverges as
\begin{align} \label{eq:Ak}
f(r) \sim  -\frac{A_k}{x} , \quad A_k = -\frac{L^2}{a d (L^2 -a)} t_k^3 . 
\end{align}
Here we have introduced a new coordinate  $x = r - t_{k}$.
The potential becomes
\begin{align}
V(r) \sim \frac{ B_k }{x^3}  , \quad B_k = - \frac{d-1}{2} \frac{A_k^2}{t_k}
\end{align}
and the  radial equation reduces to
\begin{align} \label{eq:rewithx}
\left[ \frac{d^2}{dx^2} - \frac{1}{x} \frac{d}{dx} +  \frac{\omega^2 x^2}{A_{k}^2} - \frac{B_{k}}{A_{k}^2 x }  \right] \psi = 0 .
\end{align}
Setting $u = \sqrt{\omega} x$, the equation becomes
\begin{align} \label{eq:rewithu3}
\left[ \frac{d^2}{du^2} - \frac{1}{u} \frac{d}{du} +  \frac{u^2}{A_k^2} + \frac{\epsilon_k}{u } \right] \psi = 0  ,
\end{align}
where
\begin{align}
     \epsilon_k = \frac{d-1}{2 t_{k} \sqrt{\omega}}  .
\end{align}
In the next subsection, we solve the equation to first order in $\epsilon_k$ using the Born approximation.

\subsection{Connection formula}
\label{app:born}

As mentioned above, we would like to move from contour $A$ to contour $B$ via the point $t_{-1}$. However, the approximation with $Q(z) \sim \omega$ cannot be used for the region near the point $t_{-1}$. For the region far from the point $t_{-1}$, we can simply use the WKB solutions in \eqref{eq:wkbsol}. For the regions with $r \sim t_k$, the radial equation is given by \eqref{eq:rewithu3}.
Connecting the WKB solutions in contours $A$ and $B$, we obtain how the wave function in contour $A$ is mapped to the wave function in  contour $B$.

We would like to find a solution to the equation \eqref{eq:rewithu3} to first order in $\epsilon_k$. 
At zeroth order in $\epsilon_k$, two independent solutions are given by
\begin{align}
    \psi_+^{(0)} = e^{ i u^2/(2 A_k)} , \quad 
     \psi_-^{(0)} = e^{- i u^2/ (2 A_k)} .
\end{align}
Let us consider an ansatz of the form
\begin{align}
    \psi (u) = a_\psi (u)  \psi_+^{(0)} + b_\psi (u) \psi_-^{(0)} .
\end{align}
Using the ambiguity of $a_\psi  ,b _\psi $, we assign the constraint
\begin{align} \label{eq:const}
    \frac{d a_\psi  }{du} \psi_+^{(0)}  +  \frac{d b_\psi  }{du} \psi_-^{(0)}  = 0 .
\end{align}
Noticing that 
\begin{align}
    \frac{d}{du} \psi_\pm^{(0)}  = \pm \frac{ i u }{A_k}  \psi_\pm^{(0)} {} ,
\end{align}
we find 
from the equation \eqref{eq:rewithu3} that
\begin{align} \label{eq:ab}
   \frac{d a _\psi }{du}  \frac{d}{du}\psi_+^{(0)}  + \frac{d b_\psi  }{du}  \frac{d}{du} \psi_-^{(0)} =
  \frac{iu}{A_k}  \frac{d a _\psi }{du} \psi_+^{(0)} - \frac{i u}{A_k} \frac{d b_\psi  }{du} \psi_-^{(0)}
    = - \frac{\epsilon_k}{u} \psi .
\end{align}
Solving \eqref{eq:const} and \eqref{eq:ab}, we find
\begin{align} \label{eq:variation}
     \frac{d a_\psi }{du} =  \frac{i A_k \epsilon_k}{2 u^2}\frac{\psi}{ \psi_+^{(0)} } , \quad    \frac{d b_\psi }{du} = - \frac{i A_k \epsilon_k}{2 u^2}\frac{\psi }{\psi_-^{(0)}} .
\end{align}
Notice that $(\psi_\pm^{(0)})^{-1} = \psi_\mp^{(0)}$.

We start from contour $A$  in figure~\ref{fig:RBBg}. 
We assume that the wave function along contour $A$ is given by $\psi^{A} = \psi^A_+$, where we have introduced $\psi^A_\pm $ such that $\psi^A_\pm \sim \psi_\pm^{(0)}$ at zeroth order in $\epsilon_k$.
At first order in $\epsilon_k$, the coefficients $a,b$ satisfy
\begin{align} \label{eq:derivativeab}
  \frac{d a_\psi  }{du}  =  \frac{i A_k \epsilon_k}{2 u^2} , \quad \frac{d b_\psi }{du}  =  - \frac{i A_k \epsilon_k}{2 u^2} e^{i u^2/A_k} . 
\end{align}
We set
\begin{align}
    a_\psi (u_A) = 1 , \quad b_\psi (u_A) = 0 
\end{align}
at a point $u_A$ on contour $A$.
In order to connect to the WKB wave function for large $r$, we may set $u_A \to \infty_A$. 
Integrating \eqref{eq:derivativeab}, we have
\begin{align}
    a_\psi (u) = 1 + \frac{i A_k \epsilon_k}{2} \int_{\infty_A}^u \frac{ds}{s^2} \, , \quad
    b_\psi (u) =  -  \frac{i A_k \epsilon_k}{2}\int_{\infty_A}^u \frac{ e^{i s^2 /A_k} }{s^2} ds .
\end{align}
The integrals are evaluated as
\begin{align}
    \int_{\infty_A}^u \frac{ds}{s^2} = - \frac{1}{u} 
\end{align}
and
\begin{align}
\begin{aligned}
 I_A(u) \equiv  \int_{\infty_A}^u \frac{ e^{i s^2 /A_k} }{s^2} ds &= \int^u_{\infty_A} \left[ \frac{2 i}{A} e^{i s^2/A_k}   - \frac{d}{ds}\frac{e^{i s^2/A_k}}{s}\right] ds \\
 &= - \frac{e^{i u^2/A_k}}{u} +  \frac{2 i}{A_k}\int^u_{\infty_A}  e^{i s^2/A_k} ds .
 \end{aligned}
\end{align}
For $u \to 0$, we find
\begin{align}
    I_A (u) = - \frac{1}{u} + F_A  + \frac{i}{A_k} u + \cdots, \quad F_A = - \frac{2 i}{A} \int_{0}^{\infty_A} e^{i s^2/A} ds = \sqrt{\frac{\pi}{A}} e^{- \pi i /4} .
\end{align}
We thus obtain
\begin{align} \label{eq:psiAplus}
    \psi^A_+ \sim \left[ 1  - \frac{i A_k \epsilon_k}{2 u} \right] \psi_+^{(0)} - \frac{i A_k \epsilon_k}{2} I_A (u) \psi_-^{(0)}.
\end{align}
It can be expanded near $u =0$ as
\begin{align}
    \psi^A_+ =  1 - \frac{i A_k \epsilon_k}{2} F_A - \epsilon u + \left( \frac{i}{2A_k} - \frac{\epsilon_k}{4} F_A \right) u^2 + \cdots .
\end{align}

From contour $A$ to contour $B$, we perform a clockwise rotation by $ \pi /2 $.
We compute $\psi^{B}$ to first order in $\epsilon_k$.
The non-trivial part comes from the analytic continuation of the integral $I_A(u)$.
We keep the coordinate $u$ fixed and exchange the contours from $A$ to $B$ by replacing the lower endpoint $\infty_A$ with $\infty_B$:
\begin{align}
\begin{aligned}
I_B(u)
&\equiv \int_{\infty_B}^{u}
\frac{e^{is^2/A_k}}{s^2}\,ds = -\frac{e^{iu^2/A_k}}{u}
+\frac{2i}{A_k}\int_{\infty_B}^{u}e^{is^2/A_k}\,ds \\
&
= -\frac{1}{u}-F_A+\frac{i}{A_k}u+\mathcal{O}(u^3).
\end{aligned}
\end{align}
Here we choose the asymptotic rays
$s=-\sqrt{A_k}\,y$ on contour $A$ and
$s=i\sqrt{A_k}\,y$ on contour $B$ with $y\geq0$.
From this, we have
\begin{align} \label{eq:psiBplus}
\psi_+^B
\sim
\left[
1 - \frac{iA\epsilon}{2u}
\right] \psi_+^{(0)}
 -  \frac{iA\epsilon}{2} I_B (u) \psi^{(0)}_- .
\end{align}
Near $u \to 0$, they can be expanded as
\begin{align}
\psi_+^B
&=
1 + \frac{iA_k\epsilon_k}{2} F_A - \epsilon_k u
+
\left(\frac{i}{2A_k} + \frac{\epsilon_k F_A}{4} \right) u^2
+ \cdots .
\end{align}
The wave function along contour $B$ is then expressed as
\begin{align}
    \psi^B = \alpha_\psi \psi^B_+ + \beta_\psi \psi^B_- ,
\end{align}
where
\begin{align}
     \alpha_\psi = 1 + c _\psi \epsilon_k + \mathcal{O} (\epsilon_k^2) , \quad \beta_\psi = d_\psi  \epsilon _k+ \mathcal{O} (\epsilon_k^2) .
\end{align}
As $\beta_\psi$ is of order $\epsilon_k$, we only need $\psi^B_- = \psi^{(0)}_- + \mathcal{O} (\epsilon_k) $.

We would like to connect the wave functions $\psi^A$ and $\psi^B$ near $u = 0$. Solving
\begin{align}
    \psi_+^A = (1 + c_\psi  \epsilon_k) \psi^B_+ + d _\psi \epsilon_k \psi^B_- ,
\end{align}
we find $c_\psi =0$ and 
\begin{align}
    d _\psi = - i A_k F_A = - \sqrt{\pi A_k} e^{i \pi /4} .
\end{align}
This is the connection formula between the wave functions along contours $A$ and $B$ via the point $r = t_k$.
Above we assumed that $\psi_+^A$ dominates $\psi_-^A$ along contour $A$. If $\psi_-^L$ dominates $\psi_+^L$ along a contour, say, $L$, then we should properly modify the above connection formula.
As we will not use the explicit formula below, we do not repeat the analysis here.

\subsection{QNMs with nonzero real part}

We move to the computation of the asymptotic spectrum of QNMs for the regular black brane.
Applying the connection formula in the previous subsection, we obtain the condition for the wave function satisfying the two boundary conditions at $r=r_h$ and $r= \infty$. 
In the limit $a\to 0$, the analysis should reduce to the case of a singular AdS black brane.
In that case, we have found that the QNMs are given by $\omega \propto \exp(- i \pi /d)$, see \eqref{eq:qnmsads}.
Here we assume that $\omega \propto \exp (- i \pi /d)$ even for the case of regular AdS black brane.
It will turn out that a series of QNMs exists for this type of $\omega$.
Stokes and anti-Stokes lines can be drawn as in figure~\ref{fig:RBBg}.

We start with the wave function along contour $A$.
We impose the ingoing boundary condition at the horizon $r = r_h$.
Near the horizon, the metric function 
can be expanded as (see \eqref{eq:invtemp})
\begin{align} 
    f(r) \sim \frac{4 \pi}{\beta} (r - r_h) , \quad \beta =\left . \frac{4 \pi}{\frac{d}{dr}f(r) } \right|_{r=r_h} = \frac{4 \pi L^2}{ d (L^2 m)^{1/d}} ,
\end{align}
which leads to the following behavior of tortoise coordinate defined in \eqref{eq:tortoise}:
\begin{align}
 z = \int_r^\infty \frac{dr'}{f(r')} \sim - \frac{\beta}{4 \pi} \log (r - r_h) .
\end{align}
The ingoing boundary condition requires the wave function to behave as
\begin{align}
  \psi \sim  e^{ i \omega z }\sim  (r - r_h)^{-  i \beta \omega  /(4 \pi) } 
\end{align}
near $r = r_h$. 
The wave function satisfying the condition is given by 
\begin{align}
    \psi = f_2^{(t_{-1})}
\end{align}
in terms of WKB solutions defined in \eqref{eq:wkbsol}.

We analytically continue the wave function from contour $A$ to contour $B$ via the point $r=t_{-1}$. For this purpose, we apply the connection formula obtained in the previous subsection.
The wave function along contour $A$ behaves as 
\begin{align}
    \psi = f_2^{(t_{-1})} \sim \psi_+^A
\end{align}
near $r = t_{-1}$. Here the wave function $\psi_+^A$ is given by \eqref{eq:psiAplus} up to first order in $\epsilon_{-1}$.
This wave function is mapped to the one in contour $B$ as
\begin{align} \label{eq:connectiontminus}
    \psi_+^A = \psi^B_+ + \beta_{-1} \psi^B_- + \mathcal{O} (\epsilon_{-1}^2) ,
\end{align}
where
\begin{align}
    \beta_{-1} = - i A_{-1} F_A \epsilon_{-1} = - \sqrt{\pi A_{-1}} e^{i \pi /4} \epsilon_{-1}.
\end{align}
The wave function $\psi_+^B$ in contour $B$ is given by \eqref{eq:psiBplus} up to first order in $\epsilon_{-1}$.
Away from the region near $r = t_{-1}$, the wave function is obtained by a linear combination of WKB solutions in \eqref{eq:wkbsol}. From \eqref{eq:connectiontminus}, the wave function is obtained as 
\begin{align} \label{eq:psilarger0}
    \psi \sim f_2^{(t_{-1})} + \beta_{-1} f_1^{(t_{-1})} .
\end{align}
For large but finite $r$, the wave function behaves as
\begin{align} \label{eq:psilarger}
\psi \sim  e^{ - i \omega (z_{*} - L_\text{eff}^2/r)}  + \beta_{-1} e^{ i \omega (z_{*} - L_\text{eff}^2/r)} .
\end{align}
Here we set
\begin{align}
z_* = \int_{t_{-1}}^{\infty} \frac{1}{f(r)} dr .
\end{align}

Near the AdS boundary, where $r \gtrsim |\omega|$, the WKB solutions in \eqref{eq:wkbsol} are no longer valid, since $V(z) \propto r^2$ becomes large compared to $\omega^2$ in the expression for $Q^2(z)$ defined in \eqref{eq:we}.
Instead, we can use a solution to the wave equation 
\eqref{eq:deads}.
We assign Dirichlet boundary condition $\mathcal{A} (\omega) =0$ for the expansion \eqref{eq:asymptotic} for $r \to \infty$.
The solution satisfying the boundary condition is given by
\begin{align} \label{eq:Bessel}
\psi \sim c_- r^{-1/2} J_{\nu} \left(\omega  \frac{L_\text{eff}^2}{r} \right) 
\end{align}
with a coefficient $c_-$.
Notice that
\begin{align}
    J_\nu (x) \sim \sqrt{\frac{2}{\pi x}}\cos \left( x - \frac{\nu \pi}{2} -\frac{\pi}{4}\right)
\end{align}
for $x \gg 1$. 
We connect the wave function \eqref{eq:Bessel} with that in \eqref{eq:psilarger0} by comparing the asymptotic behaviors of these wave functions.
We find that the solution exists only when
\begin{align} \label{eq:qnmrbb}
\omega_n z_* = n \pi + \left( \frac{\nu}{2} - \frac34 \right)\pi + \frac{i}{2} \log \beta_{-1} ,
\end{align}
where
\begin{align}
\beta_{-1} = \gamma_{-1} \omega^{-1/2} , \quad \gamma_{-1} = -
\sqrt{\pi A_{-1}} e^{i \pi /4} \frac{d-1}{2 t_{-1}} .
\end{align}
For large $n$, $\omega_n$ can be expanded as
\begin{align} \label{eq:omegaexpansion}
    A_\text{Q}  \omega_n \sim \pi n + B_\text{Q} \log n + C_\text{Q} ,
\end{align}
where
\begin{align} \label{eq:ABC}
 A_\text{Q} = z_* , \quad 
   B_\text{Q} = - \frac{i}{4}  , \quad C_\text{Q} = \left( \frac{\nu}{2}-\frac34 \right)\pi +\frac{i}{2}\log\gamma_{-1} -\frac{i}{4}\log\left(\frac{\pi}{z_*}\right).
\end{align}

We numerically compute the QNMs of the regular black brane using the Mathematica package provided by \cite{Jansen:2017oag}, as summarized in appendix~\ref{sec:Jansen}.
A comparison between the WKB analysis and numerical results is presented in table~\ref{tab:QNMrbb}.
We find good agreement for $A_\text{Q}$, which will be used to examine the singularity structure below. For the other parameters, $B_\text{Q}$ and $C_\text{Q}$, we find only qualitative agreement.%
\footnote{If we exclude small values of $n$ from the numerical data used for fitting, the agreement improves. A similar tendency can be observed in other cases as well.}
We have solved the wave equation only up to first order in $\epsilon_{-1}$, while the numerical computations have been performed for relatively small values of $n$.
In order to obtain quantitative agreement for $B_\text{Q}$ and $C_\text{Q}$, we need to include higher-order corrections in $\epsilon_{-1}$ or numerically compute QNMs for larger values of $n$.
\begin{table}[htbp]
   \centering
  \[
\begin{array}{|c|c|c|c|c|c|c|}
\hline
~ & \multicolumn{3}{|c|}{\text{WKB prediction}} & \multicolumn{3}{|c|}{\text{Numerical fit}}  \\
\hline
(d,a)  & A_\text{Q} & B_\text{Q} & C_\text{Q}& A_\text{Q} & B_\text{Q} & C_\text{Q} \\ \hline
(3,0.05)
& 0.63 + 1.10 i
& - 0.25 i
& 0.39 + 0.24 i
& 0.64+1.11i
& 0.10-0.17i
& 0.46+0.04i 
\\
(3,0.10)
& 0.56 + 0.97 i
& - 0.25 i
& 0.39 - 0.12 i
& 0.55+0.98i
& 0.02-0.05i
& 0.47-0.09i
\\
(4,0.05)
& 0.82 + 0.82 i
& - 0.25 i
& 1.18 + 0.43 i
& 0.82+0.82i
& 0.06-0.17i
& 1.21+0.13i
\\
(4,0.10)
& 0.74 + 0.74 i
& - 0.25 i
& 1.18 + 0.29 i
& 0.74+0.74i
& 0.09-0.10i
& 1.19+0.03i
\\
(5,0.1)
& 0.81 + 0.59 i
& - 0.25 i
& 1.96 + 0.40 i
& 0.82+0.60i
& 0.15-0.10i
& 1.90+0.07i
\\
(5,0.5)
& 0.36 + 0.26 i
& - 0.25 i
& 1.96 + 0.05 i
& 0.36+0.27i
& 0.13-0.12i
& 1.81-0.13i
\\
(6,0.1)
& 0.85 + 0.49 i
& - 0.25 i
& 2.75 + 0.49 i
& 0.85+0.50i
& 0.21-0.08i
& 2.57+0.10i
\\
(6,0.5)
& 0.39 + 0.23 i
& - 0.25 i
& 2.75 + 0.13 i
& 0.39+0.23i
& 0.21-0.10i
& 2.48-0.10i
\\
\hline
\end{array}
\]
    \caption{
    Comparison of QNMs with $\omega_n \propto \exp(-\pi i/d)$ for the regular black brane obtained from WKB analysis and numerical computations. We set $\nu=d/2$.
    }
    \label{tab:QNMrbb}

\end{table}

We find that taking the limit $a\to 0$ does not reproduce the result for the AdS black brane solution. Let us examine the reason for this discrepancy. Above, we analyzed the wave functions to first order in $1/\sqrt{\omega}$ for large $|\omega|$ and obtained the connection formula \eqref{eq:connectiontminus}.
For large $|\omega|$, the coefficient $\beta_{-1}$ scales as
\begin{align}
|\beta_{-1}|
\sim \frac{|A_{-1}|}{|t_{-1}|\sqrt{|\omega|}}.
\end{align}
With $L$ and $m$ fixed, the small-$a$ scalings are given by
\begin{align}
|t_{-1}| \propto a^{1/d}, \quad
|\beta_{-1}| \propto
|\omega|^{-1/2}a^{-(d-1)/(2d)}.
\end{align}
Thus, the perturbative treatment used to derive the connection formula is valid only when
\begin{align}
|\omega|a^{(d-1)/d} \gg 1.
\end{align}
Therefore, we cannot simply take the limit $a\to 0$ to recover the result for the AdS black brane solution. In order to properly examine this limit, we need to go beyond the first-order approximation.

\subsection{Purely imaginary QNMs}

The numerical computations suggest that there are purely imaginary QNMs in addition to those satisfying $\omega_n \propto \exp(- i \pi /d)$, which were examined in the previous subsection.
The situation is quite similar to the case of AdS-Reissner-Nordstr\"{o}m black hole, where the numerical computations in \cite{Jansen:2017oag} found purely imaginary QNMs along with those satisfying $\omega_n \propto \exp(- i \pi /d)$. The latter type has been explained from WKB analysis in \cite{Natario:2004jd}.
In this subsection, we obtain the asymptotic spectrum of purely imaginary QNMs for regular black brane from WKB analysis. In appendix~\ref{app:ChargedBH}, a similar analysis is applied to AdS-Reissner-Nordstr\"{o}m black hole.

\begin{figure}
 \centering
 \includegraphics[height=6cm]{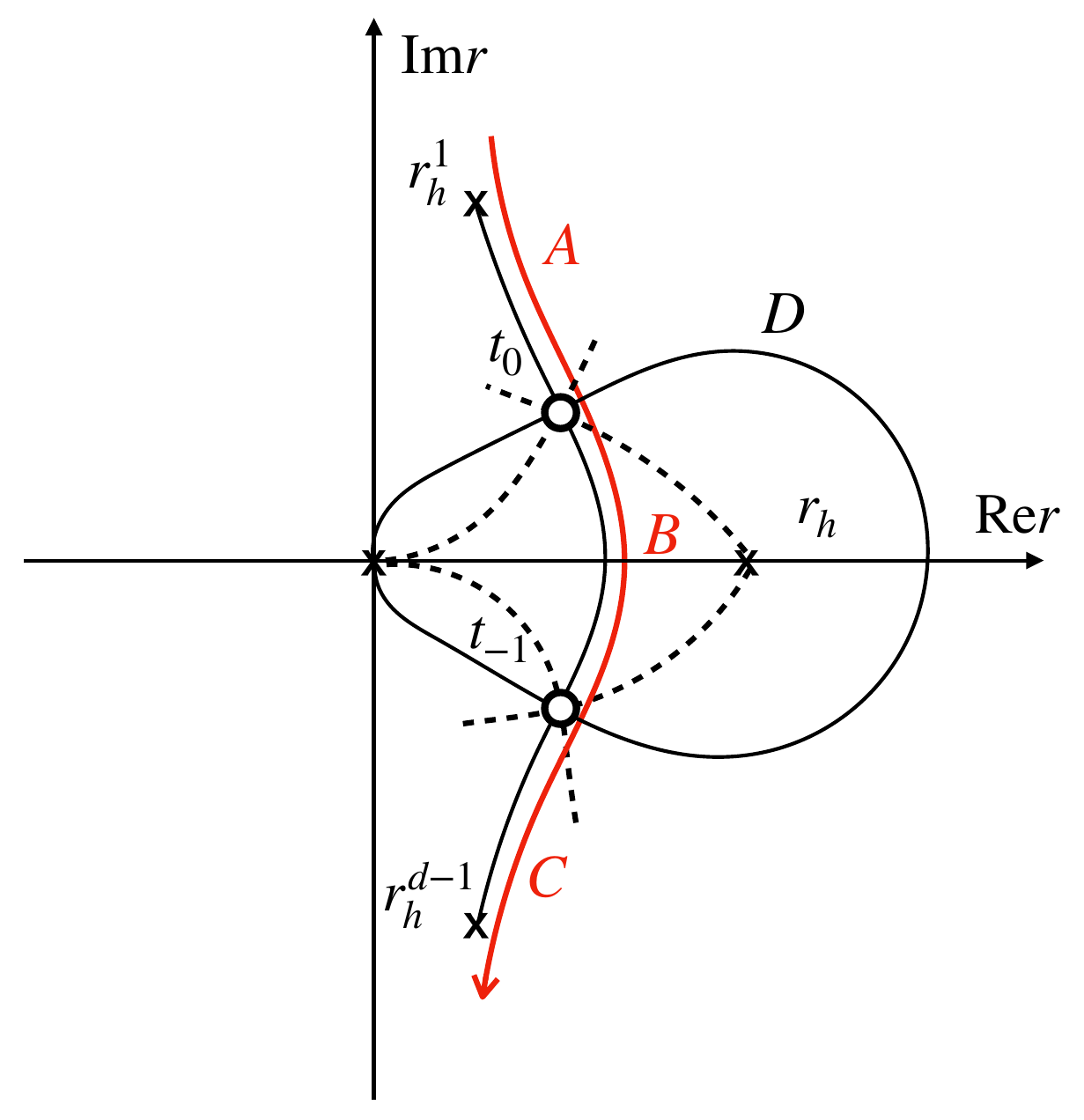}
 \caption{Stokes and anti-Stokes lines for regular black brane. Here $\omega$ is set to be pure imaginary.}
 \label{fig:RBBgp}
\end{figure}

As in the previous case, we will analytically continue the wave function along anti-Stokes lines. For pure imaginary $\omega$,  Stokes and anti-Stokes lines are drawn as in figure~\ref{fig:RBBgp}.
We start from the wave function near the AdS boundary satisfying the Dirichlet boundary condition. For large $r$, the tortoise coordinate $z$ in \eqref{eq:tortoise} is given by $z = L_\text{eff}^2/r$. Going away from the AdS boundary, the effect of potential becomes negligible and the wave function can be given by a linear combination of plane waves $\exp (\pm i \omega z)$. 
The two terms are of the same order of magnitude when $\mathrm{Im}\,\omega z = 0$, i.e., when $\mathrm{Re}\,r = 0$.
We assign the Dirichlet boundary at $r = i \infty$.
We then move the region with finite $r$, which leads to
\begin{align}
\psi = c_1 f_1^{(\infty)} + c_2 f_2^{(\infty)} .
\end{align}
Determining the coefficients $c_1$ and $c_2$ may be subtle, since we are not following an anti-Stokes line. Nevertheless, we can argue that they are of the same order of magnitude.

We  change the basis from $f_{1,2}^{(\infty)}$ to $f_{1,2}^{(t_0)}$ and put the wave function along anti-Stokes line $A$.
The change of basis provides phase factors
\begin{align}
   f_{1}^{(\infty)} = e^{-i \omega \int_{t_0}^\infty \frac{dr}{f(r)} }f_{1}^{(t_0)} , \quad 
      f_{2}^{(\infty)} = e^{i \omega \int_{t_0}^\infty \frac{dr}{f(r)} }f_{2}^{(t_0)} .
\end{align}
The integrand in the phase factors has poles at complex horizons and 
the value of integral depends on how integration contour avoids the poles. Near the complex horizons, the potential vanishes and the plane waves can be used. 
In other words, there are no Stokes phenomena near the complex horizons.
We consider a contour starting from $r = i \infty$ to anti-Stokes line $A$ by avoiding the complex horizon at $r = r^1_h$ clockwise.
We will explain the reason why we choose the integration contour later.
We define the integral as 
\begin{align}
z^{(1)}_* = \int^{\infty}_{t_0} \frac{dr}{f(r)} .
\end{align}
We can show that $z^{(1)}_* \propto \exp (- i \pi /d)$ for the choice of integration contour.
Due to the phase factors, the wave function becomes
\begin{align}
\psi = \tilde c_1  f_1^{(t_0)} + \tilde c_2   f_2^{(t_0)} 
\end{align}
with the coefficients
\begin{align}
\tilde c_1 \propto e^{-i \omega  z^{(1)}_* }  , \quad 
\tilde c_2 \propto e^{ i \omega  z^{(1)}_* }  .
\end{align}
For large large $- \text{Im} \, \omega$ with $z^{(1)}_* \propto \exp (- i \pi/d) $, we can show that 
$\tilde c_1  \ll \tilde c_2 $.
Changing the overall factor, we can set $\tilde c_1 = 0 $ and $\tilde c_2 = 1$ for simplicity.
However, we should keep in mind that these expressions are valid only up to corrections of order $\exp (-2i\omega z_*^{(1)})$.

We fist analytically continue the wave function from contour $A$ to contour $D$ via a singular point $r = t_0$.
The term $f_1^{(t_0)}$ is dominant in the region between contours $A$ and $D$, and hence no Stokes phenomenon occurs for the wave function $\psi = f_2^{(t_0)}$.
We then move to a point along contour $B$. The term $f_2^{(t_0)}$ is dominant in the region between contours $D$ and $B$.
The connection formula has been already obtained in the previous sections.
Along contour $B$, we have
\begin{align} \label{eq:pipsi}
\psi = \beta_0 f_1^{(t_0)}  +  f_2^{(t_0)} .
\end{align}
The factor $\beta_0$ is given by
\begin{align}
    \beta_0 = - i A_0 \epsilon_0 F_0 , \quad \epsilon_0 = \frac{d-1}{2 t_0 \sqrt{\omega}} , \quad F_0 = \sqrt{\frac{\pi}{A_0}} e^{-\pi i /4} .
\end{align}
Here the expressions of $t_k$ and $A_k$ can be found in \eqref{eq:zeroRBB} and \eqref{eq:Ak}, respectively.%


We move the wave function along anti-Stokes line $B$ to the region near $r =t_{-1}$. We then change the basis from $f_{1,2}^{(t_0)}$ to $f_{1,2}^{(t_{-1})}$ as
\begin{align}
\psi =  \beta_0  e^{i  \omega \gamma_* } f_1^{(t_{-1})} +  e^{ - i \omega \gamma_* } f_2^{(t_{-1})} ,
\end{align}
where 
\begin{align} \label{eq:gamma*}
\gamma_* \, (\equiv i \Gamma_*)= \int^{t_{-1}}_{t_{0}} \frac{1 }{f(r)}  dr.
\end{align}
We then analytically continue the wave function to a point along contour $C$ via a singular point $r = t_{-1}$.
Applying a similar analysis near the point $r = t_0$, the wave function along contour $C$ can be obtained as 
\begin{align}
\psi =   (\beta_0  e^{ i  \omega\gamma_* } + \beta_{-1}  e^{ - i  \omega\gamma_* } )f_1^{(t_{-1})} + e^{ - i \omega \gamma_* }  f_2^{(t_{-1})} .
\end{align}
The factor $\beta_{-1}$ is given by
\begin{align}
    \beta_{-1} = - i A_{-1} \epsilon_{-1} F_{-1} , \quad \epsilon_{-1} = \frac{d-1}{2 t_{-1} \sqrt{\omega}} , \quad F_{-1} = \sqrt{\frac{\pi}{A_{-1}}} e^{-\pi i /4} .
\end{align}

We analytically continue the wave function along contour $C$ to the vicinity of the complex horizon at $r=r_h^{d-1}$.
We would like to go to $- i \infty$ furthermore but there is no anti-Stokes line connecting the complex horizon at $r = r_h^{d-1}$ to the infinity $- i \infty$.
Following a similar argument as before, we assume that 
\begin{align}
\psi = c_1 ' f_1^{(\infty)} + c_2 'f_2^{(\infty)}
\end{align}
near $r = - i \infty$. Moving the ansatz to the region near $r = r_h^{d-1}$ along contour $C$, we have
\begin{align}
\psi = \tilde c_1 ' f_1^{(t_{-1})} + \tilde c_2  ' f_2^{(t_{-1})} , \quad \tilde c_1 ' \propto e^{i \omega  z^{(-1)}_* }  , \quad 
\tilde c_2 \propto e^{ - i \omega  z^{(-1)}_* } .
\end{align}
The integral
\begin{align}
z^{(-1)}_* \,(\equiv z_*)= \int^{\infty}_{t_{-1}} \frac{dr}{f(r)}
\end{align}
is performed along the integration contour where the complex horizon $r = r_h^{d-1}$ is avoided clockwise.
We can show that $z_*^{(-1)} \propto \exp (i \pi/d)$ and therefore $\tilde c_1 ' \ll \tilde c_2 ' $ for large $- \text{Im} \, \omega$.
We thus require that 
\begin{align}
\beta_0 e^{  i \omega \gamma_* }  + \beta_{-1}   e^{ - i \omega \gamma_* }  = 0 ,
\end{align}
which leads to
\begin{align} \label{eq:omegaexpansion2}
    A_\text{Q} \omega \sim n  \pi + C_\text{Q} \quad (n=1,2,\ldots)
\end{align}
with
\begin{align} \label{eq:AC}
A_\text{Q} =  \gamma_* , \quad 
 C_\text{Q} =  - \frac{ \pi}{2 d}  .
\end{align}
A comparison with the numerical results obtained in appendix~\ref{sec:Jansen} is presented in table \ref{tab:pmQNMrbb}.
We find good agreement for the parameter $A_\text{Q}$ between the two approaches, while only qualitative agreement is found for $C_\text{Q}$.
As before, we have included only the leading-order corrections in $1/\sqrt{\omega}$ near the singular points $r=t_0,t_{-1}$.
Moreover, we have neglected terms of order $\exp (-2i\omega z_*^{(1)})$.
To obtain better quantitative agreement, we need to deal with finite $\omega$ in the WKB analysis  or perform numerical computations for larger values of $n$.

Here we would like to comment on our choice of contour.
We start from $r=iR$ with $1\ll R\ll|\omega|$, follow the anti-Stokes lines $A$, $B$, and $C$ to reach $r=-iR$, and close the contour along the arc $r=Re^{i\theta}$ with $-\pi/2<\theta<\pi/2$.
Near the complex horizons, we choose the contour such that it encloses the outer horizon at $r=r_h$.
The monodromy around the contour is thus $\exp (-2\pi\omega/f'(r_h))$.
On the other hand, the wave function is given by a linear combination of $\exp (\pm i\omega z)$, and the dominant contribution comes from $\exp(-i\omega z)$ for large negative $\mathrm{Im}\,\omega$ and $\mathrm{Re}\,z>0$.
The coordinate $z=-\int^r dr'/f(r')$ also has a nontrivial monodromy around $r=r_h$, and the term $\exp(-i\omega z)$ yields the same monodromy $\exp(-2\pi\omega/f'(r_h))$, where $f'(r_h)=\left.\frac{d}{dr}f(r)\right|_{r=r_h}$.
This is in contrast to the asymptotically flat case, where the term $\exp(-i\omega z)$ yields the monodromy $\exp(2\pi\omega/f'(r_h))$, and matching the monodromies leads to a nontrivial quantization condition for $\omega$.
\begin{table}[htbp]
    \centering
 \[
\begin{array}{|c|c|c|c|c|}
\hline
~ & \multicolumn{2}{|c|}{\text{WKB prediction}} & \multicolumn{2}{|c|}{\text{Numerical fit}}  \\
\hline
(d,a) & A_\text{Q} & C_\text{Q} & A_\text{Q} & C_\text{Q} \\ \hline
(3,0.05)
& ~~ 0.44i ~~
& ~ - 0.52 ~ 
& ~~ 0.44i ~~ 
& ~ - 0.86 ~
\\
(3,0.10)
& 0.70i
& - 0.52
& 0.68i
& - 0.87
\\
(4,0.05)
& 0.24i
& - 0.39
& 0.23i
& -0.84
\\
(4,0.10)
& 0.39i
& - 0.39
& 0.38
& -0.82
\\
(5,0.1)
& 0.27i
& - 0.31
& 0.26i
& -0.70
\\
(5,0.5)
& 0.92i
& - 0.31
& 0.93i
& -0.16
\\
(6,0.1)
& 0.20i
& - 0.26
& 0.19i
& -0.70
\\
(6,0.5)
& 0.72i
& - 0.26
& 0.73i
& -0.13
\\
\hline
\end{array}
\]
        \caption{Comparison of purely imaginary QNMs for the regular black brane obtained from WKB analysis and numerical computations. We set $\nu = d/2$.}
    \label{tab:pmQNMrbb}
\end{table}

\section{Singularity resolution}
\label{sec:singularity}

We would like to examine the thermal two-point function, which can be obtained from the QNM spectrum as in \eqref{eq:thermalproduct}.
The asymptotic behavior of the two-point function at large $\omega$ can be determined from the asymptotic spectrum of QNMs.
We found that there are two series of QNMs.
One series satisfies $\omega_n \propto \exp(- i \pi /d)$, and its asymptotic spectrum is given by \eqref{eq:omegaexpansion} with \eqref{eq:ABC}. 
This series contains terms proportional to $\log n$, which are absent in \eqref{eq:qnmform}.
The other series consists of purely imaginary QNMs, whose asymptotic spectrum is given by \eqref{eq:omegaexpansion2} with \eqref{eq:AC}.
A series of purely imaginary QNMs also exists in the case of an AdS-Reissner-Nordstr\"{o}m black hole, and their contribution can be incorporated as in \eqref{eq:thermalproductpm}. 
In the next subsection, we first explain how to modify the previous analysis to account for the additional $\log n$ terms in \eqref{eq:omegaexpansion}. 
In subsection~\ref{sec:resolution}, we show how the resolution of the singularity can be understood from the perspective of the dual CFT two-point function.

\subsection{Singularities from two-point function}
\label{app:log}

We would like to obtain the asymptotic form of the two-point function $C(\omega)$ by applying the thermal product formula \eqref{eq:thermalproduct}. Among the QNMs, we consider only the type with $\omega_n \propto \exp(- i \pi /d)$ as in \eqref{eq:omegaexpansion} with \eqref{eq:ABC} in this subsection. The purely imaginary QNMs as in  \eqref{eq:omegaexpansion} with \eqref{eq:ABC} can be incorporated by directly applying the result in \cite{Dodelson:2025jff}.
We thus use the QNMs given by%
\footnote{Note that $\log n$ in \eqref{eq:omegaexpansion} is replaced with $\log \omega_n$ in \eqref{eq:AQsim}.}
\begin{align} \label{eq:AQsim}
 A_\text{Q} \omega_n \sim n \pi  - \frac{i}{4} \log \omega_n +  C ' _\text{Q}  ,
\end{align}
where
\begin{align} \label{eq:AQCQ}
    A_\text{Q} = z_* , \quad C_\text{Q} ' = \left( \frac{\nu}{2} - \frac{3}{4}\right) \pi + \frac{i}{2} \log \gamma_{-1} .
\end{align}
We may rewrite it as
\begin{align}\label{eq:omegaabc} 
\omega_n \sim a_\text{Q} n + b_\text{Q} \log n + c _\text{Q} ,
\end{align}
where
\begin{align}
a_\text{Q} = \frac{\pi}{A _\text{Q}} , \quad b_\text{Q} = - \frac{i}{4 A_\text{Q}} , \quad c_\text{Q} = \frac{C_\text{Q} '}{A_\text{Q}} - \frac{i}{4 A _\text{Q}} \log \frac{\pi}{A_\text{Q}} .
\end{align}
We further assume that 
\begin{align}
a_\text{Q} = r e^{- i \theta} , \quad c_\text{Q} = s e^{- i \phi}
\end{align}
 with $r > 0$, $s > 0$ and $0 < \theta < \pi /2$.
This implies that
\begin{align} \label{eq:AQ}
    A_Q = \frac{\pi}{r} e^{i \theta} .
\end{align}

Let us introduce a function of $\omega$ by
\begin{align} \label{eq:N}
 N(\omega) = \frac{1}{\pi} \left(A_\text{Q} \omega - C '_\text{Q} + \frac{i}{4} \log \omega \right) .
\end{align}
The two-point function $C(\omega)$ has poles at $\omega = \omega_n$, which may be represented by the conditions $N(\omega) = n$ with $n = 1,2,\ldots$.
We may express the two-point function $C(\omega)$ as
\begin{align}
C(\omega) = D (\omega) \bar D (\omega) ,
\end{align}
where $ D(\omega)$ $(\bar D (\omega))$ is defined such as to reproduce the structure of poles at $\omega_n , - \omega_n$ ($\omega^*_n , - \omega^*_n$).
Then the two-point function can be expressed from the structure of poles as
\begin{align} \label{eq:D}
\begin{aligned}
D (\omega ) &= E (\omega) \Gamma (1 - N (\omega)) \Gamma (1 - N (- \omega)) \\
& = \frac{\pi E (\omega)}{\sin[\pi N(\omega)]} R (\omega) ,
\end{aligned}
\end{align}
where
\begin{align} \label{eq:R}
R (\omega)  =  \frac{ \Gamma (1 - N (- \omega))}{\Gamma (N (\omega))}  .
\end{align}
The overall normalization is expressed by $E(\omega)$, which is a smooth function of $\omega$.
The factor $C(0)$ in \eqref{eq:thermalproduct} does not depend on $\omega$. However, the residues of $D(\omega)$ generically depend on $\omega$,  and the factor $E(\omega)$ is chosen so as to cancel the $\omega$-dependence.

We need the asymptotic form of $C(\omega)$ for $\omega > 0$.
For large $\omega$, the factor $R (\omega)$ can be approximated as
\begin{align}
R (\omega) = \frac{\Gamma (\frac{A_Q \omega}{\pi} + \frac{5}{4} + \frac{C_Q '}{\pi} - \frac{i}{4 \pi} \log \omega )}{\Gamma(\frac{A_Q \omega}{\pi} - \frac{C '_Q}{\pi} +\frac{i}{4 \pi} \log \omega )} \sim \left(\frac{A _Q \omega}{\pi} \right)^{\frac{5}{4} + \frac{2 C' _Q}{\pi} - \frac{i}{2 \pi} \log \omega} 
\end{align}
with $N(\omega)$ defined in \eqref{eq:N}.
We have used the Sterling's formula
\begin{align}
    \Gamma(z) \sim \sqrt{\frac{2 \pi}{z}} \left( \frac{z}{e}\right)^z
\end{align}
for large $|z|$.
Thus we find
\begin{align}
R (\omega) \bar R (\omega) \propto \omega^{\frac{5}{2} + \frac{4}{\pi} \text{Re}\, C'_\text{Q} + \frac{\theta}{\pi}} ,
\end{align}
where we have used \eqref{eq:AQ}.
Since $\text{Im}\, A_\text{Q} > 0$, we have
\begin{align} \label{eq:sinexp}
\frac{1}{\sin (\pi N(\omega))} = - 2 i \sum_{m=0}^\infty e^{i (2 m + 1) \pi N(\omega)} , \quad
\frac{1}{\sin (\pi \bar N(\omega))} =  2 i \sum_{k=0}^\infty e^{-i (2 k + 1) \pi \bar N(\omega)} .
 \end{align}
The leading contribution is thus
\begin{align}
\frac{1}{\sin (n N(\omega) ) \sin (n \bar N (\omega))} \sim 4 e^{2 \text{Im}\, C '_\text{Q} - \frac{2 \pi \sin \theta}{r} \omega} \omega^{-1/2} \propto  e^{ - \frac{2 \pi \sin \theta}{r} \omega} \omega^{-1/2} .
\end{align}
Combining the above expressions, the asymptotic behavior for $\omega \gg 1$ are given by
\begin{align} \label{eq:CEE}
    C(\omega) \propto E(\omega) \bar E(\omega)  e^{- \frac{2 \pi \sin \theta}{r} \omega} \omega^{2 + \frac{4}{\pi}  \text{Re} \, C'_\text{Q} + \frac{\theta}{\pi}} .
\end{align}

In order to extract the expression of $C(\omega)$ for large $\omega$, we need the asymptotic behavior of the factor $E(\omega)$.
Let us assume that the function $C(\omega)$ takes the form
\begin{align}
C(\omega) \sim e^{- B_\text{asy} \omega} \omega^{G_\text{asy}}
\end{align}
for large $\omega$.  
From the assumption, we have
\begin{align} \label{eq:ST}
\partial_\omega \log C(\omega) \sim - B_\text{asy} + \frac{G_\text{asy}}{\omega} .
\end{align}
Applying the formula \eqref{eq:thermalproduct}, we find 
\begin{align} \label{eq:ST2}
\partial_\omega \log C(\omega) = - \sum_{n=1}^\infty F(n) , \quad F(n) =   \frac{1}{\omega - \omega_n} +  \frac{1}{\omega + \omega_n} +  \frac{1}{\omega - \omega_n^*} +  \frac{1}{\omega + \omega_n^*}   .
\end{align}
We thus compute the asymptotic expression for the sum in \eqref{eq:ST2} and extract $B_\text{asy}$ and $G_\text{asy}$.
We use the expression for $\omega_n$ given in \eqref{eq:omegaabc}.
The case with $b_\text{Q}=0$ was already analyzed in section~5 of \cite{Dodelson:2023vrw} as reviewed below.
We extend the analysis by including the contribution from the term proportional to $b_\text{Q}$.

We evaluate the sum in  \eqref{eq:ST2} up to order $\mathcal{O}(\omega^{-2})$.
We use the expansion
\begin{align}
\frac{1}{\omega - \omega_n} = \frac{1}{\omega - a_\text{Q} n} + \frac{b_\text{Q} \log n + c_\text{Q}}{(\omega - a_\text{Q} n)^2} + \cdots ,
\end{align}
where we have substituted the expression for $\omega_n$ in \eqref{eq:omegaabc}.
In order to evaluate the sum, we apply the Euler–Maclaurin formula
\begin{align}
\sum_{n=1}^\infty F(n) = \int_1^\infty F(x) dx + \frac{1}{2}F(1) + \mathcal{O} (\omega^{-2}) .
\end{align}
Defining $F_0 (n)$ by replacing $\omega_n$ in $F (n)$ with $a_\text{Q} n$, we find
\begin{align}
\begin{aligned}
- \int_1^\infty F_0 (x) dx  - \frac{1}{2} F(1) &= - \int_0^\infty F_0 (x) dx + \int_0^1 F_0(x) dx - \frac{1}{2} F(1) \\
&=  - \frac{2 \pi \sin \theta}{r} + \frac{2} {\omega} + \mathcal{O}(\omega^{-2}) .
\end{aligned}
\end{align}
We further have
\begin{align}
\begin{aligned}
- \int_1^\infty \ dx \left[ \frac{c_\text{Q} }{(\omega - a_\text{Q} x)^2} - \frac{c_\text{Q} }{(\omega + a_\text{Q} x)^2} + \frac{c_\text{Q} ^*}{(\omega - a_\text{Q} ^* x)^2}  - \frac{c_\text{Q} ^*}{(\omega + a_\text{Q} ^* x)^2} \right]  \\
 =  \frac{4}{\omega} \text{Re} \frac{c_\text{Q} }{a_\text{Q} } + \mathcal{O } (\omega^{-2})= \frac{4 s  \cos (\theta - \phi)}{r \omega}+ \mathcal{O } (\omega^{-2}) .
\end{aligned}
\end{align}

We then examine the effects of nonzero $b_\text{Q}$.
The integrals left are of the form
\begin{align}
- \int_1^{\infty} d x \frac{b _\text{Q} \log x}{ (\omega - a_\text{Q} x)^2} =  \frac{b_\text{Q}}{a _\text{Q} \omega } \log \frac{\omega - a_\text{Q} }{- a_\text{Q} } .
\end{align}
Combining terms from $\omega_n$ and $- \omega_n$, we have
\begin{align}
\begin{aligned}
- \int_1^{\infty} d x \left[ \frac{b_\text{Q}  \log x}{ (w - a_\text{Q} x)^2} - \frac{b_\text{Q}  \log x}{ (w + a_\text{Q} x)^2} \right] &= \frac{b_\text{Q}}{a_\text{Q} \omega } \left[  \log \frac{\omega - a_\text{Q} }{- a_\text{Q} } + \log \frac{\omega + a_\text{Q} }{ a_\text{Q} } \right] \\
&\sim \frac{b_\text{Q} }{a_\text{Q}  \omega} \left[ 2 \log \frac{\omega}{r} - i (\pi - 2 \theta) \right] .
\end{aligned}
\end{align}
Adding its complex conjugate, we arrive at 
\begin{align}
 \frac{4}{ \omega}  \log \frac{\omega}{r}  \text{Re}\, \frac{b_Q }{a_Q }   + \frac{2 (\pi - 2 \theta)}{\omega} \text{Im}\, \frac{b_Q }{a_Q } + \mathcal{O} \left(\frac{\log \omega}{\omega^2}\right)  . 
\end{align}
In the current case, $b_\text{Q} /a_\text{Q}  = - i /(4 \pi)$, and hence
\begin{align}
\partial_\omega \log C(\omega) \sim - \frac{2 \pi \sin \theta}{r} + \frac{1}{\omega} \left[ \frac{3}{2} + \frac{\theta}{\pi} + \frac{4 s}r{ \cos (\phi - \theta)}\right]  ,
\end{align}
which leads to 
\begin{align} \label{eq:canonical}
 C(\omega ) \propto e^{- \frac{2 \pi \sin \theta}{r}}  \omega^{\frac{3}{2} + \frac{\theta}{\pi} + \frac{4s}{r} \cos (\phi - \theta)} ,
\end{align}
see \eqref{eq:ST}. Compared with \eqref{eq:CEE}, we have
\begin{align}
E (\omega)\bar E (\omega) \propto \omega^{- \frac12 + 4 \text{Re}\, \frac{c_\text{Q}}{a_\text{Q}} - \frac{4}{\pi} \text{Re}\, C '_Q } .
\end{align}
Using
\begin{align}
    \frac{c_\text{Q}}{a_\text{Q}} = \frac{1}{\pi} \left( C'_\text{Q} - \frac{i}{4} \log \frac{\pi}{A_\text{Q}} \right) , \qquad \log \frac{\pi}{A_\text{Q}} = \log r - i \theta ,
\end{align}
we find
\begin{align}
E (\omega)\bar E (\omega) \propto \omega ^{- \frac12 - \frac{\theta}{\pi}} .
\end{align}
The asymptotic expression of $C(\omega)$ is thus
\begin{align}
C(\omega) \propto \frac{\omega^{2 + \frac{4}{ \pi} \text{Re}\, C_Q '} }{\sin ((A_\text{Q})^* \omega - (C'_\text{Q})^* - \frac{i}{4} \log \omega) \sin (A _Q \omega - C_Q ' + \frac{i}{4} \log \omega) } .
\end{align}
Note that 
\begin{align}
 2 +   \frac{4}{\pi} \, \text{Re} \, C ' _\text{Q} =
  2 +   \frac{4}{\pi}  \left( \frac{\nu}{2} - \frac{5}{8} + \frac{1}{4 d}\right) \pi  = 
 2 \nu - \frac{1}{2} + \frac{1}{d} ,
\end{align}
where we have used \eqref{eq:AQCQ}.

\subsection{Complex null geodesics and singularity resolution}
\label{sec:resolution}

As explained above, the asymptotic from of the two-point function for $\omega > $ can be read off from the asymptotic spectrum of QNMs. In the previous subsection, we have examined the contributions from the QNMs satisfying $\omega_n \propto \exp (- i \pi/d)$. 
The other series of QNMs satisfies $\omega_n \in i\mathbb{R}$.
The asymptotic spectrum is given by \eqref{eq:omegaexpansion2} with \eqref{eq:AC}, i.e., 
\begin{align} 
     \omega_n \sim - i p n - i q ,
\end{align}
where
\begin{align}
    p = \frac{\pi}{\Gamma_*} , \quad q = - \frac{\pi }{2 d \Gamma_*} .
\end{align}
In this case, we need to multiply an additional factor \eqref{eq:thermalproductpm} to the two-point function.
As argued in \cite{Dodelson:2025jff} that the pole structure can be reproduced by the factor 
\begin{align}
   \prod_{\pm} \Gamma \left( \frac{\pm i \omega + q}{p} + 1 \right) \sim \frac{\pi}{\sinh \left(\frac{\pi \omega}{p} \right)} \left( \frac{\omega}{p}\right)^{1 + \frac{2 q}{p}} .
\end{align}
Thus the asymptotic form of the two-point function is now given by
\begin{align}
C(\omega) \propto \sum_{m,k ,\ell \geq 0} \mathcal{A}_{mk\ell} \omega^{q_{mk\ell}} e^{i t_{mk\ell} \omega} .
\end{align}
Here we set
\begin{align} \label{eq:tqA}
\begin{aligned}
&t_{mk\ell} =  \left( m + \frac12 \right)  v_+ + \left( k + \frac12 \right) v_- + \left( \ell + \frac12 \right) v_p ,\\ 
&q_{mk\ell} = 3 + \frac{4 \text{Re}\, C '_\text{Q}}{\pi}  + \frac{2q}{p} - \frac{m + k +1}{2}, \\  &\mathcal{A}_{mk} = e^{- i ((2 m+1) C'_\text{Q} - (2 k +1) (C '_\text{Q})^* )} ,
\end{aligned}
 \end{align}
where
\begin{align}
    v_ + =2  A_\text{Q} = 2 z_* , \quad v_- = - 2 (A_\text{Q})^* = - 2 (z_*)^* , \quad v_p = \frac{2 \pi i}{p} = 2 \Gamma_* i .
\end{align}
Its Fourier transform is computed as
\begin{align}
C(t) \propto \sum_{m,k,\ell \geq 0} \mathcal{A}_{mk\ell} \left[ \frac{\Gamma (q_{mk\ell} +1)}{(- i (t_{mk\ell} + t))^{q_{mk\ell}+1}}+  \frac{\Gamma (q_{mk\ell} +1)}{(- i (t_{mk\ell} - t))^{q_{mk\ell}+1}}  \right] .
\end{align}
There are singularities at $t = \pm t_{mk\ell}$ when $q_{mk\ell} + 1 > 0 $, i.e.,
\begin{align} \label{eq:qmklcond}
q_{mk\ell} + 1 = 2 \nu + 1 - \frac{m + k}{2} > 0 .
\end{align}
In the case of AdS black brane solution, the singularities lie on real sections, and this indicate that the singularities can be observed by a CFT correlator of two operators inserted at some of boundaries. 
We would like to examine whether the singularities at $t = t_{mk\ell} $ lie on real sections, which are given by
\begin{align} \label{eq:realsectionrbb}
\text{Im}\, t = \frac{\beta}{2} (1 + n)  
\end{align}
with $n \in \mathbb{Z} _{\geq 0}$.
The expression of the inverse temperature $\beta$ can be found in \eqref{eq:invtemprbb}.

For this,
we need the explicit form of $t_{mk\ell}$, which is written in terms of $v_+ , v_- , v_p $ as in \eqref{eq:tqA}.
We first consider $v_+ = 2 z_*$, where $z_*$ was introduced in \eqref{eq:ABC}.
Using the metric function in \eqref{eq:Hayward}, we find
\begin{align}
    z_* = \int_{t_{-1}}^\infty \frac{1}{f(r)}
    = e^{i\pi/d}\left(
\frac{L^2-a}{aL^2m}
\right)^{1/d} H .
\end{align}
Here we set 
\begin{align}
H = - a +   
L^2 \,F \left(
\frac{1}{d},1;
1+\frac{1}{d};
-\frac{L^2-a}{a}
\right) ,
\end{align}
where $F(a,b;c;z)$ is the hypergeometric function.
Similarly, we compute $v_p = 2 \gamma_* = 2 \Gamma_* i$ defined in \eqref{eq:AC} as
\begin{align}
\Gamma_* = - i \gamma_* = - i \int^{t_{-1}}_{t_0}\frac{dr}{f(r)}
&= \frac{  \beta}{2} - 2  H \left(
\frac{L^2-a}{aL^2m}
\right)^{1/d}
\sin\frac{\pi}{d} .
\end{align}
It can be shown that
$
0 <  \Gamma_* < \beta/2
$
for $0 < a < L^2$ and $\Gamma_* \to 0$ as $a \to 0$.
We may write $z_*$ as
\begin{align}
z_* = \frac{e^{i\pi/d}}{2 \sin \frac{\pi}{d}} \left(\frac{\beta}{2} - \Gamma_* \right) .
\end{align}
Noticing that
\begin{align}
    z_* - (z_*)^* + i \Gamma_* = \frac{i \beta}{2} , 
\end{align}
we find 
\begin{align} \label{eq:imt}
    \text{Im} \, t_{mk\ell} 
    = \frac{\beta}{2} ( 1 + m + k ) + \Gamma_* (2 \ell - m - k) . 
\end{align}

We would like to examine whether \eqref{eq:imt} satisfies the condition for real sections in  \eqref{eq:realsectionrbb}.
We assume that $\Gamma_*$ is sufficiently generic such that $\Gamma_* \neq (r/s)\beta$ for integers $r,s=1,2,\ldots$.
In the case of the AdS black brane solution, we considered a singularity at $t=t_{10}$ corresponding to a null geodesic bouncing off the black hole singularity.
We first examine what happens to this bouncing null geodesic when we replace the singular geometry with a regular one.
We thus set $(m,k)=(1,0)$.
Under the genericity assumption for $\Gamma_*$, we find that%
\footnote{
Let us consider a radial null geodesic in a geometry characterized by the metric function $f(r)$. Its arrival time is given by
$
t=2\int_{r_0}^{\infty}\frac{dr}{f(r)},
$
where $r_0$ is a zero of $1/f(r)$. For the AdS black brane, $r_0=0$ corresponds to the curvature singularity, and the arrival time is $2z_*$, with $z_*$ defined in \eqref{eq:z0}. In contrast, for a regular AdS black hole, there is no singularity at $r=0$, but zeros of $1/f(r)$ exist in the complex $r$-plane. The integration contour therefore extends into the complex plane, and the corresponding geodesic is referred to as a complex null geodesic.
}
\begin{align}
\mathrm{Im}\,t_{10\ell} \neq \frac{\beta}{2}(1+n)
\end{align}
for all possible values of $\ell$ and $n\in\mathbb{Z}$.
This implies that the corresponding singularity is replaced by a finite bump in the thermal two-point function of the dual CFT.%
\footnote{Let us suppose that the two-point function has a singularity of the form $A \propto (t - t_*)^{ - \gamma}$ with $t - t_* \in \mathbb{R}$. We introduce an imaginary part as $ t - t_* + i \epsilon$ with $\epsilon > 0$. The amplitude is now $|A|^2 \propto (t - t_* + i \epsilon)^{ - \gamma}(t - t_* - i \epsilon)^{ - \gamma} = ((t -t_*)^2 + \epsilon^2)^{- \gamma}$. Therefore, the introduction of a small imaginary part replaces a singularity with a finite bump.
}

As shown above, the singularity at $t=t_{10\ell}$ does not lie on the real sections defined by \eqref{eq:realsectionrbb}.
Similarly, we can show that the singularities at $t=t_{mk\ell}$ with odd $m+k$ do not lie on the real sections.
However, for even $m+k$ and $\ell=(m+k)/2$, we find that
\begin{align}
t_{mk\ell}=2 (m-k)\mathrm{Re}\,z_* +\frac{i\beta}{2}(1+m+k),
\end{align}
and hence these singularities lie on the real sections.
This implies that these singularities can persist even in the case of a regular black hole.
Although their real parts are given by $\text{Re}\,t_{mk\ell}=2(m-k)\text{Re}\,z_*$, we do not have a clear interpretation of these singularities in terms of real null geodesics in the geometry.
For $\ell\neq(m+k)/2$, the singularities do not lie on the real sections for generic $\Gamma_*$.
For special values of $\Gamma_*$ satisfying $\Gamma_*=(r/s)\beta$ with positive integers $r$ and $s$, some singularities can lie on the real sections.
Again, we do not have a clear physical interpretation of these singularities.

\section{Extension to spherical horizon}
\label{sec:RBH}

So far, we have focused on the case of a flat horizon.
In this section, we show that our results are not very sensitive to the horizon topology by extending the previous analysis to the case of a spherical horizon.
In the next subsection, we introduce a regular black hole with a spherical horizon and examine its properties.
In subsection~\ref{sec:WKBrbh}, we evaluate the asymptotic spectrum of QNMs using WKB analysis and confirm the results by comparing them with numerical computations described in appendix~\ref{sec:Jansen}.
In subsection~\ref{sec:singularityrbh}, we examine singularity resolution using the dual CFT two-point function obtained from the QNM spectrum.

\subsection{Geometry}

We consider a $(d+1)$-dimensional geometry described by a metric of the form
\begin{align} \label{eq:metricbh}
    ds^2 = - f(r) dt^2 + f(r)^{-1} dr^2 + r^2 d \Omega^2 ,
\end{align}
where $d\Omega^2$ denotes the metric on the $(d-1)$-dimensional unit sphere $S^{d-1}$.
The AdS-Schwarzschild black hole is described by the metric function
\begin{align} \label{eq:frbh}
    f(r) = 1 +  \frac{r^2}{L^2} - \frac{m}{r^{d-2}} .
\end{align}
The geometry has a horizon at $r = r_h$ satisfying $f(r_h) = 0$. The inverse temperature is 
\begin{align} 
\beta = \frac{1}{T} = \left.\frac{4 \pi}{  \frac{d}{dr}f(r)}  \right|_{r=r_h} , \quad \left. \frac{d}{dr}f(r)\right|_{r=r_h} = \frac{d-2}{r_h} + \frac{dr_h}{L^2} .
\end{align}
The curvature singularity is located at $r = 0$.

We would like to consider a geometry where the black hole singularity at $r=0$ is resolved. As in the previous case, we consider a regular AdS black hole of Hayward type for simplicity.
The regular geometry is described by the metric function
\begin{align} \label{eq:frrbh}
    f(r) = 1 - \frac{r^{d+2} - L^2 m r^2}{a r^d - L^2 (r^d + am)} .
\end{align}
For $m=0$, it reduces to
\begin{align}
    f(r) = 1 + \frac{r^2}{L^2 - a} = 1 + \frac{r^2}{L^2_\text{eff}} .
\end{align}
Therefore, the effective AdS radius is again given by $L_\text{eff} = \sqrt{L^2 -a}$ and the parameter region is restricted as $0 < a < L^2$.
For $m \neq 0$ and $r \to \infty$, the metric function can be expanded as
\begin{align}
    f(r) = 1 + \frac{r^2}{L^2_\text{eff}} - \frac{L^4 m}{L^4_\text{eff} r^{d-2}} + \mathcal{O} (r^{2 -2 d}) .
\end{align}
Thus the metric function approaches to \eqref{eq:frbh} with replacing $L , m$ by $L_\text{eff} , (\frac{L}{L_\text{eff}})^4 m$.

As explained in \cite{Aguayo:2025xfi}, there are inner and outer horizons $r_\pm$ $(r_- < r_+)$ satisfying $f(r_\pm) = 0$ for $m > m_c$ with a critical mass $m_c$. 
The horizons are degenerated as $r_+ =r_- = r_c$ when $m = m_c$.
We compute the value of $m_c$ from the degenerated case.
The condition for horizon $f(r_h) = 0$ leads to
\begin{align}
    m (r_h) = \frac{r^d_h (r^2_h + L^2 -a) }{L^2 (r_h^2 -a)} .
\end{align}
For $r_h = r_c$, the condition $\frac{d}{dr_h}m(r_h)|_{r_h = r_c} = 0$ should be satisfied. This leads to
\begin{align}
    d (r_c^2 -a) (r_c^2 + L^2 - a) = 2 L^2 r_c^2 .
\end{align}
The solution for $r_c > 0$ is
\begin{align}
\frac{r_c}{L} = \left[ \frac{a}{L^2} - \frac{d-2}{2d}  + \frac12 \sqrt{ \left(\frac{d-2}{d}\right)^2 + \frac{8a}{d L^2}}\right]^{1/2} .
\end{align}
In terms of $r_c$, the critical mass can be written as
\begin{align}
    m_c = \frac{2 r_c^{d+2}}{d (r_c^2 - a)^2} .
\end{align}
In the following, we assume that there are inner and outer horizons, i.e., $m > m_c$.
In particular, the inverse temperature at $r=r_\pm$ is obtained as
\begin{align} \label{eq:invtempbh}
    \beta_\pm = \frac{1}{T_\pm} = \left. \frac{4 \pi}{ \frac{d}{dr}f(r)} \right |_{r= r_\pm} , \quad
   \left. \frac{d}{dr} f(r) \right |_{r=r_\pm}= \frac{d (r_\pm^2 -a) (r_\pm^2 + L^2 -a) - 2 L^2 r_\pm^2}{L^2 r_\pm^3} .
\end{align}
Near $r=0$, the metric function becomes
\begin{align}
    f(r) =1  - \frac{r^2}{a}  + \mathcal{O} (r^{d+2}) .
\end{align}
The metric thus describes the static patch of dS spacetime.
In the case of regular black brane, the inner horizon is degenerated with the horizon of dS spacetime.
In the current case, $r=0$ corresponds to the center of dS spacetime, and the center is surrounded by the inner horizon and nothing happens there.

\subsection{Complex WKB analysis and QNMs}
\label{sec:WKBrbh}

We would like to compute the two-point function \eqref{eq:Ct0} from gravity theory on the regular black hole background. For this purpose, we need the asymptotic spectrum of QNMs, which will be computed in this subsection. The QNMs for the regular black brane were evaluated in subsection~\ref{sec:WKB}. We will see that the analysis can be extended to the present case with only slight modifications.

In order to obtain QNMs, we solve the Klein-Gordon equation for dual scalar field satisfying the Dirichlet boundary condition at the AdS boundary and the ingoing boundary condition at the outer horizon.
We decompose the scalar field as
\begin{align}
    \phi = e^{- i \omega t }  r^{\frac{1-d}{2}} 
 \psi_{\omega  k} (r)  Y_{\ell \vec m} (\Omega),
\end{align}
where $Y_{\ell \vec m}$ denotes the spherical harmonics on $S^{d-1}$.
The equation for the radial part is given by \eqref{eq:radial} 
with the potential
 \begin{align} 
 \begin{aligned}
     V(z)&=f(r)\left[\frac{\ell (\ell + d - 2)}{r^2}+ \frac{\nu^2 - \frac{d^2}{4}}{L^2} \right. \\ 
    & \left.\qquad \qquad \qquad\qquad  + \left(\frac{(d-3)(d-1)}{4}\right)\frac{f(r)}{r^2}+\left(\frac{d-1}{2} \right)\frac{\frac{d}{dr}f(r)}{r}\right] .
    \end{aligned}
 \end{align}
In the following we set $\ell = 0$.

The (complex) horizons are realized by $r = r_k$ $(k=1,2,\ldots ,d+2)$ satisfying $f(r_k) = 0$ or equivalently
\begin{align}
    r_k^{d+2} + (L^2-a)r_k^d - L^2 m r_k^2 + L^2 a m = 0 .
\end{align}
We are assuming that $m > m_c$ and the inner and outer horizons are identified as $r_+ = r_1$ and $r_- = r_2$.
The function $1/f(r)$ has zeros at $r = t_k$, where
\begin{align} \label{eq:zeroRBB2}
t_{k} = e^{(2 k + 1)\pi i/d} \left ( \frac{ a L^2 m }{ L^2 -a } \right ) ^{1/d}
\end{align}
with $k=-1,0,1,\ldots , d-2$. These are actually the same as those for regular black brane given in \eqref{eq:zeroRBB}.
The WKB wave functions are given by \eqref{eq:wkbsol0}, which can be reduced to \eqref{eq:wkbsol} for the regions with $\omega^2 \gg V(r)$.
There are several regions where the potential $V(r)$ becomes large.
One of these regions is near the AdS boundary, where $r \gtrsim |\omega|$ and the geometry can be approximated by pure AdS.
Another region is near the singular point $r=t_k$.
This region can be treated in the same way as before, since the metric function differs from the previous case only by an additional constant term $1$, whose effect is negligible near $r=t_k$.
The third region is near the center at $r=0$, where the potential also becomes large.
This region was absent in the previous analysis, but we will see that it is not relevant to our analysis.

\begin{figure}
 \centering
 \begin{minipage}[b]{0.49\columnwidth}
    \centering
 \includegraphics[height=6cm]{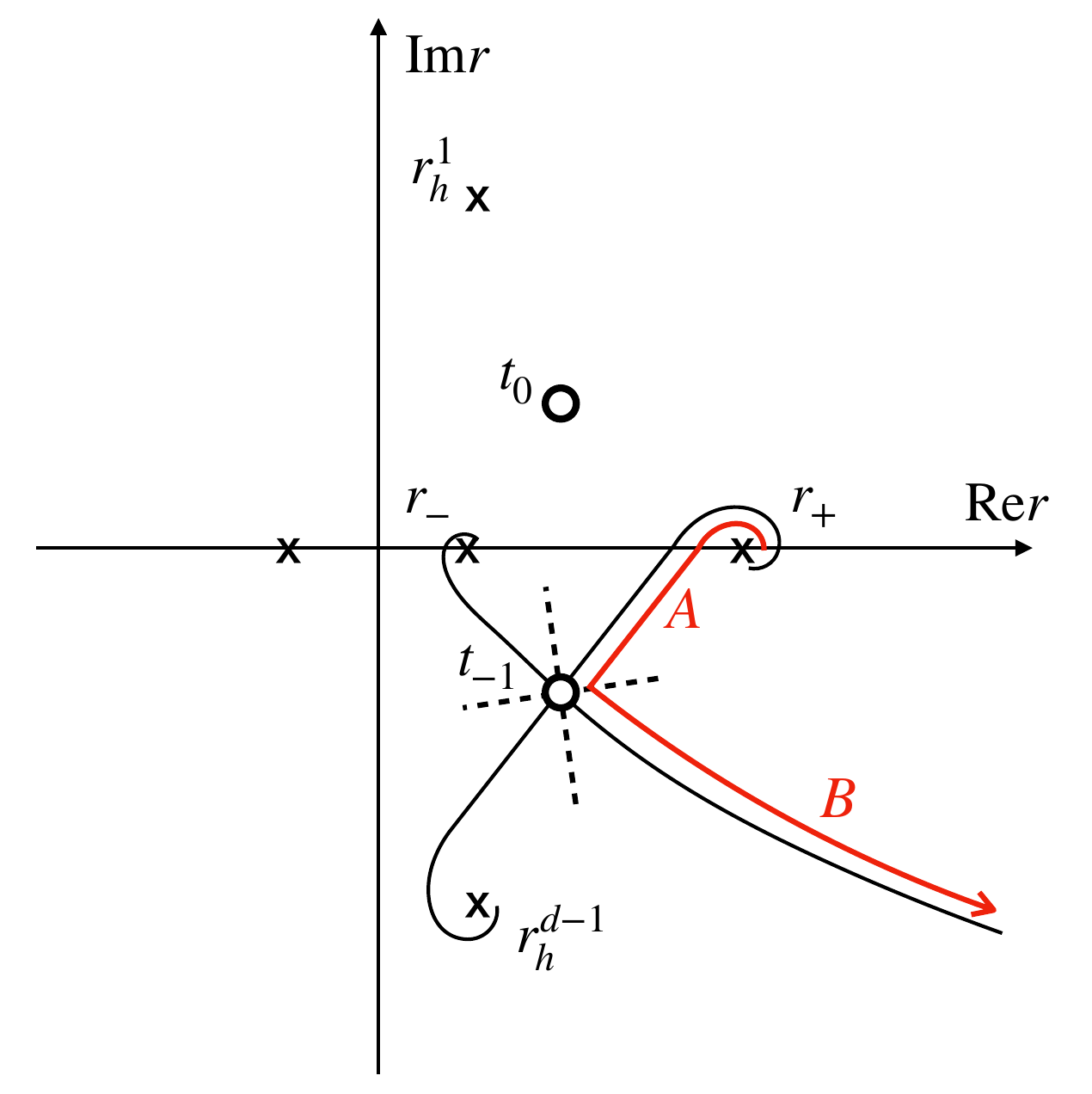}
\end{minipage}
\begin{minipage}[b]{0.49\columnwidth}
 \centering
 \includegraphics[height=6cm]{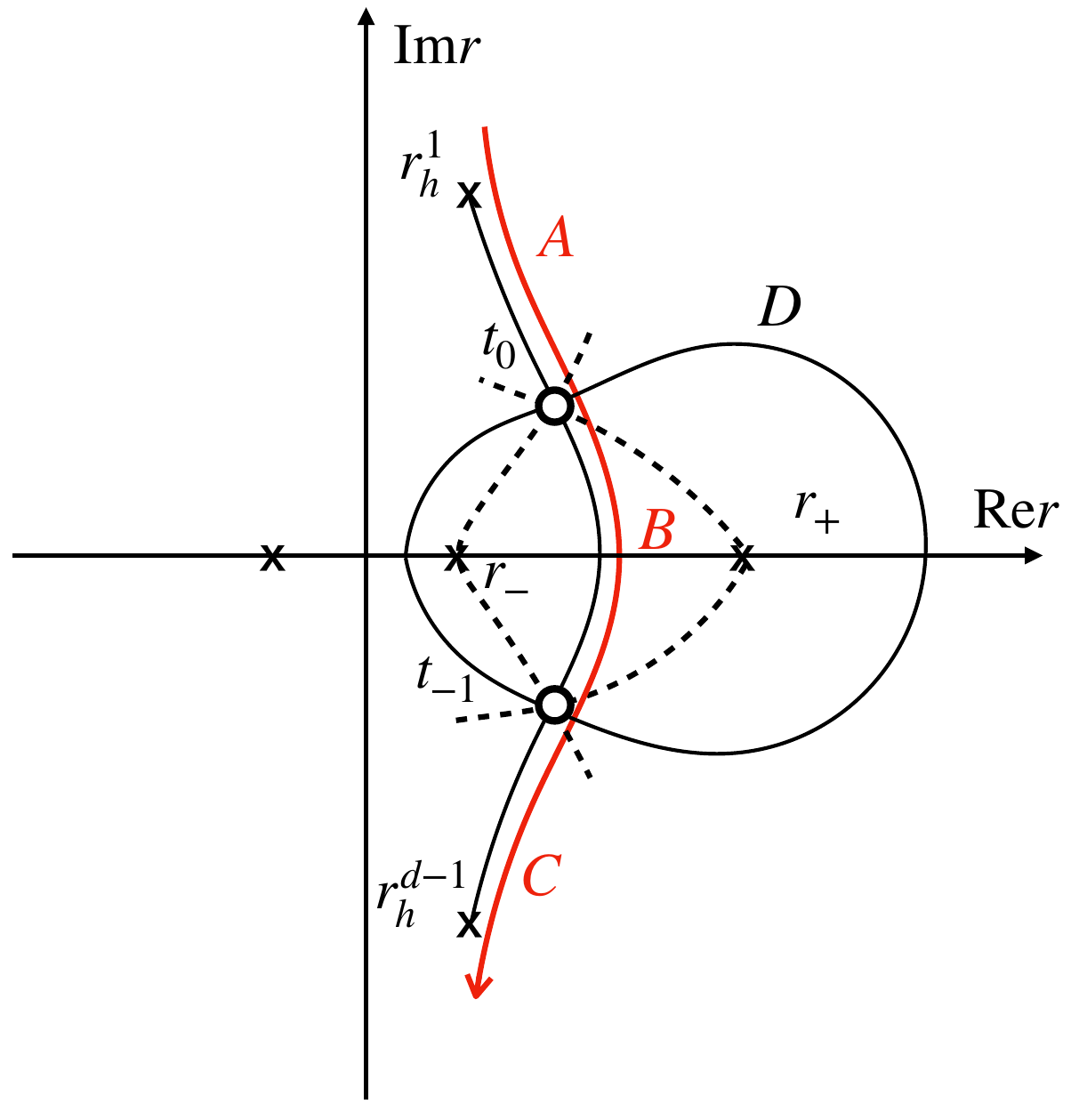}
\end{minipage}
 \caption{Stokes and anti-Stokes lines for regular black hole. 
 In the left figure we set $\omega \approx |\omega| \exp(- i \pi /d)$ and 
 in the right figure we set $\omega$ is set to be pure imaginary.}
 \label{fig:RBHg}
\end{figure}

We assign the Dirichlet boundary condition at $r \to \infty$ and the ingoing boundary condition at $r = r_+$.
We analytically continue the wave function to connect the two points along anti-Stokes lines as before.
Stokes and anti-Stokes lines can be obtained as in figure~\ref{fig:RBHg}. 
In the left panel, we assume that $\omega \approx |\omega| \exp(- i \pi /d)$. In the right panel, we assume pure imaginary $\omega$.
The positions of zeros and poles are slightly shifted but qualitative feature of (anti-)Stokes lines does not change. 
For instance, the inner horizon is located at  $r = r_-$, but it was degenerated with the center $r=0$ in the case of the regular black brane.
Now we are interested in anti-Stokes lines away from the inner horizon at $r=r_-$ and the center at $r=0$.
Therefore, the previous results can be directly applied to the case with spherical horizon with slight modifications.

Let us start from the case with  $\omega \approx |\omega| \exp(- i \pi /d)$.
The asymptotic spectrum of QNMs can be obtained as (see \eqref{eq:omegaexpansion} with \eqref{eq:ABC})
\begin{align} \label{eq:omegabh}
    A_\text{Q} \omega_n \sim \pi n + B_\text{Q} \log n + C_\text{Q} ,
\end{align}
where
\begin{align} \label{eq:ABCbh}
 A_\text{Q} = z_* , \quad 
   B_\text{Q} = - \frac{i}{4}  , \quad C_\text{Q} = \left( \frac{\nu}{2}-\frac34 \right)\pi +\frac{i}{2}\log\gamma_{-1} -\frac{i}{4}\log\left(\frac{\pi}{z_*}\right)
\end{align}
for large $n$.
The expression takes the same form as before, but the value of $z_*$ differs from the previous one since%
\footnote{In the case of regular black brane, we can show that $z_* \propto \exp( i \pi  /d)$. However, in the current case, the relation holds only approximately. This is not a problem since we use the relation only to find Stokes and anti-Stoke lines, which are not so sensitive to the exact relation.}
\begin{align}
z_* = \int_{t_{-1}}^{\infty} \frac{1}{f(r)} dr 
\end{align}
depends on the metric function.
A comparison between the analytic prediction and the numerical results obtained using the method of \cite{Jansen:2017oag} is presented in table~\ref{tab:QNMrbh}.
The parameter $A_\text{Q}$ shows good agreement between the two approaches, as in the previous case of the regular black brane.
We find slightly better agreement for the parameters $B_\text{Q}$ and $C_\text{Q}$ compared with the previous case, but we do not have a clear explanation for this improvement.
\begin{table}[htbp]
   \centering
  \[
\begin{array}{|c|c|c|c|c|c|c|}
\hline
~ & \multicolumn{3}{|c|}{\text{WKB prediction}} & \multicolumn{3}{|c|}{\text{Numerical fit}}  \\
\hline
(d,a)  & A_\text{Q} & B_\text{Q} & C_\text{Q} & A_\text{Q} & B_\text{Q} & C_\text{Q} \\ \hline
(3,0.01)
& 1.24 + 0.84 i
& - 0.25 i
& 0.51 + 0.54 i
&1.24+0.83i
&-0.08-0.30i
&0.68+0.48i
\\
(3,0.03)
& 1.11 + 0.80 i
& - 0.25 i
& 0.50 + 0.34 i
&1.14+0.82i
&-0.04-0.28i
&0.62+0.33i
\\
(4,0.05)
& 0.93 + 0.51 i
& - 0.25 i
& 1.25 + 0.41 i
&0.93+0.51i
&-0.05-0.29i
&1.44+0.43i
\\
(4,0.10)
& 0.84 + 0.49 i
& - 0.25 i
& 1.24 + 0.27 i
&0.84+0.49i
&-0.03-0.29i
&1.37+0.33i
\\
(5,0.1)
& 0.81+ 0.36 i
& - 0.25 i
& 2.01 + 0.37 i
&0.81+0.36i
&-0.04-0.30i
&2.21+0.43i
\\
(5,0.15)
& 0.74 + 0.35 i
& - 0.25 i
& 2.01 + 0.29 i
&0.75+0.35i
&-0.02-0.30i
&2.14+0.36i
\\
(6,0.1)
& 0.79 + 0.29 i
& - 0.25 i
& 2.80 + 0.45 i
&0.79+0.29i
&-0.05-0.30i
&3.03+0.51i
\\
(6,0.2)
& 0.68 + 0.27 i
& - 0.25 i
& 2.79 + 0.30 i
&0.68+0.27i
& + 0.00-0.30i
&2.89+0.38i
\\
\hline
\end{array}
\]
    \caption{Comparison QNMs with nonzero real part for regular black hole obtained by WKB analysis and numerical computations. We set $\nu = d/2$.}
    \label{tab:QNMrbh}
\end{table}

For purely imaginary QNMs, we have (see \eqref{eq:omegaexpansion2} with \eqref{eq:AC})
\begin{align} \label{eq:omegabh2}
   A_\text{Q} \omega  \sim n  \pi + C_\text{Q}  \quad (n=1,2,\ldots)
\end{align}
with
\begin{align} \label{eq:ACbh}
A_\text{Q} =  \gamma_* , \quad 
 C_\text{Q} =  - \frac{ \pi}{2 d}  .
\end{align}
Again, the expression takes the same form as before, but it differs because
\begin{align}
\gamma_* = \int^{t_{-1}}_{t_{0}} \frac{1 }{f(r)}  dr
\end{align}
depends on the metric function.
A comparison between the analytic prediction and the numerical results obtained using the method of \cite{Jansen:2017oag} is presented in table \ref{tab:pmQNMrbh}.
As in the case of $\omega \approx |\omega| \exp (-\pi i/d)$, we find good agreement for the parameter $A_\text{Q}$ and slightly better agreement for $C_\text{Q}$ compared with the case of the regular black brane.
\begin{table}[htbp]
    \centering
 \[
\begin{array}{|c|c|c|c|c|}
\hline
~ & \multicolumn{2}{|c|}{\text{WKB prediction}} & \multicolumn{2}{|c|}{\text{Numerical fit}}  \\
\hline
(d,a) & A_\text{Q} & C_\text{Q} & A_\text{Q} & C_\text{Q} \\ \hline
(3,0.01)
& ~~0.36 i ~~
& ~ - 0.52 ~
& ~~ 0.35i ~~
& ~ - 0.83 ~
\\
(3,0.03)
& 3.41 i
& - 0.52
& 3.44i
& - 0.49
\\
(4,0.05)
& 0.51 i
& - 0.39
& 0.51i
& -0.58
\\
(4,0.10)
& 2.51 i
& - 0.39
& 2.53i
& -0.24
\\
(5,0.1)
& 0.59 i
& - 0.31
& 0.59i
& -0.41
\\
(5,0.15)
& 1.42 i
& - 0.31
& 1.43 i
& -0.18
\\
(6,0.1)
& 0.34 i
& - 0.26
& 0.34 i
& -0.44
\\
(6,0.2)
& 1.24 i
& - 0.26
& 1.25 i
& -0.05
\\
\hline
\end{array}
\]
       \caption{Comparison of purely imaginary QNMs for regular black hole obtained from WKB analysis and numerical computations. We set $\nu = d/2$.}
    \label{tab:pmQNMrbh}
\end{table}

\subsection{Singularity resolution}
\label{sec:singularityrbh}

As reviewed in section~\ref{sec:2pt}, singularities can be observed in the two-point function of a CFT dual to an AdS black brane solution.
Furthermore, in section~\ref{sec:singularity}, we have shown that the singularities at $t=t_{10\ell}$ do not lie on the real sections defined by \eqref{eq:realsectionrbb}  for a CFT dual to a regular AdS black brane.
We have argued that this provides an indication of singularity resolution.
In this subsection, we extend the analysis to the case of a regular AdS black hole with a spherical horizon.

For the regular AdS black hole, we have assumed that there are two horizons at $r=r_\pm$, with inverse temperatures $\beta_\pm$ given by \eqref{eq:invtempbh}.
We consider the maximally extended geometry and use a complex time coordinate $t$. When crossing the inner and outer horizons, the time coordinate undergoes shifts of $\pm i\beta_-/4$ and $\pm i\beta_+/4$, respectively.
This implies that the real sections are described by%
\footnote{We are interested in correlation functions of operators inserted at some of the boundaries. In order to connect two boundaries, we need to cross an even number of inner or outer horizons.}
\begin{align} \label{eq:realsectionbh}
\mathrm{Im}\,t = (1+n)\frac{\beta_+}{2} + m\frac{\beta_-}{2}
\end{align}
with $n,m\in\mathbb{Z}$.
For more details, see subsection~2.2 of \cite{Dodelson:2025jff}.
We thus need to examine whether the singularities at $t=t_{mk\ell}$ in \eqref{eq:tqA} lie on the real sections defined by \eqref{eq:realsectionbh}.

In order to have the explicit expression of $t_{mk\ell}$, we need to compute $z_*$ and $\gamma_*$ by using the metric function in \eqref{eq:frrbh}.
For the integrals, it is convenient to express 
\begin{align}
    \frac{1}{f(r)} = \sum_{k=1}^{d+2} \frac{1}{f'(r_k)} \frac{1}{r - r_k} , \quad f'(r_k) = \left. \frac{d}{dr} f(r) \right|_{r = r_k} ,
\end{align}
where $r_k$ are the positions of complex horizons satisfying $f(r_k)=0$.
Using this form, we compute the integral $z_*$ as
\begin{align}
    z_* = - \sum_{k=1}^{d+2} \frac{1}{f'(r_k)} \log \left( t_{-1}-r_{k} \right)
\end{align}
and the integral $\gamma_*$ as
\begin{align}
    \gamma_* = \sum_{k=1}^{d+2} \frac{1}{f'(r_k)} \log \left( \frac{t_{-1}-r_{k}}{t_0 - r_{k}} \right) .
\end{align}
For the logarithmic functions, we need to fix their branches according to the integration contours. Since $t_0=(t_{-1})^*$ and the points $t_k$ always occur in complex conjugate pairs, we can show that $\gamma_*=-(\gamma_*)^*$.
This implies that $\gamma_*$ is purely imaginary, so we can write $\gamma_*=i\Gamma_*$ with $\Gamma_*\in\mathbb{R}$.
The integration contour for $z_*$ is the anti-Stokes line $B$ in the left panel of figure~\ref{fig:RBHg}. Let us consider a closed contour $L_*$ formed by connecting $B$, a contour joining $t_0$ and $t_{-1}$ through the region between the inner and outer horizons, the complex conjugate of $B$, and then closing the contour at infinity.
The contour $L_*$ encloses the outer horizon at $r=r_+$, leading to the identity
\begin{align}
z_*-(z_*)^*+i\Gamma_*=\frac{i\beta_+}{2}.
\end{align}
Furthermore, it can be shown that
$
0<\Gamma_*<\beta_+ / 2
$
for $0<a<L^2$ and $\Gamma_* \to 0 $ as $a \to 0$.

We would like to examine whether the singularities at $t=t_{mk\ell}$ satisfy the condition for the real sections in  \eqref{eq:realsectionbh}.
We find that
\begin{align} \label{eq:imt2}
\mathrm{Im} \, t_{mk\ell}
= \frac{\beta_+}{2}(1+m+k) + \Gamma_*(2\ell-m-k),
\end{align}
which is quite similar to the result for the regular black brane in \eqref{eq:imt}.
Thus, we reach a conclusion similar to that in the previous case.
We consider a generic value of $\Gamma_*$ satisfying $r\Gamma_* \neq s\beta_+/2 + u\beta_-/2$ for nonnegative integers $r,s,u$, with $r\neq 0$.
We find that the singularities at $t=t_{mk\ell}$ with odd $m+k$, including those at $t=t_{10\ell}$, do not lie on the real sections defined by \eqref{eq:realsectionbh}.
Moreover, for even $m+k$ and $\ell=(m+k)/2$, we have
\begin{align}
t_{mk\ell} = 2 (m-k)\mathrm{Re}\,z_* + (1+m+k)\frac{i\beta_+}{2},
\end{align}
which means that these singularities lie on the real sections.
For $\ell\neq(m+k)/2$, the condition for the real sections is not satisfied.

\section{Complex null geodesic from OPE}
\label{sec:OPE}

We have obtained the two-point function $C(t)$ in \eqref{eq:Ct0} or equivalently its Fourier transform $C(\omega)$, where $C(\omega)$ can be read off from the QNMs using the thermal product formula \eqref{eq:thermalproduct}. The asymptotic spectrum of QNMs is computed using WKB analysis and the result is confirmed by comparison with numerical computations.
We have observed that $C(t)$ exhibits a singularity associated with a complex null geodesic in the regular black brane/hole geometry.
In this section, we confirm this conclusion from the OPE structure of the thermal two-point function:
\begin{align}\label{eq:Ct01}
G(\tau) \equiv C \left (t = - i \tau + i \frac{\beta }{2} \right)= \left \langle \mathcal{O} \left( - i \tau ,0 \right) \mathcal{O}(0 ,0) \right \rangle_\beta .
\end{align}
We consider a regular black brane without singularity described by the metric \eqref{eq:metric} with \eqref{eq:Hayward}.
From now on we set $L_* = 1$ (i.e., $L^2 = 1 + a$) so that the asymptotic AdS condition with unit AdS radius is satisfied.  We further set $d=4$ and $m=1$ while keeping $a$ arbitrary.

Applying the OPEs  of $\mathcal{O} (\tau) \mathcal{O} (0)$ in terms of operators $\mathcal{O}_n$ with conformal dimensions $\Delta_n$, the two point function can be decomposed as
\begin{align}
G(\tau) = \sum_n C_n \langle \mathcal{O}_n \rangle_\beta \tau^{\Delta_n - 2 \Delta} .
\end{align}
In the classical limit, the bulk fields dual to $\mathcal{O}_n$ are argued to be the composites of multiple gravitons and the composites of two scalar fields.%
\footnote{See \cite{Komargodski:2026oob} for the effects of quantum corrections.}
These fields correspond to multi-stress-tensors $T^n$ and double-trace operators $\mathcal{O} \partial \cdots \partial \mathcal{O}$, respectively. Thus the two-point function can be decomposed into two parts as
\begin{align}
G(\tau) = G_\text{T} (\tau) + G_{[\mathcal{O} \mathcal{O}]} (\tau) ,
\end{align}
where $G_\text{T} (\tau)$ and $G_{[\mathcal{O} \mathcal{O}]} (\tau)$ arise from the exchanges of multi-stress-tensors and the double-trace operators, respectively.
In the following, we focus on $G_\text{T} (\tau)$ and compute it from the bulk theory following the analysis of \cite{Fitzpatrick:2019zqz}. See, e.g.,  \cite{Ceplak:2024bja,Buric:2025fye,Ceplak:2025dds,Araya:2026shz,Giombi:2026kdz} for applications to the singular AdS black branes.

As usual, we solve the Klein-Gordon equation for a bulk scalar field $\phi$, which is given by
\begin{align}
( \square - M^2 ) \phi = 0 .
\end{align}
The mass of the scalar field is denoted by $M$, which satisfies $M^2 = \Delta (\Delta - d)$ as before.
Following \cite{Fitzpatrick:2019zqz}, we introduce new coordinates defined by
\begin{align}
\rho^2 = r^2 \vec x^2 , \quad w^2 = 1 + r^2 (\tau^2 + \vec x^2) 
\end{align}
and write
\begin{align}
\phi (w , \rho , r) = \left( \frac{r}{w^2}\right)^\Delta \psi (w , \rho , r) .
\end{align}
The CFT correlator can be read off as
\begin{align} \label{eq:Gtau}
G(\tau) = \frac{1}{|\tau|^{2 \Delta}} \lim_{r \to \infty} \psi .
\end{align}
We decompose the scalar field as $\psi = \psi_\text{T} + \psi_{[\mathcal{O} \mathcal{O}]}$. The contribution to $G_\text{T}$ comes from $\psi_\text{T}$, which can be expanded as
 \begin{align} \label{eq:psiT}
\psi_\text{T} = \sum_{i=0}^\infty \sum_{j=0}^i \sum_{k = - i}^{ 2 i - j} a_{j,k}^i \frac{\rho^{2 j} w^{2 k}}{r^{4 i}} 
\end{align}
with $a^0_{0,0} = 1$.
From \eqref{eq:Gtau}, we obtain 
\begin{align} \label{eq:GT}
G_\text{T} (\tau) = \frac{1}{|\tau|^{2 \Delta}} \sum_{n=0}^\infty  a_n \tau^{4 n} 
\end{align}
with $a_n \equiv a_{0,2n}^n$.

\begin{figure}
 \centering
 \includegraphics[height=6cm]{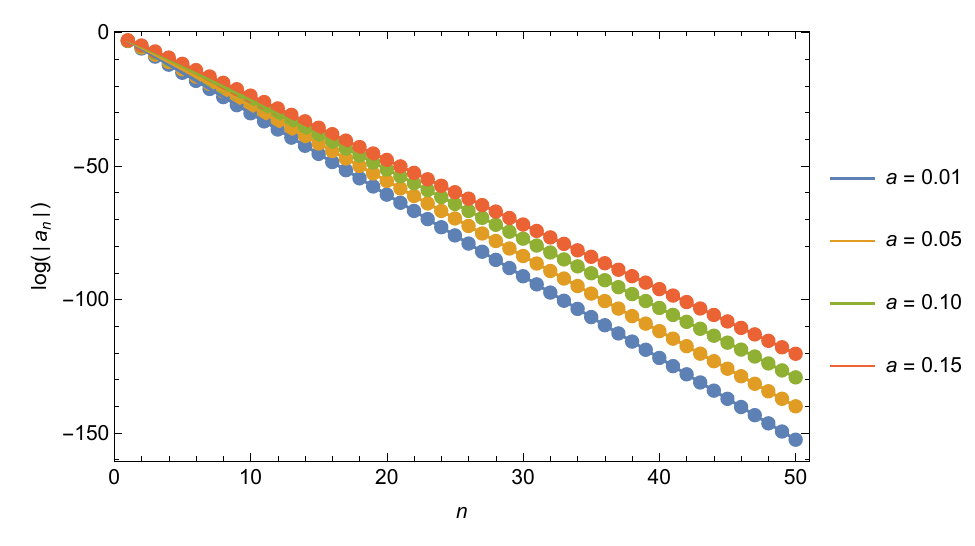}
 \caption{The numerical data for the coefficients $a_n$. We set $\Delta = 31/20$ and $a = 0.01,0.05,0.1,0.15$.
  }
 \label{fig:OPE_rbh}
\end{figure}

We would like to solve the Klein-Gordon equation for $\psi_\text{T}$ in \eqref{eq:psiT}. 
The equation can be reduced to a set of algebraic equations for $a_{j,k}^i $ as (see \cite{Buric:2025fye} for more details)
\begin{align} \label{eq:algebraic}
\sum_{(b,c,d) \in S } a^{I -b}_{J -c , K-d} C^{I -b , J -c , K-d}_{b,c,d} = 0 
\end{align}
for all $I,J,K$.
Here $S$ is given by
\begin{align}
\begin{aligned}
S &= \{(0, 0, 2), (0, 1, 1), (0, 1, 2), (1, 0, 2), (1, 1, 0), (1, 1, 
    1), (1, 1, 2), (1, 2, 0), (2, 0, 2), \\& \quad (2, 1, 0), (2, 1, 1), (2, 1, 
    2),  (2, 2, 0), (3, 0, 2), (3, 1, 0), (3, 1, 1), (3, 1, 2), (3, 2, 
    0)\} 
    \end{aligned}
\end{align}
and nonzero components of $C^{i,j,k}_{b,c,d}$ are listed in appendix~\ref{app:C}.
The coefficients  $a_{j,k}^i $  can be obtained by solving the algebraic equations.
The numerical results for $a_n = a^{n}_{0,2n}$ are shown in figure~\ref{fig:OPE_rbh} for $\Delta = 31/20$ and $a = 0.01,0.05,0.1,0.15$.%
We find that the coefficients $a_n$ behave as
\begin{align} \label{eq:fittings}
a_n \sim \frac{C_\text{T} n^{b_\text{T}}}{(-1)^n A_\text{T}^{4n}}
\end{align}
for large $n$.

We examine the singularity structure of $G_\text{T}(\tau)$.
Using the asymptotic expression of $a_n$ in \eqref{eq:fittings}, we find
\begin{align}
    G_\text{T}(\tau) \sim \frac{C_\text{T}}{|\tau|^{2 \Delta}}\sum_{n=1}n^{b_\text{T}} z^{n} = \frac{C_\text{T}}{|\tau|^{2 \Delta}} L_{- b_\text{T}} (z) ,
\end{align}
where $z = - \tau ^4 / A_\text{T} ^4 $.
Here $L_{\alpha}(z)$ denotes the polylogarithm.
The numerical fits suggest that $b_\text{T}>-1$. In this case, the polylogarithm has the following asymptotic behavior as $z\to 1$:
\begin{align}
    L_{- b_\text{T}} (z) \sim \Gamma(1 + b_\text{T}) (- \log z)^{-1-b_\text{T}} .
\end{align}
Using $(- \log z) \sim (1 - z)$, we thus find
\begin{align}
    G_\text{T}(\tau) \sim \frac{C_\text{T} \Gamma(1 + b_\text{T})}{|\tau|^{2 \Delta}} \left( 1 + \frac{\tau^4}{A_\text{T}^4} \right)^{-1 -b_\text{T}} .
\end{align}
The conformal block has a singularity at $\tau = \tau_*$ with
\begin{align}
  - i   \tau_* = e^{ i \pi /4} A_\text{T} ,
\end{align}
and we have chosen one of the fourth roots of $(-1)^{1/4}$.
This value is expected to coincide with the arrival time of a complex null geodesic in the geometry of regular black brane,
\begin{align} \label{eq:tauc}
- i \tau_c = 2 \int_{t_{-1}}^\infty \frac{dr}{f(r)} . 
\end{align}
The numerical estimates of $A_\text{T}$ obtained from the fits are presented in table~\ref{tab:B}
and we find good agreement between $|A_\text{T}|$ obtained from the numerical fit and the absolute value of the integral in \eqref{eq:tauc}.
\begin{table}[htbp]
    \centering
    \begin{tabular}{|c|c|c|}
        \hline
        $a$  & $A_\text{T}$ &  $ | \tau_c | $ \\
        \hline
        0.01 & 2.1524 & 2.1541\\
        \hline
        0.05 & 2.0267 & 2.0237\\
        \hline
        0.1 & 1.9210& 1.9167 \\ \hline
        0.15 & 1.8387 &1.8339 \\ \hline
    \end{tabular}
    \caption{
    Comparison of $|A_\text{T}|$ obtained from numerical fits and $|\tau_c|$ computed from the integral for $\Delta=31/20$ and $a=0.01,0.05,0.1,0.15$.
    }
    \label{tab:B}
\end{table}

\section{Conclusion and discussion}
\label{sec:conclusion}

In this paper, we have computed the thermal two-point function of scalar operators \eqref{eq:Ct0} in a CFT dual to a regular black brane solution. We found that the correlator has singularities at $t = \pm t_{mk\ell}$ for sufficiently small $m,k$ satisfying \eqref{eq:qmklcond}.
Here, the expression for $t_{mk\ell}$ is given in \eqref{eq:tqA}. We have shown that the singularities at $t_{10\ell}$ do not lie on the real sections since $\text{Im}\, t_{10\ell} \neq (1 + n) \frac{\beta}{2}$ with $n \in \mathbb{Z}$, where $\beta$ is the inverse temperature.
We argued that this provides an indication of the resolution of black hole singularity. We also found that $t_{mk\ell}$ with even $m+k$ and $\ell = (m+k)/2$ lie on the real sections.
For special choices of parameters such that  $\Gamma_*/\beta$ is rational, some values of $(m,k,\ell)$ lead to $\text{Im}\, t_{mk\ell} = (1 + n) \beta/2$. It would be interesting to investigate whether these singularities on the real sections can be interpreted in terms of real null geodesics in the regular black brane geometry.

We used the thermal product formula \eqref{eq:thermalproduct} to compute the two-point function, which requires knowledge of the QNM spectrum. We computed the asymptotic spectrum of QNMs using WKB analysis and confirmed the results by comparison with numerical computations. 
There are two types of QNMs. One type consists of QNMs with nonzero real parts, which play an important role in our analysis of singularities. 
The other type consists of purely imaginary QNMs, which required us to develop a technique for their analysis. In order to check its validity, we applied the technique to compute purely imaginary QNMs for the AdS-Reissner-Nordstr\"{o}m black hole in appendix~\ref{app:ChargedBH}. We also extended the analysis to the case of a regular AdS black hole with a spherical horizon and confirmed that our conclusion is insensitive to the topology of the black hole horizon and the AdS boundary.
We have also examined the OPE structure of the two-point function. 
We showed that $G_\text{T}$ arising from the exchanges of multi-stress-tensors exhibits a singularity associated with a complex null geodesic, providing further support for our analysis based on the thermal product formula.

In this paper, we have focused on regular AdS black holes of the Hayward type for simplicity. However, other types of regular black holes have also been studied in \cite{Hennigar:2025yqm,Aguayo:2025xfi}. We can also construct regular black holes with charges and/or angular momentum. It is interesting to extend the current analysis to more generic cases and identify generic features associated with the resolution of black hole singularities.
One of the main technical achievements is the determination of the asymptotic spectrum of QNMs in regular black holes using WKB analysis. We have analyzed only the leading order contribution in $1/\sqrt{\omega}$, and it is important to examine higher-order corrections in order to improve the agreement between the WKB predictions and numerical results. The QNMs of black holes themselves are important quantities with many other applications. 
It would also be interesting to apply the current analysis to other cases, such as asymptotically flat regular black holes.
We have considered a specific type of regular AdS black hole, which can be realized as a solution of quasi-topological gravity as shown in \cite{Bueno:2024dgm}.
We would like to identify explicit examples of CFTs dual to these regular black holes and investigate the resolution of black hole singularities purely from the CFT perspective.

\acknowledgments

We are grateful to Javier Moreno and Kotaro Tamaoka for useful discussions. 
ChatGPT was used to assist with language editing and discussions. 
The work of H.\,Y.\,C. is supported in part by Ministry of Science and Technology (MOST) through the grant 114-2112-M-002-022-. 
The work of Y.\,H. is supported by JSPS KAKENHI Grant Number JP23K25867. The work of Y.\,K. is supported by JSPS KAKENHI Grant Numbers JP23KK0048.

\appendix

\section{Stokes lines of regular AdS black branes/holes}
\label{app:Stokes}

In this appendix, we investigate the Stokes structures relevant to our WKB analysis.

\subsection{Stokes structure}

For the WKB solutions~\eqref{eq:wkbsol} around a turning point $r=t\in \mathbb{C}$, a Stokes line emanating from $t$ is given by
\begin{align}
\label{eq:stokes}
   \mathrm{Re}\left[\omega\int^r_t\frac{1}{f(r')}dr'\right]=0.
\end{align}
Parametrizing the curve as $r=r(\lambda)$ and differentiating eq.\,\eqref{eq:stokes} by $\lambda$, we obtain the ordinary differential equation (ODE),
\begin{align}
\label{eq:stokesODE}
   \frac{d}{d\lambda}r=\pm i\mathcal{F}^*,
\end{align}
where $\mathcal{F}(r):=\omega f^{-1}(r)$.
We can draw Stokes lines by numerically integrating the ODE.
Stokes lines, which emanate from a turning point, terminate at a singular point or the same/another turning point.
The $\pm$ sign is due to the freedom of reparametrization $\lambda\to -\lambda$ and should be chosen so that the curve extends away from $t$.

There are multiple Stokes lines from each turning point.
As the initial condition of the ODE, we need to specify the direction in which a Stokes line emanates.
Let us assume that a turning point $t$ is a repeated root of multiplicity $n$ of $1/f(r)$.
Define coordinates $\bar r e^{i\bar \theta}=r-t$ with $\bar\theta\in[0,2\pi)$ around $t$.
Letting $f_\text{n}(r)$ and $f_\text{d}(r)$ be the polynomials of the meromorphic function $f(r)=f_\text{n}(r)/f_\text{d}(r)$, the condition~\eqref{eq:stokes} becomes
\begin{align}
   \mathrm{Re}\left[\omega\frac{f_\text{d}^{(n)}(t)}{f_\text{n}(t)}\frac{1}{(n+1)!} \bar r^{n+1} e^{i(n+1)\bar\theta}\right]=0,
\end{align}
around the turning point.
This gives the condition for $\bar\theta$ as
\begin{align}
   (n+1)\bar\theta+\arg(\omega\alpha)=\left(\frac12+m\right)\pi,\quad \alpha\equiv\frac{f_\text{d}^{(n)}(t)}{f_\text{n}(t)},\quad m\in \mathbb{Z},
\end{align}
with $2(n+1)$ solutions.
Thus, there are $2(n+1)$ Stokes lines from $t$, corresponding to the directions $\bar\theta$.
In particular, if the turning point is a single root of $f_\text{d}(r)$, four Stokes lines emanate from it.
As the WKB solutions $f_{1,2}$ have the phase,
\begin{align}
    \pm \omega\int_t^rf(r')dr'
    =\pm[(n+1)\bar\theta+\arg(\omega\alpha)]
    =\pm\left(\frac12+m\right)\pi,
\end{align}
$f_1$ is exponentially large and dominant for an odd $m$ while $f_2$ is dominant for an even $m$.
The ODE~\eqref{eq:stokesODE} can be solved with the initial condition $r(0)=\epsilon e^{i\bar\theta}$ for some small $\epsilon$.

Similarly, anti-Stokes lines around a turning point $t$ are given by the condition,
\begin{align}
\label{eq:antistokes}
   \mathrm{Im}\left[\omega\int^r_t\frac{1}{f(r')}dr'\right]=0.
\end{align}
The ODE for the curves is
\begin{align}
   \frac{d}{d\lambda}r=\pm \mathcal{F}^*.
\end{align}
The expansion $\bar r e^{i\bar \theta}=r-t$ around $t$ gives
\begin{align}
    (n+1)\bar\theta+\arg(\omega\alpha)=m\pi,\quad \alpha\equiv\frac{f_\text{d}^{(n)}(t)}{f_\text{n}(t)},\quad m\in \mathbb{Z}.
\end{align}
There are $2(n+1)$ solutions for $\bar\theta$ and corresponding anti-Stokes lines.
In particular, if the turning point is a single root of $f_\text{d}(r)$, four anti-Stokes lines emanate from it.

\begin{figure}[h]
\centering
\includegraphics[height=6cm]{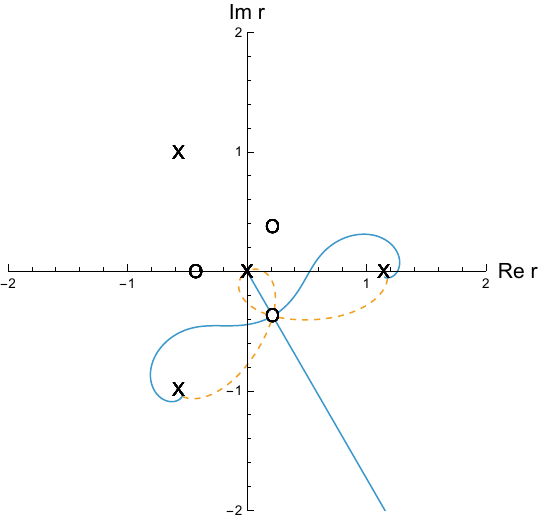}
\includegraphics[height=6cm]{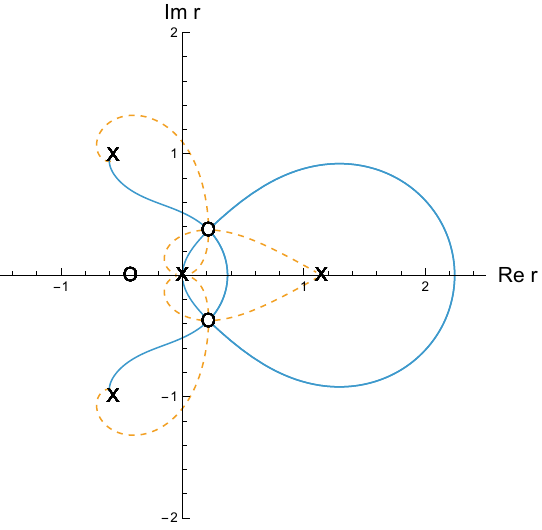}
\caption{The Stokes (dashed) and anti-Stokes (solid) lines of the regular black brane for $d=3,\ m=1.5,\ a=0.05$. The `o' and `x' denote the turning points and the singularities of $f(r)$, respectively.
We assumed $\omega\propto \exp (-i\pi/d )$ (left) and $\omega\propto \exp (-i\pi/2)$ (right).
}
\label{fig:stokes-rbb-d3}
\end{figure}

\begin{figure}[h]
\centering
\includegraphics[height=4cm]{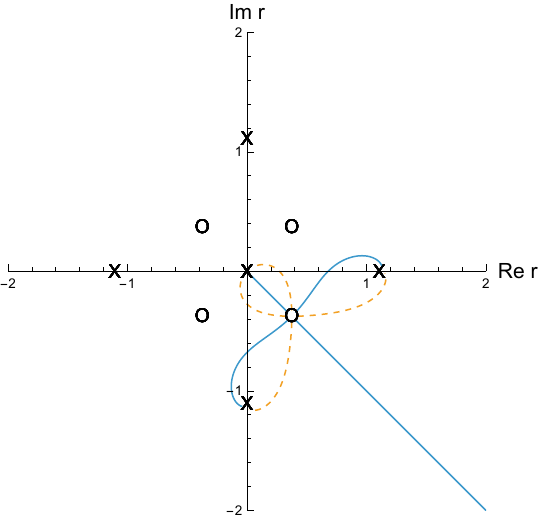}
\includegraphics[height=4cm]{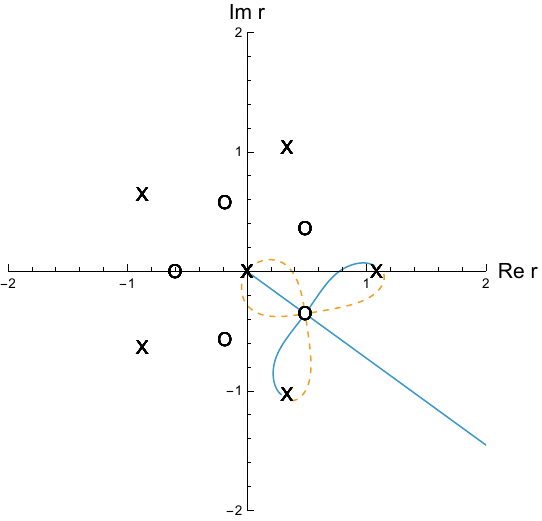}
\includegraphics[height=4cm]{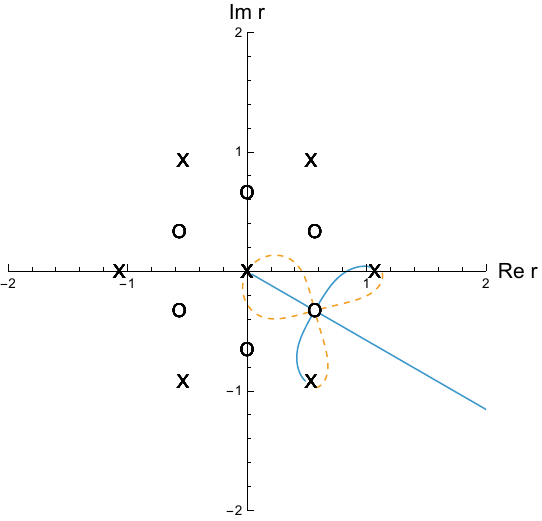}\\
\includegraphics[height=4cm]{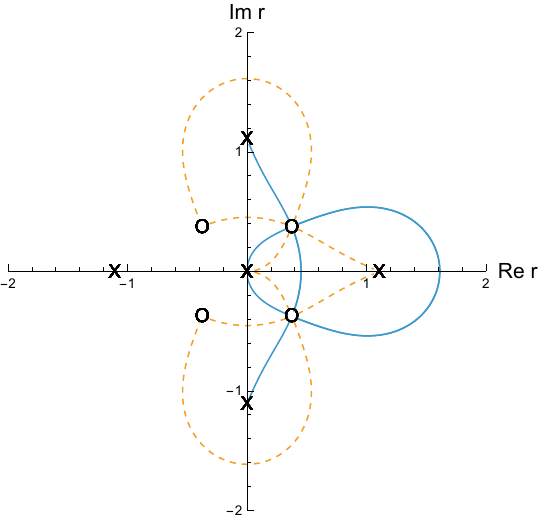}
\includegraphics[height=4cm]{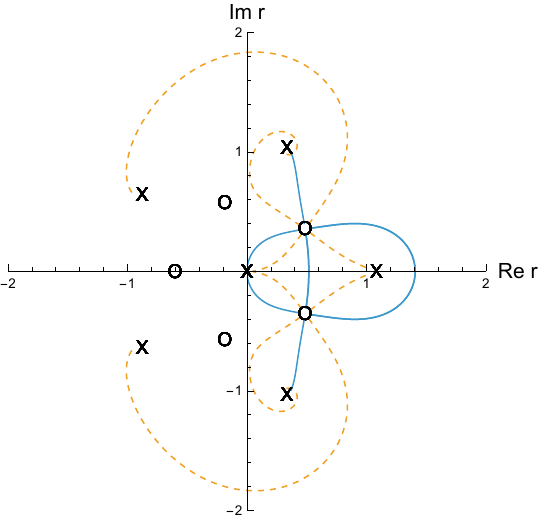}
\includegraphics[height=4cm]{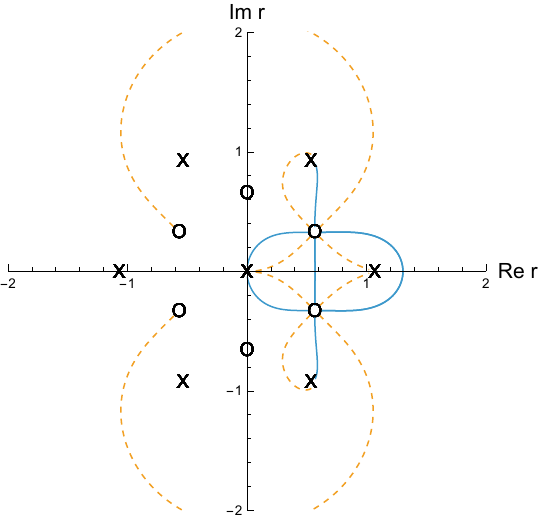}
\caption{The Stokes structure of the regular black brane for $d=4,5,6$ with $m=1.5,\ a=0.05$. $\omega\propto \exp(-i\pi/d)$ for the upper panels and $\omega\propto \exp(-i\pi/2)$ for lower.
}
\label{fig:stokes-rbb-d456}
\end{figure}

We show the representative cases of the Stokes structures focusing on the part relevant to our purpose.
For the regular black branes, $f_\text{d}(r)=a r^d-L^2(r^d+a m)$ has $d$ single roots on $\mathbb{C}$ corresponding to the turning points. $f_\text{n}(r)=-(r^{d+2}-L^2mr^2)$ has $d$ single roots corresponding to the (complex) horizons and one repeated root at the origin. Stokes and anti-Stokes lines, which emanate from a turning point, terminate at one of these singularities or infinity.
Figure~\ref{fig:stokes-rbb-d3} shows the Stokes structure relevant to one of the turning points for $d=3$ with the assumption $\omega\propto \exp (-i\pi/d)$ and $\omega\propto \exp(-i\pi/2)$.
The qualitatively same structure can be observed for different $d,\ m$ and $a$.
Figure~\ref{fig:stokes-rbb-d456} shows the $d=4,5,6$ cases.
In all the cases, $f_2$ is dominant on the Stokes lines into the singular point $r=r_h$ on the positive real axis.

\begin{figure}[h]
\centering
\includegraphics[height=6cm]{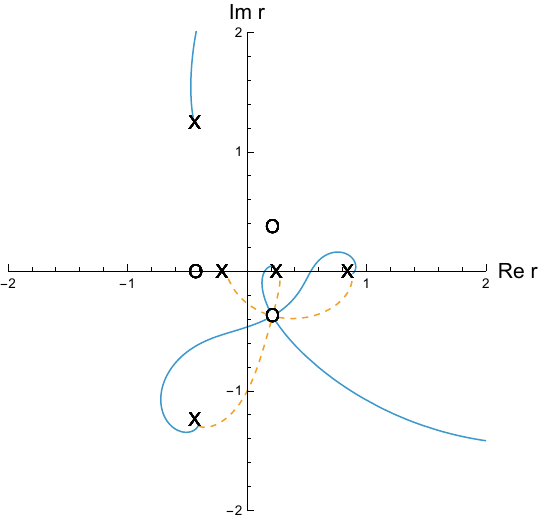}
\includegraphics[height=6cm]{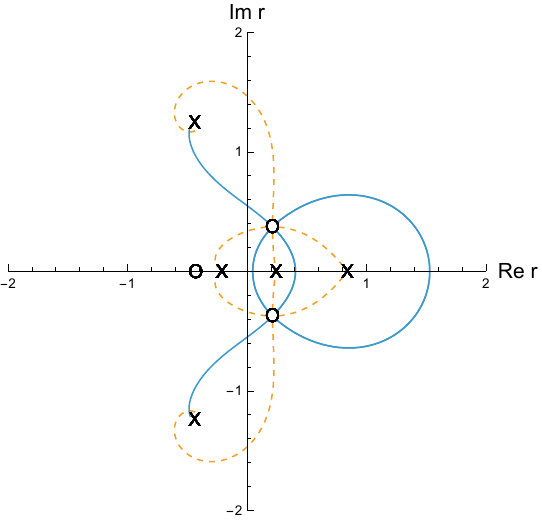}
\caption{The Stokes (dashed) and anti-Stokes (solid) lines of the regular black hole for $d=3,\ m=1.5,\ a=0.05$. The `o' and `x' denote the turning points and the singularities of $f(r)$, respectively.
We assumed $\omega\propto \exp(-i\pi/d)$ (left) and $\omega\propto \exp(-i\pi/2)$ (right). For the left case, the anti-Stokes line extending toward infinity finally comes back to the singular point in the second quadrant.
}
\label{fig:stokes-rbh-d3}
\end{figure}

\begin{figure}[h]
\centering
\includegraphics[height=4cm]{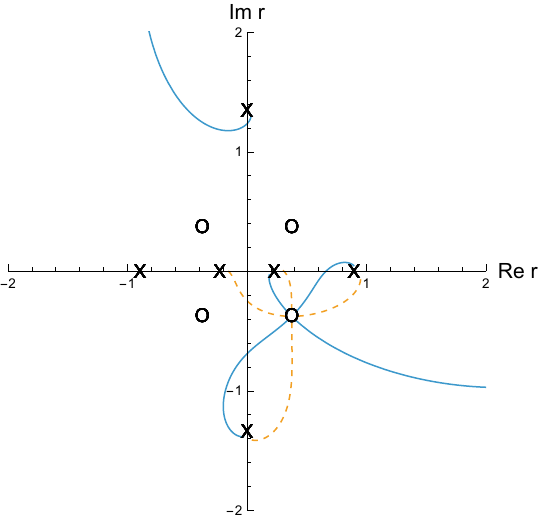}
\includegraphics[height=4cm]{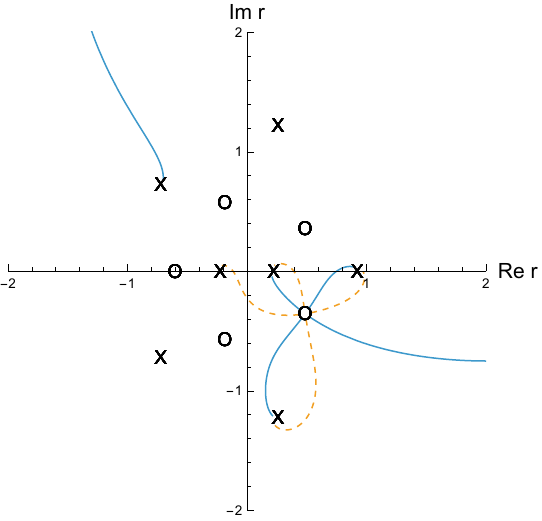}
\includegraphics[height=4cm]{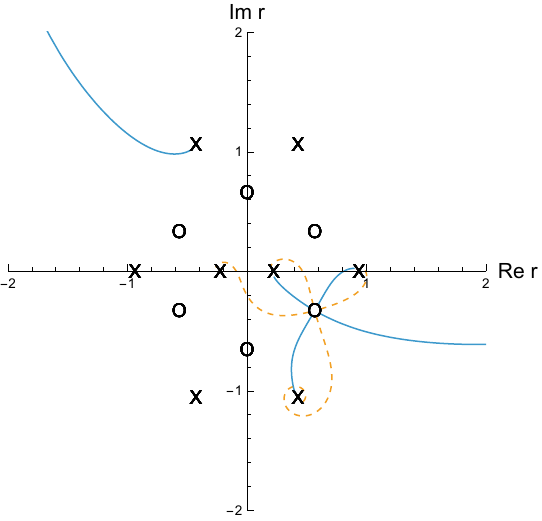}\\
\includegraphics[height=4cm]{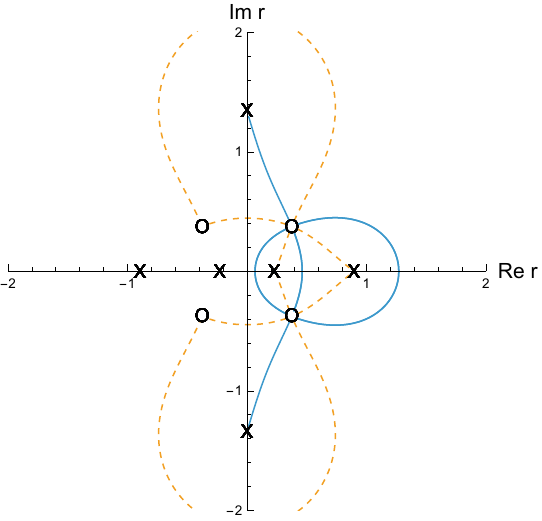}
\includegraphics[height=4cm]{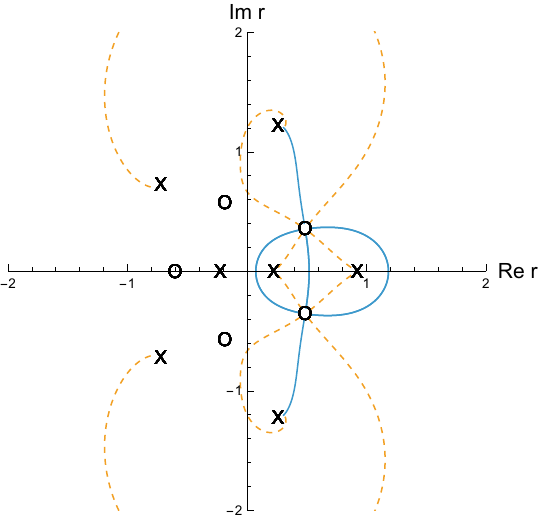}
\includegraphics[height=4cm]{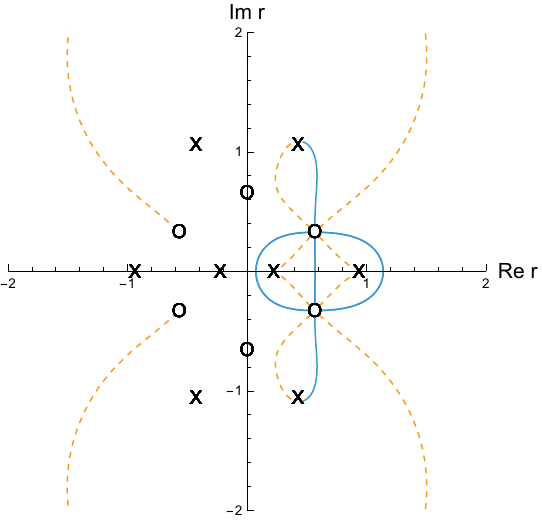}
\caption{The Stokes structure of the regular black hole for $d=4,5,6$ with $m=1.5,\ a=0.05$. $\omega\propto \exp(-i\pi/d)$ for the upper panels and $\omega\propto \exp(-i\pi/2)$ for the lower ones.
}
\label{fig:stokes-rbh-d456}
\end{figure}

For the regular black holes, $f_\text{d}(r)=a r^d-L^2(r^d+a m)$ has $d$ single roots.
$f_\text{n}(r)=a r^d-L^2(r^d+a m)-(r^{d+2}-L^2mr^2)$ has $d$ single roots corresponding to the outer (complex) horizons and two roots corresponding to the inner (complex) horizons.
Figure~\ref{fig:stokes-rbh-d3} shows the Stokes structure for $d=3$ with the assumption $\omega\propto \exp(-i\pi/d)$ and $\omega\propto \exp(-i\pi/2)$.
Figure~\ref{fig:stokes-rbh-d456} shows the $d=4,5,6$ cases of the regular black holes.
In all the cases, $f_2$ is dominant on the Stokes lines into the singular point, $r=r_h$, on the positive real axis.

\subsection{Branch cut around horizon}
\begin{figure}[h]
\centering
\includegraphics[height=5cm]{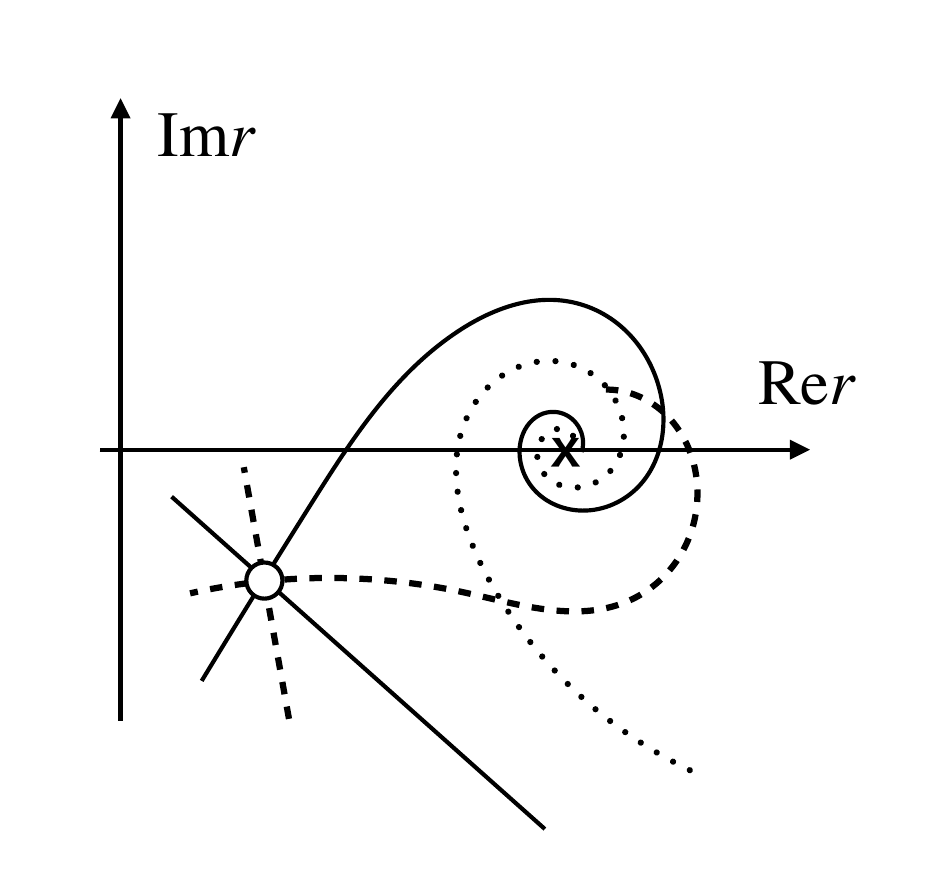}
\caption{The Stokes line (dashed), anti-Stokes line (solid), and branch cut (dotted) around the horizon.
}
\label{fig:log_branch}
\end{figure}

As shown in the left panel of figure~\ref{fig:stokes-rbb-d3} and other cases, Stokes and anti-Stokes lines for a complex $\omega$ are sucked into the singular point at a horizon.
Since the singular point is logarithmic, the lines infinitely wind around it and the rotational directions are opposite for the Stokes and anti-Stokes lines.
The two lines look crossing each other, however, they cannot cross each other at a point other than turning points.
Thus, they are sucked into the singular point in different Riemann sheets.
To analytically continue the WKB solution along the anti-Stokes line, we take the branch cut from the singular point so that it does not cut the anti-Stokes line but the Stokes line as in figure~\ref{fig:log_branch}.
The other end point of the cut can be taken at either infinity, another singular point, or another turning point without cutting the contour for the analytic continuation.

For a pure imaginary $\omega$, we take a branch cut from the real horizon to infinity $r=\infty$ along the real axis.
Since we close our contour in figures~\ref{fig:RBBgp} and~\ref{fig:RBHg} by attaching a great semicircle to the right side, the contour crosses the branch cut once and the analytic continuation leads to the nontrivial monodromy.

\section{Numerical results of QNMs}
\label{sec:Jansen}
A Mathematica package for numerical computation of QNMs in generic black objects are provided by~\cite{Jansen:2017oag}.
The package is based on the Pseudo-spectral method and assumes the ingoing boundary condition at the event horizon.
Adopting the Eddinton-Finkelstein coordinates, this condition is translated to regularity of the wave function as $r\to r_h$ because the other mode  (i.e., outgoing from horizon) is infinitely oscillating there.
In an asymptotically AdS geometry, the Dirichlet boundary condition is also translated to regularity at $r\to\infty$ by an appropriate choice of the wave function.
We compute the QNMs of the regular black branes/holes using this package.

\begin{figure}[h]
 \centering
 \includegraphics[height=5cm]{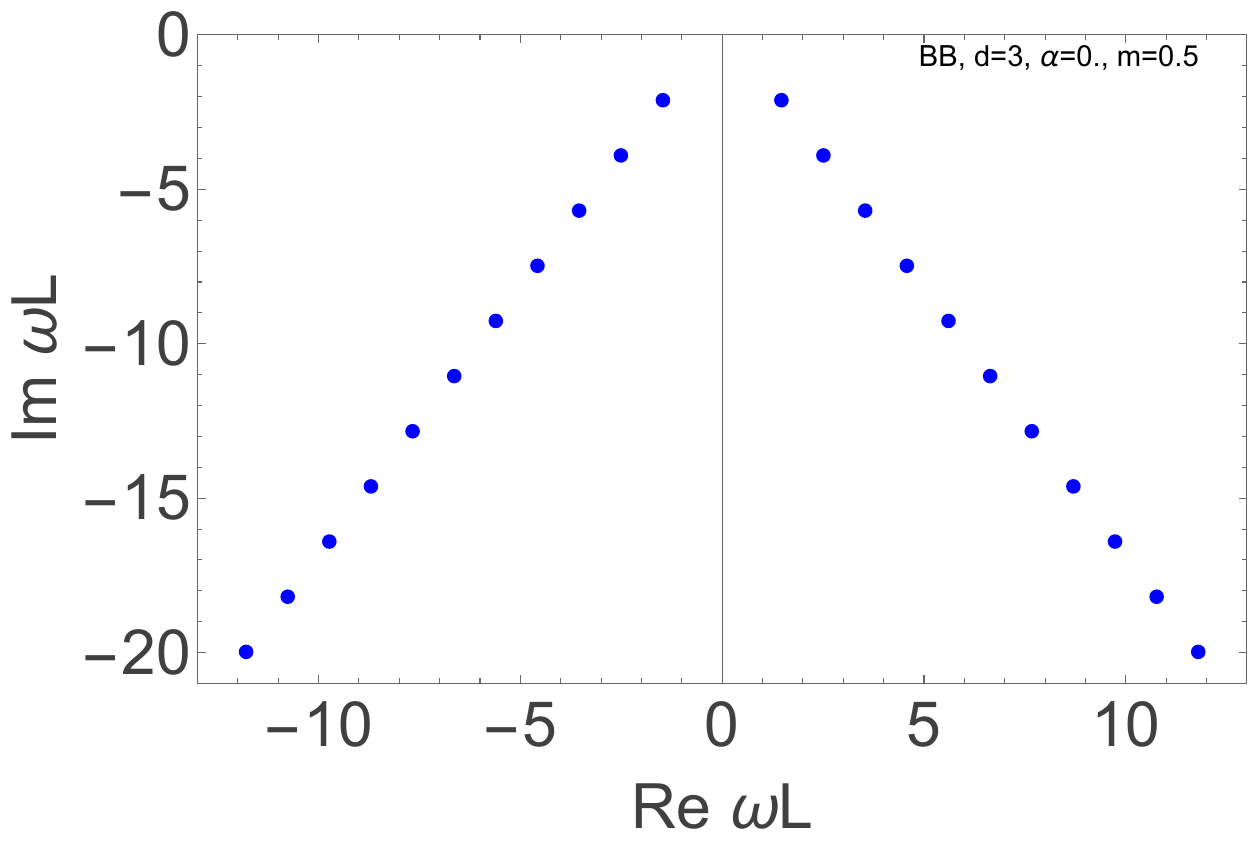}
 \includegraphics[height=5cm]{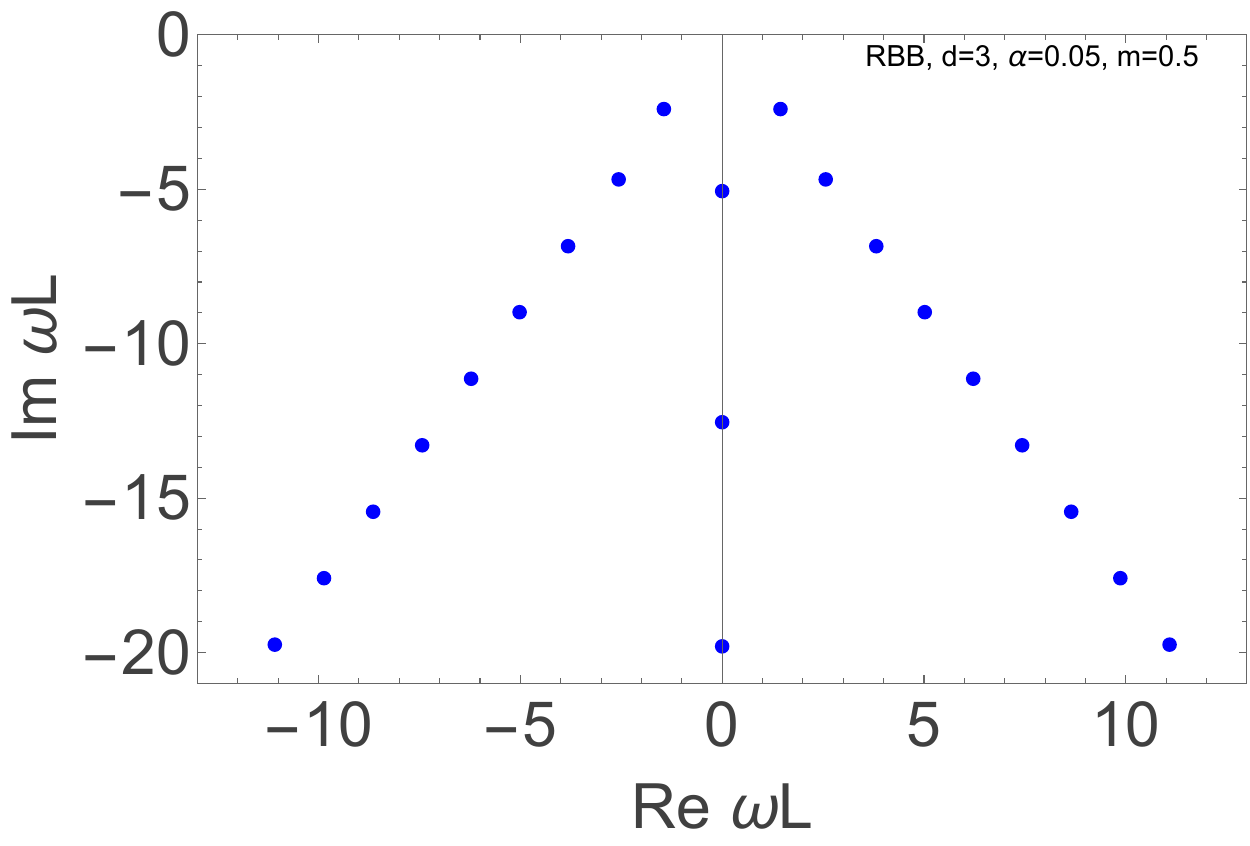}
 \includegraphics[height=5cm]{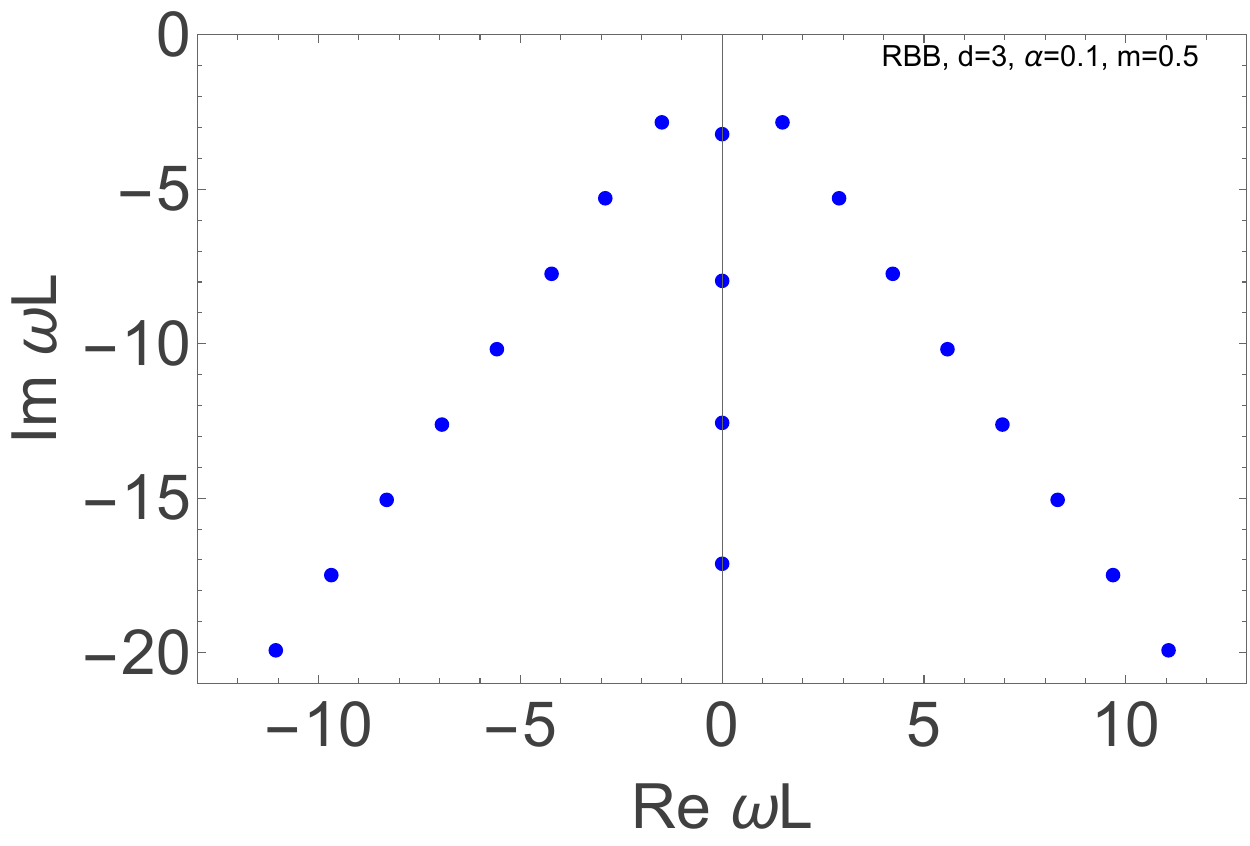}
 \includegraphics[height=5cm]{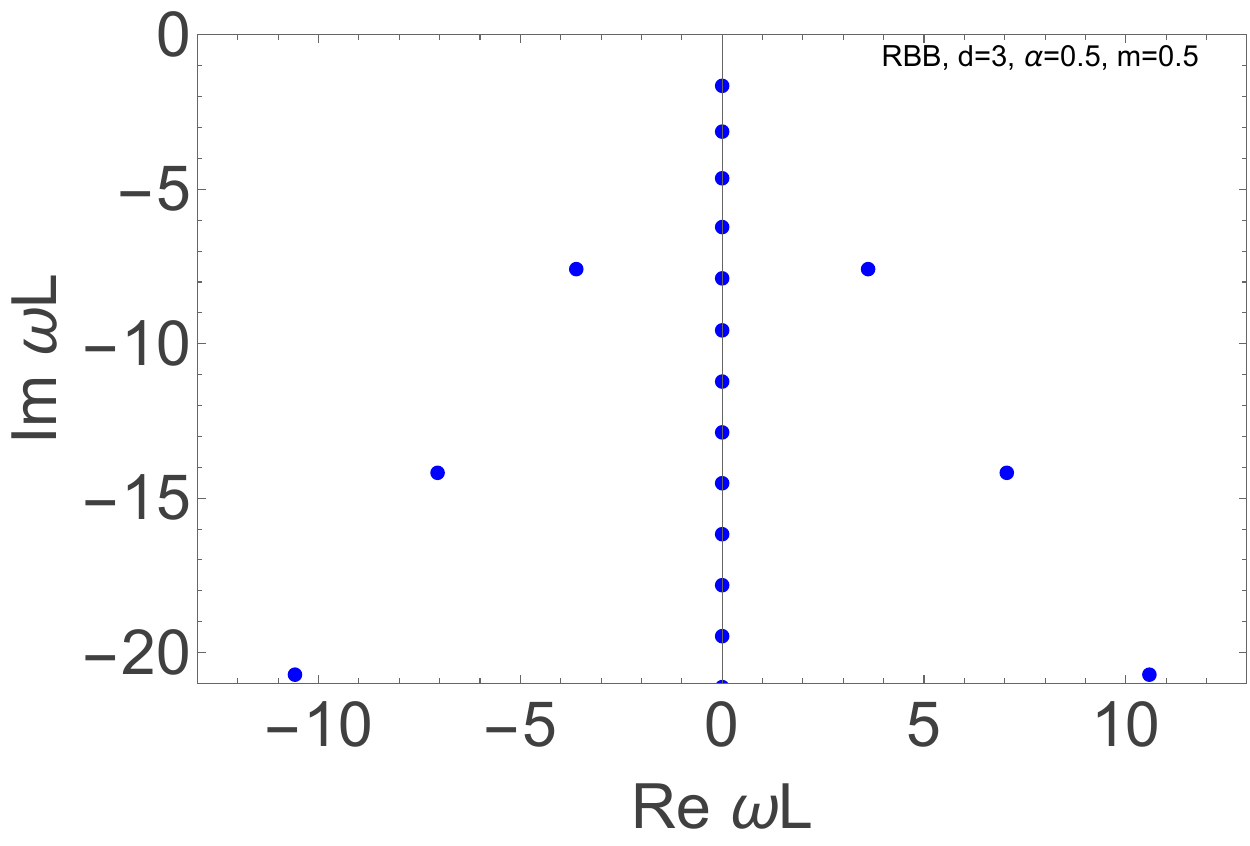}
 \caption{QNMs for the regular black brane with $d=3, \ m=0.5$ and $\alpha=0,0.05,0.1,0.5$.
 }
 \label{fig:qnm-rbb-d3-alpha}
\end{figure}

\begin{figure}[h]
 \centering
 \includegraphics[height=5cm]{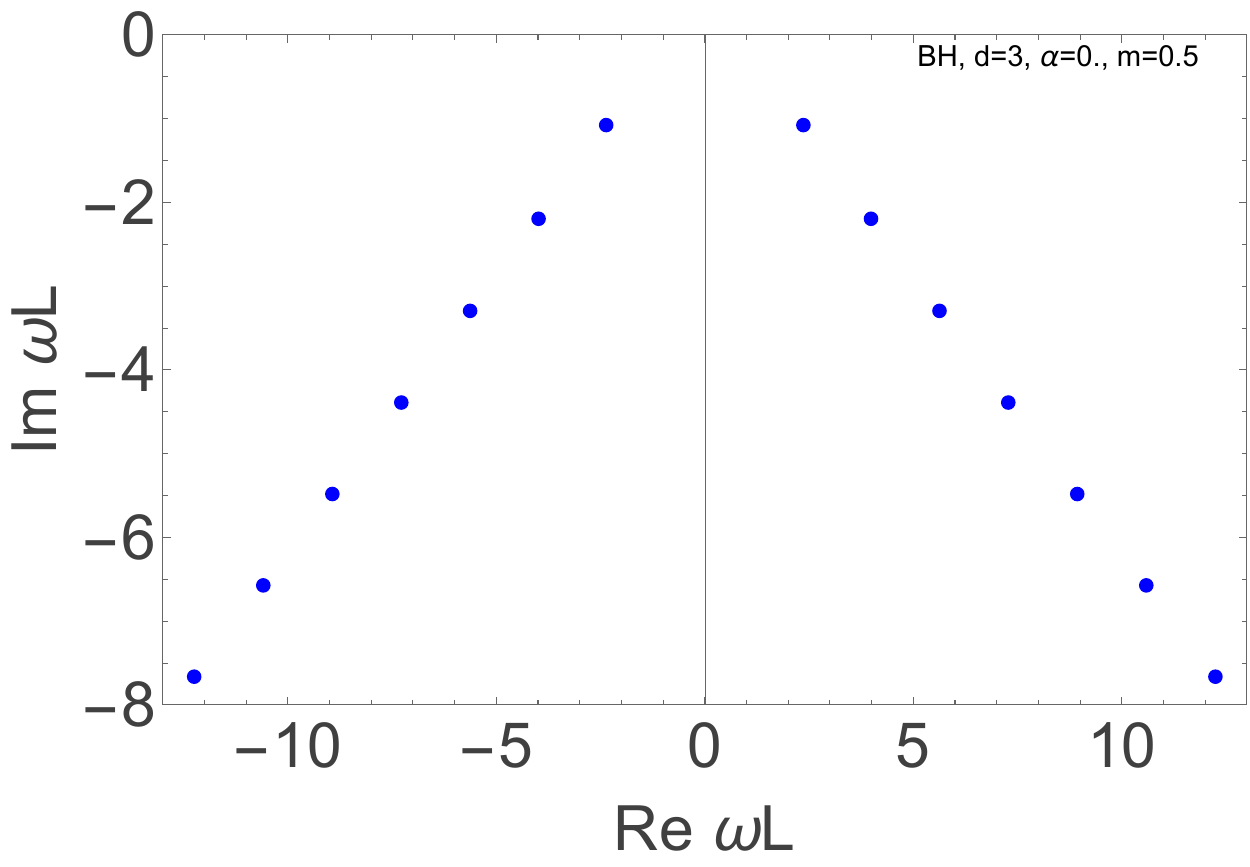}
 \includegraphics[height=5cm]{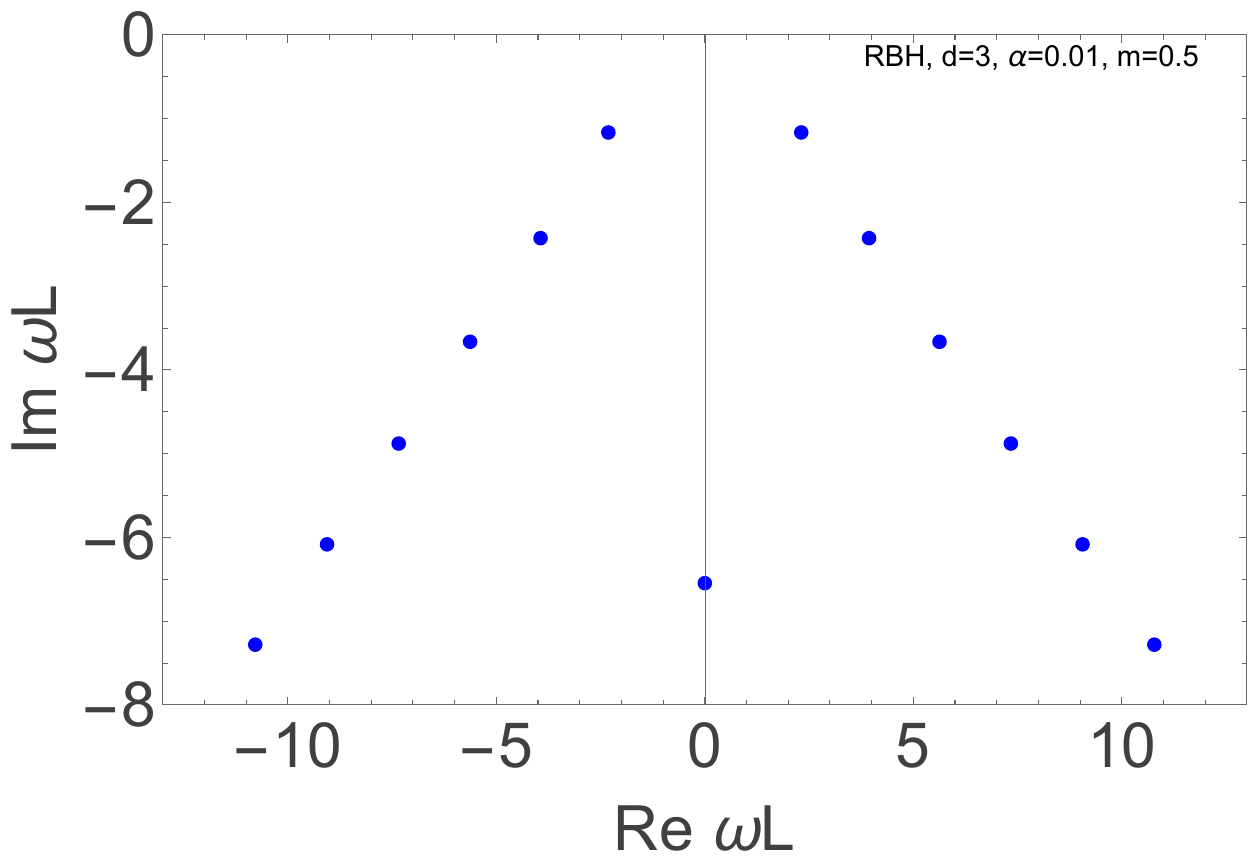}
 \includegraphics[height=5cm]{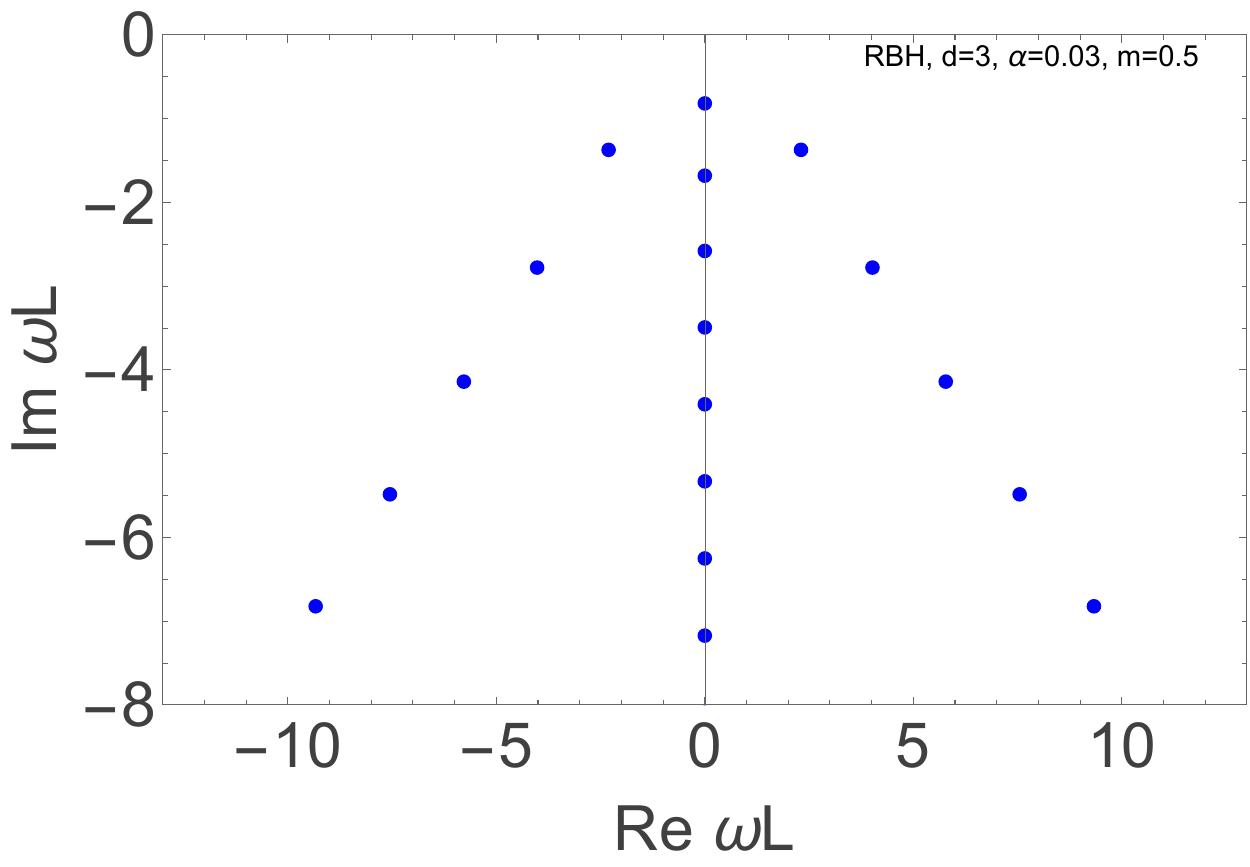}
 \caption{QNMs for the regular black hole with $d=3, \ m=0.5$ and $\alpha=0,0.01,0.03$.
 }
 \label{fig:qnm-rbh-d3-alpha}
\end{figure}

The results are shown in figures~\ref{fig:qnm-rbb-d3-alpha} and \ref{fig:qnm-rbh-d3-alpha} including their characteristic purely imaginary modes.
We also find that the purely imaginary modes of regular black branes become arbitrarily dense for larger $a\to L^2$ and sparse for smaller $a \to 0$.
This dependency is common for other $d$.
For the regular black holes, the similar behavior can be observed, however, the upper bound of $a$ is its extremal value such that the inner and outer horizons degenerate.
The QNMs for several parameters are listed in tables.~\ref{tbl:qnm-bb-d34}-\ref{tbl:qnm-bh-d56}. The complex modes have their counterparts with opposite sign in the real part.
\begin{table}
\begin{tabular}{|c|c|c|c|c|}
\hline
$n$  & $d=3,\ a=0.05$ & $d=3,\ a=0.1$ & $d=4,\ a=0.05$ & $d=4,\ a=0.1$    \\
\hline
1 & $1.44448-2.40137i$ & $1.49476-2.83177i$ & $2.74968-2.58293i$ & $2.9275-2.90966i$ \\
2 & $2.56518-4.67847i$ & $2.8962-5.28989i$ & $4.60833-4.57621i$ & $5.04102-5.15479i$ \\
3 & $3.81824-6.84186i$ & $4.22669-7.73531i$ & $6.50261-6.56747i$ & $7.15555-7.32374i$ \\
4 & $5.01901-8.97789i$ & $5.58183-10.1771i$ & $8.41871-8.52969i$ & $9.25798-9.48101i$ \\
5 & $6.22015-11.1331i$ & $6.94363-12.617i$ & $10.333-10.4722i$ & $11.3624-11.6358i$ \\
6 & $7.4327-13.2879i$ & $8.31167-15.055i$ & $12.2425-12.4086i$ & $13.4701-13.7876i$ \\
7 & $8.64798-15.4396i$ & $9.68409-17.4916i$ & $14.1503-14.3442i$ & $15.5804-15.9369i$ \\
8 & $9.86516-17.5913i$ & $11.0599-19.9272i$ & $16.0586-16.2797i$ & $17.6927-18.0845i$ \\
9 & $11.0847-19.7426i$ & $12.4383-22.3618i$ & $17.9683-18.2149i$ & $19.8069-20.2306i$ \\
10 & $12.3059-21.8934i$ &  & $19.8792-20.1493i$ & $21.9225-22.3755i$ \\
\hline
1 & $-5.05841i$ & $-3.2142i$ & $-9.74854i$ & $-5.96071i$ \\
2 & $-12.5414i$ & $-7.96359i$ & $-23.8626i$ & $-14.3044i$ \\
3 & $-19.7985i$ & $-12.5618i$ & $-37.4908i$ & $-22.436i$ \\
4 & $-26.993i$ & $-17.124i$ & $-50.9794i$ & $-30.4946i$ \\
5 & $-34.1605i$ & $-21.6699i$ & $-64.4075i$ &  \\
6 & $-41.3135i$ & $-26.2073i$ &  &  \\
7 & $-48.458i$ &  &  &  \\
 %
\hline\end{tabular}
\caption{$\omega L$ of the regular black brane with $m= 0.5$ and $d=3,4$.}
\label{tbl:qnm-bb-d34}\end{table}
\begin{table}
\begin{tabular}{|c|c|c|c|c|}
\hline
$n$  & $d=5,\ a=0.1$ & $d=5,\ a=0.5$ & $d=6,\ a=0.1$ & $d=6,\ a=0.5$    \\
\hline
1 & $4.07064-2.88148i$ & $8.72756-6.76625i$ & $5.05218-2.81892i$ & $10.5121-6.38945i$ \\
2 & $6.62662-4.84015i$ & $14.4627-11.1825i$ & $7.90896-4.54087i$ & $16.689-10.151i$ \\
3 & $9.16918-6.74504i$ & $20.1351-15.4865i$ & $10.7172-6.21656i$ & $22.7371-13.7915i$ \\
4 & $11.6996-8.62344i$ & $25.7962-19.7442i$ & $13.5089-7.8666i$ & $28.7479-17.3823i$ \\
5 & $14.223-10.4925i$ & $31.4565-23.9764i$ & $16.2906-9.50318i$ & $34.7445-20.9458i$ \\
6 & $16.7443-12.3575i$ & $37.1187-28.1923i$ & $19.0664-11.133i$ & $40.7352-24.492i$ \\
7 & $19.2656-14.2199i$ & $42.7833-32.3969i$ & $21.8393-12.759i$ & $46.7233-28.0262i$ \\
8 & $21.7875-16.0802i$ & $48.4503-36.5931i$ & $24.6107-14.3825i$ & $52.7104-31.5516i$ \\
9 & $24.3101-17.9388i$ & $54.1195-40.783i$ & $27.3815-16.0041i$ & $58.6973-35.0701i$ \\
10 & $26.8335-19.796i$ & $59.7908-44.9677i$ & $30.1521-17.6243i$ & $64.6844-38.5833i$ \\
\hline
1 & $-9.0201i$ & $-3.406i$ & $-12.3036i$ & $-4.33439i$ \\
2 & $-21.5044i$ & $-6.60185i$ & $-29.2579i$ & $-8.45052i$ \\
3 & $-33.6284i$ & $-9.91947i$ & $-45.6687i$ & $-12.7021i$ \\
4 & $-45.6308i$ & $-13.305i$ & $-61.9011i$ & $-17.0166i$ \\
5 & $-57.579i$ & $-16.7116i$ & $-78.0545i$ & $-21.3523i$ \\
6 & $-69.4982i$ & $-20.1252i$ & $-94.166i$ & $-25.6963i$ \\
7 & $-81.3999i$ & $-23.5419i$ &  & $-30.0438i$ \\
 \hline\end{tabular}
\caption{$\omega L$ of the regular black brane with $m=0.5$ and $d=5,6$. 
}
\label{tbl:qnm-bb-d56}\end{table}
\begin{table}
\begin{tabular}{|c|c|c|c|c|}
\hline
$n$  & $d=3,\ a=0.01$ & $d=3,\ a=0.03$ & $d=4,\ a=0.05$ & $d=4,\ a=0.1$    \\
\hline
1 & $2.31329-1.16488i$ & $2.30713-1.37214i$ & $3.98631-1.72451i$ & $4.18095-2.05163i$ \\
2 & $3.93949-2.42657i$ & $4.02139-2.7778i$ & $6.4604-3.31041i$ & $6.84732-3.85661i$ \\
3 & $5.62888-3.6637i$ & $5.77825-4.13975i$ & $8.98772-4.82426i$ & $9.56597-5.58659i$ \\
4 & $7.34083-4.87895i$ & $7.55117-5.48523i$ & $11.535-6.31263i$ & $12.3067-7.28757i$ \\
5 & $9.05924-6.08236i$ & $9.33248-6.82205i$ & $14.0936-7.78797i$ & $15.0595-8.97324i$ \\
6 & $10.7807-7.2803i$ & $11.1181-8.15339i$ & $16.6592-9.25529i$ & $17.8199-10.6495i$ \\
7 & $12.5046-8.47521i$ &  & $19.2297-10.7172i$ & $20.5854-12.3194i$ \\
8 & $14.2304-9.668i$ &  & $21.8037-12.1752i$ & $23.3476-13.9907i$ \\
9 & $15.9579-10.8591i$ &  & $24.3804-13.6303i$ &  \\
10 & $17.6867-12.049i$ &  & $26.9592-15.0831i$ &  \\
\hline
1 & $-6.54598i$ & $-0.818098i$ & $-5.00239i$ & $-1.20248i$ \\
2 & $-15.5106i$ & $-1.68062i$ & $-11.1256i$ & $-2.39376i$ \\
3 & $-24.437i$ & $-2.57958i$ & $-17.2983i$ & $-3.61247i$ \\
4 & $-33.3255i$ & $-3.49247i$ & $-23.4623i$ & $-4.84898i$ \\
5 &  & $-4.4101i$ & $-29.6173i$ & $-6.09397i$ \\
6 &  & $-5.32992i$ & $-35.7658i$ & $-7.34306i$ \\
7 &  & $-6.25088i$ & $-41.9096i$ & $-8.59434i$ \\
 \hline\end{tabular}
\caption{$\omega L$ of the regular black hole with $m=0.5$ and $d=3,4$.}
\label{tbl:qnm-bh-d34}\end{table}
\begin{table}
\begin{tabular}{|c|c|c|c|c|}
\hline
$n$  & $d=5,\ a=0.1$ & $d=5,\ a=0.15$ & $d=6,\ a=0.1$ & $d=6,\ a=0.2$    \\
\hline
1 & $5.70015-2.01894i$ & $6.00254-2.31558i$ & $7.07388-1.93973i$ & $7.8528-2.50807i$ \\
2 & $8.80817-3.67233i$ & $9.34092-4.15606i$ & $10.4542-3.44354i$ & $11.7492-4.3358i$ \\
3 & $11.9644-5.23596i$ & $12.7254-5.89914i$ & $13.8791-4.8469i$ & $15.6772-6.04826i$ \\
4 & $15.1424-6.76559i$ & $16.1314-7.60438i$ & $17.3242-6.2122i$ & $19.6226-7.71556i$ \\
5 & $18.3328-8.27737i$ & $19.5497-9.28961i$ & $20.781-7.55783i$ & $23.5786-9.35889i$ \\
6 & $21.5312-9.7782i$ & $22.9759-10.9625i$ & $24.2457-8.89147i$ & $27.5418-10.9875i$ \\
7 & $24.735-11.2716i$ & $26.4076-12.6272i$ & $27.7159-10.2171i$ & $31.51-12.6061i$ \\
8 & $27.9429-12.7598i$ & $29.8432-14.2858i$ & $31.1901-11.5369i$ & $35.482-14.2175i$ \\
9 & $31.1538-14.2439i$ & $33.2818-15.94i$ & $34.6674-12.8524i$ & $39.4569-15.8235i$ \\
10 & $34.3672-15.725i$ &  & $38.1471-14.1645i$ & $43.4342-17.4253i$ \\
\hline
1 & $-4.69887i$ & $-2.16237i$ & $-8.01022i$ & $-2.56104i$ \\
2 & $-9.93335i$ & $-4.28237i$ & $-17.2666i$ & $-4.99694i$ \\
3 & $-15.2688i$ & $-6.45208i$ & $-26.5944i$ & $-7.4714i$ \\
4 & $-20.6171i$ & $-8.64801i$ & $-35.909i$ & $-9.97393i$ \\
5 & $-25.9659i$ & $-10.8557i$ & $-45.2102i$ & $-12.4914i$ \\
6 & $-31.313i$ & $-13.0691i$ &  & $-15.0168i$ \\
7 & $-36.6583i$ & $-15.2854i$ &  & $-17.5468i$ \\
 \hline\end{tabular}
\caption{$\omega L$ of the regular black hole with $m=0.5$ and $d=5,6$.}
\label{tbl:qnm-bh-d56}\end{table}

\section{AdS-Reissner-Nordstr{\"o}m black hole}
\label{app:ChargedBH}

\begin{figure}
 \centering
 \includegraphics[height=6cm]{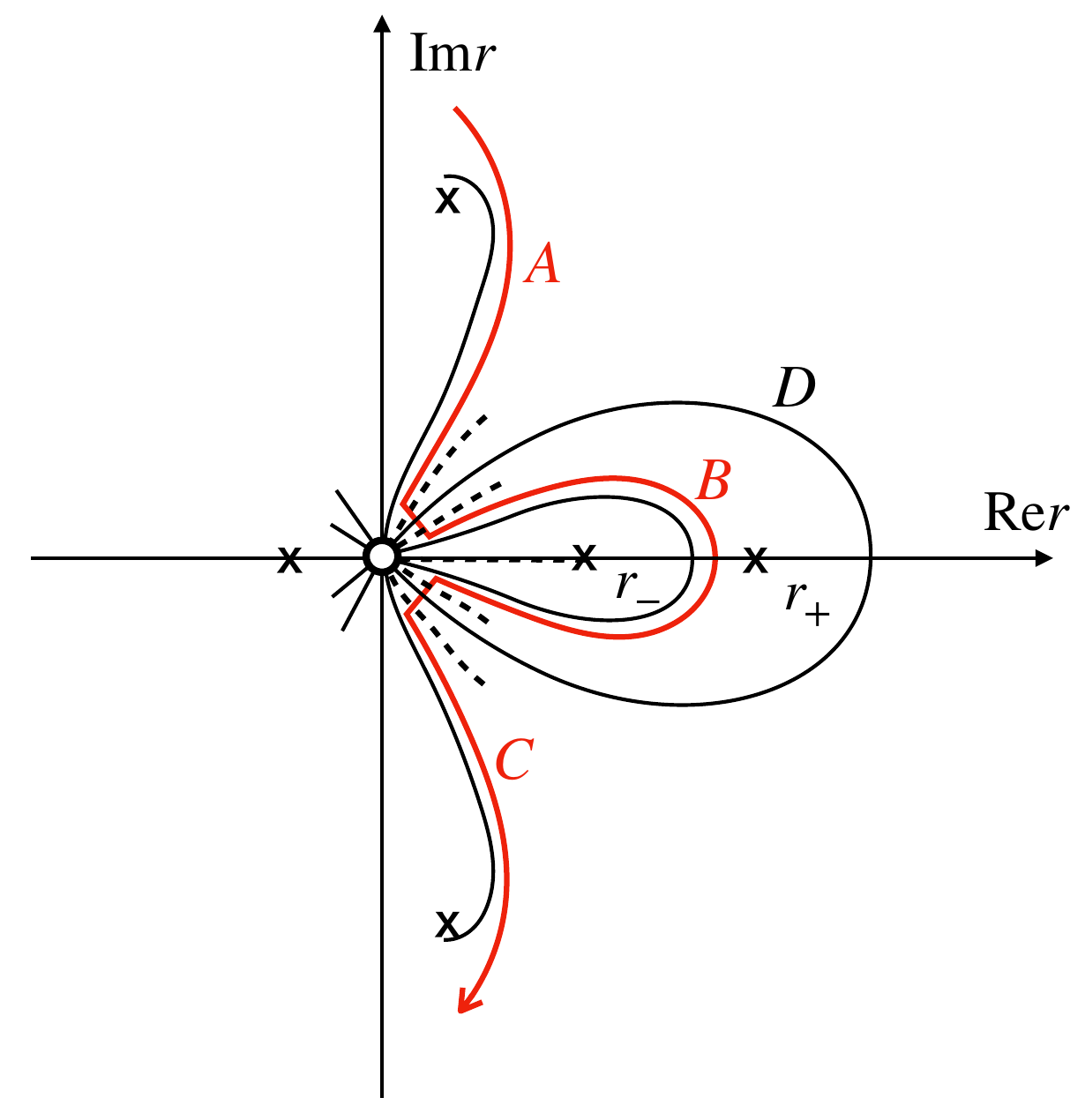}
 \caption{
 Stokes and anti-Stokes lines for the AdS-Reissner-Nordstr\"{o}m black hole with purely imaginary $\omega$.
 }
 \label{fig:CAdSB}
\end{figure}

The QNMs of the AdS-Reissner-Nordstr\"{o}m black hole with large $|\mathrm{Im}\,\omega|$ and $|\mathrm{Re}\,\omega|$ were computed in \cite{Natario:2004jd} using the method of \cite{Motl:2003cd}. In \cite{Jansen:2017oag}, a series of purely imaginary QNMs was found numerically. In this appendix, we derive this series using the method of \cite{Motl:2003cd}.
We use the metric given by \eqref{eq:metricbh}
with
\begin{align} \label{eq:fr3}
f(r) = 1 - \frac{2 m}{r^{d - 2}} + \frac{q}{r^{2 d - 4}} + r^2.
\end{align}
There are outer and inner horizons located at $r=r_+$ and $r = r_{-}$, respectively. 
The zero of $1/f(r)$ is the location of black hole singularity at $r = 0$.
Stokes and anti-Stokes lines for the wave functions can be found as in figure~\ref{fig:CAdSB}.

Let us start from contour $A$ near $r = 0$. Along this contour, the tortoise coordinate behaves as
\begin{align}
 \hat z \equiv \int_{0}^r \frac{dr'}{f(r')} \sim \frac{r^{2d-3}}{(2 d - 3) q} .
\end{align}
Since the potential in the radial equation becomes 
\begin{align}
    V[r] \sim \frac{j^2 -1}{4 \hat z^2} ,
\end{align}
where $j = (d-2)/(2 d -3)$,
the solution is given by%
\footnote{In this case, we can eliminate the explicit dependence on $\omega$ from the differential equation by introducing a new variable $x=\omega\hat z$. Thus, we can obtain the wave function to all orders in $1/\omega$ for finite $\omega\hat z$. This is different from the case of regular black branes/holes, where the differential equations can only be treated perturbatively in $1/\sqrt{\omega}$.
}
\begin{align} \label{eq:BpBm}
    \psi (r) =
    b_+ \sqrt{2 \pi \omega \hat z} J_{j/2} (\omega \hat z) + b_- \sqrt{2 \pi \omega \hat z} J_{ - j/2} (\omega \hat z) .
\end{align}

For $|\omega \hat z| \gg 1$, the wave function has the asymptotic form
\begin{align}
\begin{aligned}
    \psi (r) &\sim 2 b_+ \cos (\omega \hat z - \alpha_+) + 2 b_- \cos (\omega \hat z - \alpha_-) \\
     &= (b_+ e^{- i \alpha_+ }+ b_- e^{- i \alpha_-}) e^{i \omega \hat z} + (b_+ e^{i \alpha_+} + b_- e^{i \alpha_-}) e^{- i \omega \hat z} ,
     \end{aligned}
\end{align}
where
\begin{align}
\alpha_\pm = \frac{\pi}{4} (1 \pm j) .
\end{align}
Here, we have used the asymptotic expansion of the Bessel function,
\begin{align}
    J_\nu (x) \sim \sqrt{\frac{2}{\pi x}} \cos \left( x - \frac{\nu \pi}{2} - \frac{\pi}{4}\right) .
\end{align}
In the same region, the WKB wave function takes the form
\begin{align}
    \psi (r) \sim \frac{c_+}{\sqrt{Q(r)}} e^{i \int_0^{r} \frac{Q(r')dr'}{f(r')}} + \frac{c_-}{\sqrt{Q(r)}} e^{ -i \int_0^{r} \frac{Q(r')dr'}{f(r')}} .
\end{align}
For $|\omega|^2\gg |V(r)|$, the integrals in the exponents can be approximated as
\begin{align}
    \pm i \int_0^{r} \frac{Q(r')dr'}{f(r')} \sim \pm i \omega \int_0^r \frac{dr'}{f(r')} = \pm i \omega \hat z .
\end{align}

In order to match the expression in \eqref{eq:BpBm}, we set
\begin{align}
    c_+ = b_+ e^{- i \alpha_+ }+ b_- e^{- i \alpha_-} , \quad 
    c_- = b_+ e^{i \alpha_+} + b_- e^{i \alpha_-} .
\end{align}
We would like to continue the wave function to the infinity $r = i \infty$ to impose the Dirichlet boundary condition.
However, there is no anti-Stokes line directly connecting the origin $r=0$ to the infinity, so we cannot directly track the coefficients of the two terms in the WKB wave function.
Nevertheless, for large $r$, the wave function should behave as
\begin{align} \label{eq:RNpsi2}
    \psi (z) \sim d_+ e^{-i \omega z} + d_-  e^{ i \omega z} ,
\end{align}
where $d_+$ and $d_-$ are of the same order of magnitude when the exponential factor $\exp(2i\omega z_*)$ is taken into account.
Here, we have defined
\begin{align}
 z \equiv  \int_r^{\infty} \frac{dr'}{f(r')} , \quad 
    z_* = \int_0^{\infty} \frac{dr'}{f(r')} .
\end{align}
In particular, we have $\hat z = z_* - z$.
We thus require
\begin{align}
    c_+ e^{i \omega z_*} \sim c_- e^{-i \omega z_*} .
\end{align}
Since $-\mathrm{Im}\,\omega\gg1$ and $\mathrm{Re}\,z_* >0$, we can approximately impose
\begin{align} \label{eq:RNcond1}
    c_+ = b_+ e^{- i \alpha_+} + b_- e^{- i \alpha_-} = 0 .
\end{align}
This is the condition we impose on the wave function along contour $A$.

We started near $r = 0$ along contour $A$. We continue the wave function to the region near $r = 0$ along contour $B$. Then the wave function acquires phase factors, which leads to
\begin{align} \label{eq:RNasym1}
\begin{aligned}
    \psi (r) &\sim 2 b_+ e^{- 4 i \alpha_+} \cos (\omega \hat z - \alpha_+) + 2 b_- e^{- 4 i \alpha_-} \cos (\omega \hat z - \alpha_-) \\
   & = (b_+ e^{-5 i \alpha_+} + b_- e^{-5 i \alpha_-}) e^{i \omega \hat z} + (b_+ e^{-3 i \alpha_+ }+ b_- e^{-3 i \alpha_-}) e^{-i \omega \hat z} .
   \end{aligned}
\end{align}
Here we have used 
\begin{align}
\begin{aligned}
    \sqrt{2 \pi e^{- 2 \pi i } \omega \hat z } J_{\pm j/2} (e^{- 2 \pi i} \omega \hat z) &= e^{- \frac{2 \pi i}{2} (1 \pm j )} \sqrt{2 \pi \omega \hat z} J_{\pm j/2} (\omega \hat z) \\
    &\sim 2 e^{- 4 i \alpha_\pm} \cos (\omega \hat z - \alpha_\pm) .
    \end{aligned}
\end{align}
We now consider a contour that goes around the inner horizon at $r=r_-$. This results in a shift
\begin{align}
    \delta = - \left. \frac{2 \pi i}{\frac{d}{dr}f(r)} \right|_{r = r_-} = \frac{\pi i}{k^-} , \quad k^- = - \left. \frac12 \frac{d}{dr}f(r) \right |_{r = r_-}.
\end{align}
After going around the inner horizon, the solution can be written near $r \sim 0$ as
\begin{align} \label{eq:RNsol1}
    \psi(r) \sim c_+ \sqrt{2 \pi \omega (\hat z - \delta)} J_{j/2} (\omega (\hat z - \delta)) + c_- \sqrt{2 \pi \omega (\hat z - \delta)} J_{-j/2} (\omega (\hat z - \delta)) ,
 \end{align}
which has the following asymptotic form for $\omega (\hat z - \delta) \ll -1$:
\begin{align} \label{eq:RNasym2}
\begin{aligned}
    \psi (r) &\sim 2 c_+ \cos (\omega (\hat z - \delta) + \alpha_+) + 2 c_- \cos (\omega (\hat z - \delta) + \alpha_- ) \\
    &\sim (c_+e^{i\alpha_+}+c_-e^{i\alpha_-})e^{-i\omega\delta}e^{i\omega\hat z} +(c_+e^{-i\alpha_+}+c_-e^{-i\alpha_-})e^{i\omega\delta}e^{-i\omega\hat z}.
    \end{aligned}
\end{align}
Comparing \eqref{eq:RNasym1} and \eqref{eq:RNasym2}, we obtain the matching conditions
\begin{align} \label{eq:RNcond2}
\begin{aligned}
&b_+ e^{- 5 i \alpha_+} + b_- e^{- 5 i \alpha_-} =(c_+e^{i\alpha_+}+c_-e^{i\alpha_-})e^{-i\omega\delta},\\
&b_+ e^{ - 3 i \alpha_+ }+ b_- e^{ - 3 i \alpha_-}  =(c_+e^{-i\alpha_+}+c_-e^{-i\alpha_-})e^{i\omega\delta}.
\end{aligned}
\end{align}
After a clockwise rotation by $2\pi$ in the $\hat z$ plane around $r=0$, the solution \eqref{eq:RNsol1}  acquires phase factors and takes the form
\begin{align}
\begin{aligned}
    \psi (r) &\sim 2 c_+ e^{- 4 i \alpha_+ } \cos (\omega (\hat z - \delta) + \alpha_+) + 2 c_- e^{- 4 i \alpha_- } \cos (\omega (\hat z - \delta) + \alpha_-) \\ 
    &=(c_+e^{-3i\alpha_+}+c_-e^{-3i\alpha_-})e^{-i\omega\delta}e^{i\omega\hat z} +(c_+e^{-5i\alpha_+}+c_-e^{-5i\alpha_-})e^{i\omega\delta}e^{-i\omega\hat z}
    \end{aligned}
\end{align}
for $\omega (\hat z - \delta) < 0$.
Here we have used 
\begin{align}
\begin{aligned}
    \sqrt{2 \pi e^{- 2 \pi i } \omega (\hat z - \delta) } J_{\pm j/2} (e^{- 2 \pi i} \omega (\hat z - \delta )) &= e^{- \frac{2 \pi i}{2} (1 \pm j )} \sqrt{2 \pi \omega (\hat z - \delta )} J_{\pm j/2} (\omega (\hat z - \delta )) \\
    &\sim 2 e^{- 4 i \alpha_\pm} \cos (\omega (\hat z - \delta) + \alpha_\pm) .
    \end{aligned}
\end{align}

We again compare this expression with the solution at large $r$, which should behave as
\begin{align} \label{eq:RNpsi1}
    \psi (z) \sim \hat d_+ e^{-i \omega z } + \hat d_-  e^{ i \omega z } ,
\end{align}
where the ratio $\hat d_+/\hat d_-$ does not contain exponentially large or small factors of $\exp(2i\omega z_*)$.
We thus obtain another condition,
\begin{align} 
    (c_+ e^{-3 i \alpha_+}  + c_- e^{-3 i \alpha_-}  ) e^{- i \omega \delta }e^{ i \omega  z_* }  \sim  (c_+ e^{- 5 i \alpha_+} + c_- e^{ - 5 i \alpha_-} ) e^{ i \omega \delta}e^{ - i \omega  z_* } .
\end{align}
Hence, to leading order, we can impose
\begin{align}\label{eq:RNcond3}
    c_+ e^{- 3 i \alpha_+}  + c_- e^{ - 3 i \alpha_-}  = 0 .
\end{align}

We require nontrivial solutions to \eqref{eq:RNcond1}, \eqref{eq:RNcond2} and \eqref{eq:RNcond3}. This leads to the following condition on $\omega$:
\begin{align}
    A_\text{Q} \omega \sim \pi n + C_\text{Q},
\end{align}
where
\begin{align}
    A_\text{Q} = \delta ,  \quad C_\text{Q} = - \frac{\pi}{2} .
\end{align}
Here we have used $\alpha_+ + \alpha_- = \pi /2$.
Purely imaginary QNMs has been obtained numerically in \cite{Jansen:2017oag}. We have repeated the numerical computations for different parameter values and compared the results with our analytic prediction in table~\ref{tab:qnm-imag-rnbh}. We find good agreement for the linear term in $n$. However, the agreement for the constant term is only qualitative. We have neglected contributions of order $\exp(2 i \omega z_*)$, which may be the main source of the discrepancy. In order to obtain better agreement, we would need either an analytic result valid at finite $\omega$ or numerical computations for much larger $n$.
\begin{table}[htbp]
    \centering
 \[
\begin{array}{|c|c|c|c|c|}
\hline
~ & \multicolumn{2}{|c|}{\text{WKB prediction}} & \multicolumn{2}{|c|}{\text{Numerical fit}}  \\
\hline
(d,a) & A_\text{Q} & C_\text{Q} & A_\text{Q} & C_\text{Q} \\ \hline
(3, 0.4)
& ~~0.46 i ~~
& ~ - 1.57 ~
& ~~ 0.47 i ~~
& ~ - 1.25 ~
\\
(3, 0.45)
& 2.01 i
& - 1.57
& 2.07 i
& - 0.91
\\
(4, 0.4)
& 0.76i
& - 1.57
& 0.77 i
& -1.04
\\
(4, 0.41)
& 1.04 i
& - 1.57
& 1.06 i
& -0.98
\\
(5, 0.4)
& 1.13 i
& - 1.57
& 1.16 i
& -0.87
\\
(5, 0.38)
& 0.55 i
& - 1.57
& 0.56 i
& -1.03
\\
(6, 0.39)
& 0.97 i
& - 1.57
& 1.00 i
& -0.81
\\
(6, 0.38)
& 0.62 i
& - 1.57
& 0.63 i
& -0.96
\\
\hline
\end{array}
\]
       \caption{Comparison of purely imaginary QNMs of the AdS-Reissner–Nordstr\"{o}m black hole obtained from WKB analysis and numerical computations. We set $L=1$ and $m=0.5$.}
    \label{tab:qnm-imag-rnbh}
\end{table}

\section{Nonzero coefficients \texorpdfstring{$C^{i,j,k}_{b,c,d}$}{C(i,j,k)_(b,c,d)}}
\label{app:C}

In section~\ref{sec:OPE}, we compute the conformal block $G_\text{T} (\tau)$ in \eqref{eq:GT}, where the coefficients $a_n = a^n_{0,2n}$ are obtained by solving the set of algebraic equations \eqref{eq:algebraic}.
In table \ref{tab:assocC}, we list nonzero components of $C^{i,j,k}_{b,c,d}$ appearing in these equation.
\begin{table}[htbp]
\centering
\begin{tabular}{|c|l|}
\hline
$(b,c,d)$ & $C_{b,c,d}^{i,j,k}$ \\
\hline
$(0,0,2)$ & $2 j (2 j+1)$ \\
\hline
$(0,1,1)$ & $4 (4 i-k) (k-\Delta )$ \\
\hline
$(0,1,2)$ & $4 (2 i-j-k) (\Delta +2 i-j-k-2)$ \\
\hline
$(1,0,2)$ & $2 \left(2 a ^2+a -1\right) j (2 j+1)$ \\
\hline
$(1,1,0)$ & $-8 (a +1)^2 \left(\Delta ^2+\Delta +k^2-2 \Delta  k-k\right)$ \\
\hline
$(1,1,1)$ & $2 (a +1) (k-\Delta ) (-4 a  \Delta -3 a -4 \Delta +8 (a -2) i+4 (a +1) j+2 a  k+8 k-3)$ \\
\hline
$(1,1,2)$ & 
$
\begin{aligned}
&(a +1) \left(-a  \Delta ^2-\Delta ^2+16 (a -2) i^2
\right. \\
&
\left.
-8 i (-a  \Delta +4 a +2 \Delta +2 (a -2) j+2 (a -2) k-2) 
+4 (a -2) j^2
\right. \\
&
\left.
+4 j (-a  \Delta +4 a +2 \Delta +2 (a -2) k-2)+4 a  k^2-8 k^2-4 a  \Delta  k
\right. \\
&
\left.
+16 a  k+8 \Delta  k-8 k\right)
\end{aligned}
$ 
\\
\hline
$(1,2,0)$ & $-4 (a +1)^2 \left(\Delta ^2+\Delta +k^2-2 \Delta  k-k\right)$ \\
\hline
$(2,0,2)$ & $2 (a -2) a  (a +1)^2 j (2 j+1)$ \\
\hline
$(2,1,0)$ & $-4 (a +1)^3 (3 a -1) \left(\Delta ^2+\Delta +k^2-2 \Delta  k-k\right)$ \\
\hline
$(2,1,1)$ & $
\begin{aligned}
&4 (a +1)^2 (k-\Delta ) \left(-3 a ^2 \Delta -2 a  \Delta +a +\Delta +(4-8 a ) i
\right. \\
&
\left.
+2 \left(a ^2-1\right) j+3 a ^2 k+4 a  k-2 k+1\right)
\end{aligned}$ \\
\hline
$(2,1,2)$ & $
\begin{aligned}
&-(a +1)^2 \left(a ^2 \Delta ^2-4 a ^2 \Delta -4 a  \Delta -\Delta ^2+16 (2 a -1) i^2
\right.
\\ &
\left.
-8 i (-2 a  \Delta +6 a +\Delta 
+(4 a -2) j+4 a  k-2 k)+(8 a -4) j^2
\right.
\\ &
\left.
+4 j (-2 a  \Delta
+6 a +\Delta +4 a  k-2 k)+8 a  k^2-4 k^2-8 a  \Delta  k+24 a  k+4 \Delta  k\right)
\end{aligned}
$ \\
\hline
$(2,2,0)$ & $-8 a  (a +1)^3 \left(\Delta ^2+\Delta +k^2-2 \Delta  k-k\right)$ \\
\hline
$(3,0,2)$ & $-2 a ^2 (a +1)^3 j (2 j+1)$ \\
\hline
$(3,1,0)$ & $-4 (a -1) a  (a +1)^4 \left(\Delta ^2+\Delta +k^2-2 \Delta  k-k\right)$ \\
\hline
$(3,1,1)$ & $2 a  (a +1)^3 (k-\Delta ) \left(-2 a ^2 \Delta -a ^2-3 a +2 \Delta +8 i-4 (a +1) j+2 a ^2 k-4 k-2\right)$ \\
\hline
$(3,1,2)$ & $
\begin{aligned}
& a  (a +1)^3 \left(a  \Delta ^2-4 a  \Delta +\Delta ^2-4 \Delta +16 i^2
-8 i (-\Delta +2 j+2 k+2)+4 j^2
\right.
\\ &
\left.
+j (-4 \Delta +8 k+8)+4 k^2-4 \Delta  k+8 k\right)
\end{aligned}
$ \\
\hline
$(3,2,0)$ & $-4 a ^2 (a +1)^4 \left(\Delta ^2+\Delta +k^2-2 \Delta  k-k\right)$ \\
\hline
\end{tabular}%
\caption{Nonzero components of $C_{a,b,c}^{i,j,k}$ in \eqref{eq:algebraic}.}
\label{tab:assocC}
\end{table}

\bibliographystyle{JHEP}
\bibliography{blackhole}

\end{document}